\pdfoutput=1 
\documentclass[11pt,a4paper]{article}
\usepackage{jheppub}
\usepackage{amsmath,amssymb}
\usepackage{amsmath,amsfonts}
\usepackage{xcolor,colortbl}
\usepackage{cancel}
\usepackage{mathrsfs}
\usepackage{graphicx}
\usepackage{hepunits}
\usepackage{feynmf}
\usepackage{enumerate}
\usepackage{subfig}
\usepackage{slashed}
\usepackage{placeins}
\usepackage{url}
\usepackage{soul}
\usepackage[utf8x]{inputenc}
\usepackage{lipsum,placeins}
\usepackage{comment}

\usepackage{tikz}
\usetikzlibrary{arrows,shapes.misc,decorations.markings,decorations.pathmorphing,positioning,intersections}
\tikzset{cross/.style={cross out, draw=black, minimum size=2*(#1-\pgflinewidth), inner sep=0pt, outer sep=0pt},
cross/.default={1.5mm}}
\tikzset{mydash/.style={dashed, dash pattern=on 4pt off 5pt}}
\tikzset{
  vertex/.style={draw,shape=circle,fill=black,minimum size=3pt,inner sep=0pt},
  cross/.style={cross out, draw=black,thick, minimum size=6pt, inner sep=0pt, outer sep=0pt},
  external/.style={inner sep=2pt},
  plabel/.style={inner sep=2pt},
  blob/.style={circle,fill=black!20,minimum size=0.7cm,draw,thick},
  whiteblob/.style={circle,fill=white,minimum size=1.0cm,draw,thick},
  effective/.style={rectangle,fill=black!20,minimum size=0.5cm,draw,thick},
  vev/.style={shape=vev,draw,inner sep=2pt,thick},
  mass/.style={shape=cross,draw,thick},
  rscalar/.style={dashed,thick},
  mfermion/.style={thick},
  scalar/.style={postaction={decorate}, decoration={markings,mark=at position .55 with {\arrow{latex}}},dashed,thick},
  ooscalar/.style={postaction={decorate}, decoration={markings,mark=at position .7 with {\arrow{latex}}},dashed,thick},
  fermion/.style={postaction={decorate}, decoration={markings,mark=at position .55 with {\arrow{latex}}},thick},
  majfermion/.style={postaction={decorate}, decoration={markings,mark=at position .7 with {\arrow{latex}}},thick},
  oofermion/.style={postaction={decorate}, decoration={markings,mark=at position .85 with {\arrow{latex}}, mark=at position .35 with {\arrowreversed{latex}}},thick},
  iifermion/.style={postaction={decorate}, decoration={markings,mark=at position .35 with {\arrowreversed{latex}}, mark=at position .85 with {\arrow{latex}}},thick},
  gaugeboson/.style={decorate, decoration={snake},thick},
  gluon/.style={decorate, decoration={coil,amplitude=4pt, segment length=5pt},thick},
  photon/.style={decorate, decoration={snake},thick},
  dashdot/.style={dash pattern=on .4pt off 3pt on 4pt off 3pt,thick}}

\def\eDMEFT{e\textsc{DMeft}}
\newcommand{\Sc}{{\cal S}}

\preprint{MPP-2020-91}

\begin{document}
\title{Model-independent constraints with extended dark matter EFT}

\author[a]{Tommi Alanne,} 
\author[b,c]{Giorgio Arcadi,} 
\author[a]{Florian Goertz,}
\author[a]{Valentin Tenorth,}
\author[d]{Stefan Vogl}

\affiliation[a]{Max-Planck-Institut f{\"u}r Kernphysik\\ Saupfercheckweg 1, 69117 Heidelberg, Germany}
\affiliation[b]{Dipartimento di Matematica e Fisica\\ Universit\`a di Roma 3, Via della Vasca Navale 84, 00146, Roma, Italy}
\affiliation[c]{INFN  Sezione  Roma  Tre}
\affiliation[d]{Max-Planck-Institut f{\"u}r Physik \\  F\"ohringer Ring 6, 80805 M\"unchen, Germany}

\emailAdd{tommi.alanne@mpi-hd.mpg.de}
\emailAdd{giorgio.arcadi@uniroma3.it}
\emailAdd{fgoertz@mpi-hd.mpg.de}
\emailAdd{valentin.tenorth@mpi-hd.mpg.de}
\emailAdd{stefan.vogl@mpp.mpg.de}

\abstract{We systematically
explore the phenomenology of the recently proposed extended dark matter effective field theory (\eDMEFT), which allows for a consistent effective description of DM scenarios across different energy scales. The framework remains applicable at collider energies and is capable of reproducing the correct relic abundance by including a dynamical mediator particle to the dark sector, while maintaining correlations dictated by gauge invariance in a `model-independent' way.
Taking into account present and future constraints from direct- and indirect-detection experiments, from collider searches for missing energy and for scalar resonances in vector-boson, di-jet, and Higgs-pair final states, as well as from the relic abundance as measured by Planck, we determine viable regions in the parameter space, both for scalar and pseudoscalar mediator. In particular, we point out regions where cancellations in the direct-detection cross section appear leading to allowed islands for scalar mediators that could be missed in a naive simplified-model approach, but are present in the full $D=5$ effective theory, as well
as a general opening of the parameter space 
due to consistently considering all operators at a given mass dimension. Thus, canonical WIMP-like scenarios can survive even the next generation of direct-detection experiments in different mass regimes, while potentially becoming testable at the high-luminosity LHC.}

\maketitle

\newpage
\section{Introduction}
Unveiling the nature of dark matter (DM) is one of the most important tasks in fundamental physics.
To make progress here it is essential to combine all available information coming from experiments that operate at largely different energy scales 
in a consistent framework with as little theory bias as possible. The extended dark matter effective field theory (\eDMEFT) \cite{Alanne:2017oqj}
offers a setup for the joint interpretation of DM direct detection, indirect detection and various collider searches while overcoming the drawbacks of other `model-independent' approaches to DM 
phenomenology, such as conventional DM effective field theory (EFT) \cite{Shepherd:2009sa,Beltran:2010ww,Goodman:2010yf,Goodman:2010ku,Bai:2010hh} or simplified DM models \cite{Alwall:2008ag,deSimone:2014pda,Abdallah:2014hon,Buckley:2014fba,Harris:2014hga,Abdallah:2015ter,DeSimone:2016fbz}.
It keeps both the DM particle and the mediator between the dark 
sector and the Standard Model (SM) as propagating degrees of freedom and thus remains valid at colliders, where DM EFT becomes problematic. Moreover, it is flexible enough to reproduce the correct DM abundance with a cutoff safely above the electroweak scale. 
In addition, in contrast to naive simplified models, it is a proper (order-by-order) renormalizable field theory, where higher-dimensional operators allow to incorporate effects from additional new physics (NP), and gauge invariance (including induced correlations) stays intact, while the stringent (model-dependent) connections between different observables that arise in second generation simplified models such as 2HDM+(pseudo)scalar constructions \cite{Bauer:2017ota,Abe:2018bpo,Arcadi:2020gge} or realistic $Z'$ models \cite{Duerr:2016tmh,Ismail:2016tod}
can be lifted.
In fact, it is conceivable that the new sector is rather rich, while both the DM field and the mediator are significantly lighter than the remaining NP states,
which justifies to capture effects of the latter via higher-dimensional operators in the \eDMEFT. Interestingly, for fermionic dark matter and a scalar or pseudoscalar mediator, the leading corrections appear at $D=5$. Compared to the Standard-Model EFT, where all contributions except the Weinberg operator for neutrino mass arise at $D\geq6$  \cite{Buchmuller:1985jz,Hagiwara:1993ck,Grzadkowski:2010es,deFlorian:2016spz}, this leads to a drastically reduced number of free coefficients.

In this work, we comprehensively explore the phenomenology of the \eDMEFT\ framework, considering first the predicted relic abundance and present and future constraints from direct- and indirect-detection experiments. These are then confronted with bounds from collider searches, where we take into account monojet searches for missing energy, but also resonances searches, which are directly sensitive to the mediator particle, including vector-boson, di-jet, and di-Higgs final states.
We in turn determine viable areas in the parameter space, both for scalar and pseudoscalar mediator, and identify regions where cancellations in the direct-detection cross section appear. These can lead to allowed spaces for scalar mediators that could be missed in a simplified-model approach but are present in the \eDMEFT\ and could result in LHC discoveries.

This article is organized as follows. In sec.~\ref{sec:S}, we first briefly review the \eDMEFT\ for scalar mediator and then survey, in sec.~\ref{sec:DMS}, its DM phenomenology, including the relic abundance and direct-detection signatures, where we provide expressions for the relevant cross sections. In sec.~\ref{sec:ColS} we turn to collider observables and give already exclusions for the final states listed above. Consequently, in an increasing level of complexity we systematically explore, in sec.~\ref{sec:ResS}, the parameter space of the \eDMEFT\ taking into account all relevant constraints discussed so far. Here we start with simple, portal-like combinations of operators present within the framework and end with scans approaching the full \eDMEFT. We repeat our analysis for a pseudoscalar mediator in sec.~\ref{sec:PS} before  presenting our conclusions in sec.~\ref{sec:Conc}.

\section{eDM{\scriptsize EFT} for a scalar mediator}
\label{sec:S}

The EFT for the case of a (Dirac) fermionic DM state, $\chi$, 
and a scalar mediator, $\Sc$, is described, at $D\leq 5$, by the Lagrangian \citep{Alanne:2017oqj}
\begin{align}
\label{eq:lagrangian_scalar}
 \mathcal{L}_{\rm eff}^{\Sc \chi} &=  \mathcal{L}_{\rm SM} +
    \frac{1}{2}\partial_\mu \Sc \partial^\mu \Sc 
    - \frac 1 2 \mu_S^2 {\Sc}^2 + \bar \chi i \slashed{\partial}\chi- m_\chi \bar \chi \chi \nonumber \\
& +\lambda_{S1}^{\prime} v^3 {\Sc}-\frac{\lambda^\prime_{S}}{2 \sqrt 2} v {\Sc}^3 - \frac{\lambda_{S}}{4} {\Sc}^4 
-\lambda^\prime_{HS} v |H|^2 {\Sc} 
	- \lambda_{HS} |H|^2 {\Sc}^2 \nonumber \\
    &- y_S \Sc  
	\bar \chi_L \chi_R -\frac{y_S^{(2)} {\Sc}^2 + y_H^{(2)} |H|^2 
	}{\Lambda} 
	\bar{\chi}_L \chi_R +\mathrm{h.c.}\\
	&- \frac{\Sc}{\Lambda} \left[ c_{\lambda S} {\Sc}^4 
    +c_{HS} |H|^2 \Sc^2 
    +c_{\lambda H} |H|^4 \right] \nonumber\\
    &- \frac{\Sc}{\Lambda} \left[(y_d^S)^{ij} \bar{Q}_{\mathrm{L}}^i H d_{\mathrm{R}}^j 
	+ (y_u^S)^{ij}\bar{Q}_{\mathrm{L}}^i\tilde{H}u_{\mathrm{R}}^j+(y_\ell^S)^{ij} \bar{L}_{\mathrm{L}}^i H \ell_{\mathrm{R}}^j +\mathrm{h.c.}\right]\nonumber\\
	&-\frac{\Sc}{\Lambda}\left[C_{BB}^S B_{\mu\nu} B^{\mu\nu}+ C_{WW}^S W^{I\mu\nu} W_{\mu\nu}^I +C_{GG}^S G^{a\mu\nu}G_{\mu\nu}^a\right]\,. \nonumber
\end{align}
Here, ${\cal L}_{\rm SM}$ is the renormalizable Standard Model (SM) Lagrangian and $H\!=\!\frac{1}{\sqrt{2}}{(0,v\!+\!\varphi)}^T$ the SM Higgs doublet (in unitary gauge) with $v=246\,$GeV the vacuum expectation value (vev). 
Moreover, $Q_L$, $u_R$ and $d_R$  ($L_L$ and $l_R$) denote the left-handed and right-handed quarks (leptons), while the field-strength tensors of SM gauge group before electroweak symmetry breaking are written as $B_{\mu\nu}$, $W^I_{\mu\nu}$ and $G^a_{\mu\nu}$.  The generic mass-suppression scale of the higher-dimensional operators is parametrized by $\Lambda$, and each operator is associated with a coefficient that fixes its interaction strength.\footnote{Note that $\Lambda$ should not be included when counting the number of free parameters of the EFT since it always appears in combination with the coefficients.} We assume that $\Sc$ does not develop a vev and remain agnostic about the origin of the new physics scale. Moreover, we neglect an invariant term $\bar \chi_L \sigma_{\mu\nu} \chi_R B^{\mu\nu}$, which would lead to direct DM interactions with photons and $Z$ bosons and  therefore is phenomenologically constrained to be tiny.
Clearly this Lagrangian is not limited to a description of DM, and a scalar-singlet extended SM EFT forms a subset of the \eDMEFT. We will focus mainly on the implications for DM; a recent detailed study of the EFT for a pure scalar singlet extension can be found in \cite{Adhikari:2020vqo}
(see also \cite{Kamenik:2016tuv,Franceschini:2016gxv,Gripaios:2016xuo,Carmona:2016qgo,Bauer:2016hcu} for earlier works on the singlet-extended SM EFT).

In the following we will comment on the terms that will turn out to be most relevant for the phenomenology. 
There are already some new physics contributions at $D\leq 4$. Of particular interest among them are the $\bar{\chi}\chi \Sc$ Yukawa term that couples the DM to the mediator, and the interactions between $\Sc$ and $H$. Together these interactions provide a minimal connection between the DM and the SM.
In addition there are a number of higher-dimensional operators that couple SM fields to the dark sector. They can be separated into three broad subgroups:
First,
there are the $\bar{\chi}\chi \Sc^2$ and the $\bar{\chi}\chi |H|^2$ terms.   Structurally these terms appear very similar but their phenomenological consequences are going to be very different. While the first mediates an additional interaction with $\Sc$ that might change the dynamics within the new physics sector, the second term provides a direct link between the DM and the SM which circumvents the mediator completely. 
Next, there are new physics extensions of the SM Yukawa interactions that couple $\Sc$ and $H$ to SM fermions.
Allowing the most general flavor structure leads to a large number of operators of this type. 
Unless stated otherwise, we will nevertheless assume the matrices $y_{U,D,S}^S$ to be diagonal in the basis of diagonal SM-Yukawa couplings:
\begin{align}
   & (y_u^S)^{ij} \to \mbox{diag}\left(y_u^S,y_c^S,y_t^S\right)\nonumber\\ 
   & (y_d^S)^{ij} \to \mbox{diag}\left(y_d^S,y_s^S,y_b^S\right)\nonumber\\
   & (y_\ell^S)^{ij} \to \mbox{diag}\left(y_e^S,y_\mu^S,y_\tau^S\right)
\end{align}
in order to avoid the insurgence of dangerous flavor violation.
Motivated by minimal-flavor-violation (MFV) \cite{DAmbrosio:2002vsn} we further impose that our diagonal Yukawa-like matrices reproduce the hierarchy of the SM-fermion masses and thus follow the relation $y_f^S\propto \frac{m_f}{v}$. 
Finally, there are 
the couplings between the mediator and the SM gauge bosons. Typically, interactions of this kind arise at the loop level in theories with additional matter fields charged under the SM gauge group. With this UV completion in mind, we can extract the loop factors and the gauge coupling from the corresponding Wilson coefficients
\begin{align}
\label{eq:loop_coefficient}
C_{GG}^S &=\frac{1}{16\pi^2}\ g_s^2 c_{G}^S\nonumber\\
C_{BB}^S &=\frac{1}{16\pi^2}\ g^{\prime 2} c_{B}^S \nonumber\\
C_{WW}^S &=\frac{1}{16\pi^2}\ g^2 c_{W}^S\,.
\end{align}

Moreover, for the phenomenological study presented below, it is convenient to use the linear combinations that correspond to effective couplings with $W^{+}W^{-}$, $ZZ$, $Z\gamma$ and $\gamma \gamma$ states. They read
\begin{align}
    C^S_{W^+W^-} &=2 C^S_{WW}\nonumber\\
    C^S_{ZZ} &=c_W^2 C^S_{WW}+s_W^2 C^S_{BB}\nonumber\\
    C^S_{Z\gamma} &=2 c_W s_W (C^S_{WW}-C^S_{BB})\nonumber\\
    C^S_{\gamma \gamma} &=s_W^2 C^S_{WW}+c_W^2 C^S_{BB}\,,
\end{align}
where $c_W\equiv \cos\theta_W$ and $s_W\equiv \sin\theta_W$, with $\theta_W$ being the Weinberg angle.

In the following, we will explore the parameter space spanned by the Lagrangian~\eqref{eq:lagrangian_scalar} and identify interesting regions that could be missed in conventional setups taking into account the relevant experimental constraints. We will make the FeynRules implementation of the \eDMEFT\  model used for this study publicly available at~\cite{eDMEFT_FeynRules}.

\subsection{DM phenomenology}
\label{sec:DMS}

Before illustrating the main aspects of DM phenomenology for the \eDMEFT\ with scalar mediator in the next subsections, we briefly review a well known effect of the $D=4$ operators, namely the Higgs-mediator mixing induced by $\lambda_{HS}^\prime$.

After electroweak symmetry breaking, the trilinear coupling $\lambda_{HS}^\prime$ induces an off-diagonal contribution in the scalar mass matrix which leads to mixing between the SM Higgs field, $\varphi$, and $\Sc$. This mixing can be described by an angle $\theta$ defined by:
\begin{align}
\label{eq:rotation}
    \left( \begin{array}{c} h \\ S \end{array} \right) =  
    \left( \begin{array}{cc} \cos\theta & \sin\theta \\ -\sin\theta & \cos\theta \end{array} \right)
    \left( \begin{array}{c} \varphi \\ \Sc \end{array} \right),
\end{align}
with
\begin{equation}
\label{eq:mixinghS}
    \tan 2\theta=\frac{2 \lambda_{HS}^\prime v^2}{M_{\varphi}^2 -M_{\Sc}^2}\,,
\end{equation}
where $\lambda_H$ is the coefficient of the quartic Higgs operator and $M_{\varphi}$ and $M_{\Sc}$ denote the masses of the scalar fields in the absence of mixing.
The two physical masses are given by
\begin{equation}
    m_{h/S}^2 = \frac{1}{2}(M_{\varphi}^2 +M_{\Sc}^2) \mp \frac{M_{\Sc}^2-M_{\varphi}^2}{2\,\cos 2\theta} \,.
\end{equation}
We identify $h$ as the SM-like Higgs state and do not make any assumptions about the ordering of the scalar mass eigenstates.

The mixing in combination with the $y_S$ Yukawa  will generate a coupling between the DM and the SM sector, described by the Lagrangians
\begin{align}
    & \mathcal{L}^{\rm mix}_{\rm SM}=({h c_\theta - S s_\theta}) \left[\frac{2 M_W^2}{v} W^{+}_\mu W^{\mu -}+ \frac{M_Z^2}{v} Z^\mu Z_\mu-\sum_f \frac{m_f}{v} \bar f f \right],\nonumber\\
    & \mathcal{L}^{\rm mix}_{\rm DM}= - ({h s_\theta + S c_\theta}) y_S \bar \chi \chi\,,
\end{align}
as well as a Lagrangian for the trilinear couplings between the scalar fields 
\begin{equation}
    \mathcal{L}_{\rm scal}^{\rm mix} = -\frac{v}{2} \bigg[ \kappa_{hhh}~h^3+ \kappa_{hhS} s_\theta~h^2 S+ \kappa_{hSS} c_\theta ~h S^2 + \kappa_{SSS}~ S^3 \bigg]\,,
\end{equation}
where ${c_\theta\equiv\cos\theta},\, {s_\theta\equiv\sin\theta}$.
The explicit results for the couplings $k_{ijk}$ can be straightforwardly derived, see e.g.~\cite{Arcadi:2017kky}.  This represents a  peculiar case, since it can be realized from renormalizable interactions. As these interactions are not suppressed by the scale of the higher-dimensional operators, it is natual to assume that they will generically dominate over effects that arise at $D=5$. However, the  
LHC Higgs data constrains this kind of interaction severely \cite{Robens:2015gla,Falkowski:2015iwa}. 
We will therefore treat mixing and effective operators on the same footing and include both in our analysis.

With the above expressions at hand, we now turn to the DM phenomenology of the scenario
beginning with the relic abundance obtained within the \eDMEFT.

\subsubsection{Relic density}

Throughout this work we assume that the DM was produced by thermal freeze-out. 
An approximate condition to generate the correct relic density is a thermally averaged annihilation cross section $\langle \sigma v\rangle \approx 2 \times 10^{-26}{\mbox{cm}}^3 {\mbox{s}}^{-1} \equiv\!\sigma_v^0$\, \cite{Steigman:2012nb}. In the following we will briefly discuss the most relevant annihilation channels and give some approximate results for the cross section using the velocity expansion $\langle \sigma v \rangle \approx a+b v_\chi^2$ in order to build up some intuition for the most relevant contributions. Note that the velocity expansion is not a reliable approximation in some phenomenologically relevant regimes~\cite{Griest:1990kh}; for our numerical study we solve the freeze-out equations in the full model with  \verb|micrOMEGAs|~\cite{Belanger:2013oya} and do not rely on these analytic estimates. 

A number of channels can contribute significantly to the annihilation rate.
The DM can annihilate into SM fermions through $s$-channel exchange of the mediator field $S$ or the Higgs.  The effective operator induced cross sections scale as $y_S^2 (y_f^S)^2 v^2/\Lambda^2$ or $(y_H^{(2)})^2 y_f^2 v^2/\Lambda^2$, where $y_f$ is the corresponding SM-like Yukawa coupling. 
Neglecting the Higgs portal interaction and mixing, the leading contribution in the velocity expansion is
\begin{align}
    \langle \sigma v\rangle_{ff} &\approx \frac{N_c }{8\pi}\, \frac{ v^2}{\Lambda^2}\, \frac{y_S^2 (y^S_f)^2 m_\chi^2}{(m_S^2-4 m_\chi^2)^2}\ v_\chi^2\,\nonumber\\
    &\approx \left \{
    \begin{array}{cc}
    2.5 \times 10^{-3} \,\sigma_v^0 N_c \Big(\frac{3\,{\rm TeV}}{\Lambda}\Big)^2 \Big(\frac{m_\chi}{100\,{\rm GeV}}\Big)^2 {\left(\frac{500\,{\rm GeV}}{m_S}\right)}^4 y_S^2 (y_f^S)^2,      &m_\chi \ll \frac{m_S}{2}\\
    0.1\, \sigma_v^0 N_c \Big(\frac{3\,{\rm TeV}}{\Lambda}\Big)^2 \Big(\frac{100\,{\rm GeV}}{m_\chi}\Big)^2 y_S^2 (y_f^S)^2,    &m_\chi \gg \frac{m_S}{2} \,,
    \end{array} \right.
\end{align}
where we used $v_\chi^2 \approx 0.1$. 
If the coefficients $C_{VV}^S,\, V=G,B,W$, (or $y_H^{(2)}$) are sizable, the DM can also annihilate, through $s$-channel exchange of $S$ (or the Higgs), into gauge bosons. 
Taking annihilations into gluon pairs as an example, the cross section can be numerically estimated as
\begin{align}
    \langle \sigma v \rangle_{gg} &\approx \frac{1}{\pi \Lambda^2}\, \frac{ (C_{GG}^S)^2 y_S^2  m_\chi^4}{(m_S^2-4 m_\chi^2)^2} v_\chi^2\nonumber\\ 
    & \approx \left \{ \begin{array}{cc}
    3.2 \times 10^{-3}  \, \sigma_v^0\, \Big(\frac{3\,{\rm TeV}}{\Lambda}\Big)^2 \Big(\frac{m_\chi}{100\,{\rm GeV}}\Big)^4 {\left(\frac{500\,{\rm GeV}}{m_S}\right)}^4 \left(\frac{\alpha_S c^S_{G}}{4\pi}\right)^2\, y_S^2 \,,    &  m_\chi \ll \frac{m_S}{2} \\
    0.1\, \sigma_v^0\, \Big(\frac{3\,{\rm TeV}}{\Lambda}\Big)^2 \Big(\frac{\alpha_S c^S_{G}}{4\pi}\Big)^2\, y_S^2\,,    & m_\chi \gg \frac{m_S}{2} \,.
    \end{array}\right.
\end{align}  
From the equations above we see that the annihilation cross section into gluons remains in general below the thermally favored value, unless a rather low scale and sizable couplings enhance the annihilation rate considerably.

As the DM mass increases, the $hh$, $hS$ and $SS$ channels become kinematically allowed.
These are particularly interesting since these annihilations can be realized even at $D=4$. Therefore, the cross section is not suppressed by $\Lambda$. For example, the minimal contribution to the annihilation into $SS$ final states is given by
\begin{equation}
    \langle \sigma v \rangle_{SS} \approx \frac{3}{64 \pi}\, \frac{y_S^4}{m_\chi^2}\, v_\chi^2 \approx 87.5\, \sigma_v^0\, {\left(\frac{100\,{\rm GeV}}{m_\chi}\right)}^2\, y_S^4
\end{equation}
for $m_S \ll m_\chi$. The other scalar final states can also be realized without higher-dimensional operators. The corresponding cross sections are, however, proportional to powers of $s_\theta$, which is suppressed by the constraints from Higgs physics. 

Finally, an intriguing new option arises due to the presence of 
the coupling $y_S^{(2)}$, which allows for the annihilation
into mediators without inducing $s$-channel interactions with SM
particles in the presence of $C_{VV}^S,\,y_f^S,$ or $\lambda_{HS}^\prime$. For $m_S\ll m_\chi$ the corresponding annihilation cross section is approximately
\begin{equation}
    \langle \sigma v \rangle_{SS} \approx \frac{1}{64 \pi} \left(\frac{y_S^{(2)}}{\Lambda}\right)^2 v_\chi^2 \,\simeq 3.2 \times 10^{-2}\, \sigma_v^0\, {\left(\frac{3\,{\rm TeV}}{\Lambda}\right)}^2 (y_S^{(2)})^2.
\label{eq:xsec_XXSS}
\end{equation}
Even though this contribution is naturally suppressed by a factor $m_\chi^2/\Lambda^2$ relative to the one from the $\bar{\chi} \chi S$ coupling, it can dominate the annihilation channels for $y_S < 1$.  

We notice that for all annihilation channels, the first non-zero term in the cross section is the $p$-wave ($\propto v_\chi^2$) one. 
This leads to a generic suppression of the DM annihilation cross section, since $v_\chi^2 \approx 0.1$ at freeze-out, so that the observed relic density requires larger couplings than in the $s$-wave case. 
Another important remark is the overall scaling of the cross sections with $1/\Lambda^2$, with the notable exception of the annihilation into $SS$ final states. 
The cutoff suppression originates from the coupling of the DM with the Higgs boson or from the coupling of the scalar $S$ to the SM fermions or gauge bosons. We notice here a very relevant difference between the \eDMEFT\ setup and the `simplified' models. In the former case a gauge invariant construction imposes a Higgs insertion or some mixing between the scalar mediator and the Higgs. This implies a suppression factor, proportional to $v/\Lambda$ or a mixing angle, for the couplings of the mediator with SM fermions. In contrast, simplified models frequently consider couplings of arbitrary size (limited only by perturbativity) between the mediator and SM fermions. This has relevant phenomenological implications since it implies that DM simplified models allow for (spuriously) larger annihilation cross-sections of the DM into SM fermions, which does not happen in an appropriately used EFT.
An enhancement of the annihilation cross section occurs also in the pole region, $m_\chi \simeq m_S/2$. This regime is best investigated numerically since the velocity expansion is not a good approximation here.

\subsubsection{Direct detection}

One of the  main constraints on DM comes from direct-detection (DD) experiments that test the strength of the DM-nucleus interaction. Conventional DD experiments aim to observe the recoil of a nucleus hit by a DM particle in a low-background environment; for a detailed introduction to DD see for example \cite{Undagoitia:2015gya}. Typical detectors of this kind, e.g. XENON1T \cite{Aprile:2018dbl}, DEAP-3600 \cite{Ajaj:2019imk}, PandaX \cite{Cui:2017nnn}, LUX \cite{Akerib:2016vxi} or DarkSide \cite{Agnes:2018fwg}  are mostly sensitive to  spin-independent (SI) interactions of DM with nucleons.   These are  induced at the microscopic level by diagrams with $t$-channel exchange of $h$ and $S$ between DM and the constituents of the nucleons, i.e.  quarks and gluons. The resulting DM-proton cross section can be written as
\begin{equation}
\label{eq:DDfull}
    \sigma_{\chi p}^{\rm SI}=\frac{\mu_{\chi p}^2}{\pi}\,  \frac{m_p^2}{\Lambda^2} \left[\sum_{q=u,d,s,c,b,t} f_q^p\left(\frac{g_{H\chi \chi} g_{Hqq}}{m_h^2}+\frac{g_{S\chi\chi} g_{Sqq}}{m_S^2}\right)
    -\frac{2}{9} \frac{g_{S\chi \chi} c_{G}^S}{m_S^2} f_{TG}\right]^2\,,
\end{equation}
where $\mu_{\chi p}$ is the reduced mass of the DM-proton system and $m_p$ is the proton mass. We have adopted generic expressions for the couplings of the $h,S$ states with pairs of DM particles and SM quarks. In absence of mixing between the $h$ and $S$ fields they simply read $g_{H\chi \chi}=y_H^{(2)}$, $g_{S \chi \chi}=y_S$, $g_{Hqq}=1$ and $g_{Sqq}=y_q^S \frac{v}{\sqrt{2} m_q}$, while they will be more complicated in the presence of mixing. 
The parameters $f_q^p$ are the structure functions of the proton with $f_c^p=f_b^p=f_t^p=\frac{2}{27}f_{TG}$, $f_{TG}=1-\sum_{q=u,d,s}f_q^p$ . We have adopted the default assignation of \verb|micrOMEGAs|~\cite{Belanger:2013oya} for these coefficients. The cross section for interactions with neutrons can be obtained by replacing the proton with the neutron mass and substituting the appropriate values for the structure functions $f^n$ instead of $f^p$ in the expression above. 

The different terms in the SI cross section, eq.~\eqref{eq:DDfull}, do not have the same sign. Therefore, destructive interference between the different contributions is possible. A perfect cancellation leads to a so-called blind spot in which DD experiments are unable to probe the DM. While blind spots are known in the DM literature, so far they have been found when combining different types of mediators~\cite{Cheung:2012qy,Huang:2014xua,Berlin:2015wwa,Choudhury:2017lxb}, for example in supersymmetry. Here we point out a new kind of blind spot in the EFT context, where the effect is more subtle (i.e.~between different operators featuring only one type of mediator).

It is instructive to consider the conditions for the occurrence of blind spots for simple limiting cases. For example, in the absence of mixing and with a Higgs-DM coupling $g_{H\chi\chi}=0$ and a scalar mediator coupled only with the gluons and the top quark, we arrive at a blind spot for
\begin{equation}
    c_{G}^S=\frac{y_t^S v}{3 \sqrt{2}m_t}\,.
\end{equation}
The case in which the couplings of the scalar mediator with the SM quarks are exclusively induced by the mixing with the Higgs doublet is only slightly more complicated and we arrive at
\begin{equation}
    c_{G}^S=\sqrt{2}\ \frac{f_u^p+f_d^p+f_s^p+\frac{2}{27}f_{TG}}{f_{TG}}\, \frac{\Lambda}{v}\, \frac{\left(m_h^2-m_S^2\right)c_\theta s_\theta}{m_h^2 c_\theta^2-m_S^2 s_\theta^2} \,.
\end{equation}
If only mixing is present a natural blind spot also arises for $m_S=m_h$.
Blind spots can also be realized in more general scenarios but the analytic conditions become cumbersome and do not add significantly to the understanding. Therefore, we do not report them explicitly.

\subsubsection{Indirect detection}

Indirect detection (ID) is searching for cosmic rays and photons produced by residual DM annihilations happening in the Universe today. 
In order to assess the potential implications of this kind of search for DM, the velocity expansion of the annihilation cross section is of great interest. The typical velocity of DM in astrophysical structure today is $\mathcal{O}(10^{-3})$ whereas the typical velocity at freeze-out is $\approx 0.3$. Consequently, higher-order terms in the velocity expansion are strongly suppressed today, and only $s$-wave annihilations lead to an annihilation rate in the ballpark of the canonical cross section for a thermal relic of $\langle \sigma v \rangle \approx 2\times 10^{-26} \;\mbox{cm}^3/\mbox{s}$.  

As already pointed out, all the relevant DM annihilation channels of the scalar \eDMEFT\ feature a velocity-dependent annihilation cross section. Consequently, for a scalar mediator, the impact of ID constraints is expected to be marginal and will not be considered in this section.
We will discuss ID in greater detail in sec.~\ref{sec:PS}.

\subsection{Collider signals}
\label{sec:ColS}

The framework under consideration is characterized by different kinds of collider signatures whose relative relevance depends on the values of the different effective couplings. Most of these signatures are associated to the (single) production and subsequent decay of the mediator, $S$. 
Taking the MFV-inspired ansatz detailed above for the dimension-five couplings of $S$ with the SM quarks, the main production channels of the new mediator will be gluon fusion through the effective coupling $C^S_{GG}$, through mixing with the Higgs boson, or via a $y_t^S$ induced top loop. Production through vector-boson fusion (VBF) would also be possible in the presence of sizable $C_{BB}^S$ and $C_{WW}^S$ couplings.
The process of resonant production, through gluon fusion, and subsequent decay of $S$ can be approximated through the expression
\begin{equation}
    \sigma (pp \to ij)=\mathcal{K}\ \frac{\pi^2}{8s\, m_S}\ I_{GG}(m_S)\, \Gamma(S \to gg)\, \mbox{Br}(S\to ij) \,.
    \label{eq:res_production}
\end{equation}
The decay width for the effective contact interaction is given by 
\begin{equation}
    \Gamma(S \to gg)= \frac{2 m_S^3}{\pi \Lambda^2}\, \left(\frac{\alpha_s c_{G}^S}{4 \pi}\right)^2 \,,
\end{equation}
and for the top loop by
\begin{equation}
    \Gamma(S\to gg) = \frac{\alpha_s^2}{16\pi^3}\ m_{S} \frac{v^2}{\Lambda^2}\, \sum_q (y^S_q)^2 \, F_S\left(\tau_{q,S}\right) ,
    \label{eq:loop_width}
\end{equation}
where  $\tau_{q,S}=4m_q^2/m_S^2$ and the loop function agrees with the one familiar from SM-Higgs physics~\cite{Gunion:1989we,Djouadi:2005gi}
\begin{equation}
    F_S(x) = x^2 \left|1+(1-x)\arctan^2\frac{1}{\sqrt{x-1}} \right|^2 \,.
\end{equation}
The structure of the proton is taken into account by the parton luminosity factor
\begin{equation}
    I_{GG}=\int_{\frac{m_S^2}{s}}^1 \frac{dx}{x}f_G (x) f_{G}\left(\frac{m_S^2}{s x}\right)\,,
\end{equation}
where $f_G$ is the parton-distribution function (PDF) for gluons in the proton, $\sqrt{s}$ is the center of mass energy (13 TeV for the current LHC run), while the factor $\mathcal{K}$ accounts for eventual higher-order corrections. We take $\mathcal{K}= 1.5$ which is known to describe the NLO corrections to Higgs production with masses in the range $100-1000$ GeV to good accuracy~\cite{Spira:1995rr}.

After production, the particle $S$ can decay into four classes of final states: (i) DM pairs; (ii) pairs of SM fermions; (iii) pairs of gauge bosons; (iv) $hh$. 
The DM pair production processes can be tagged only if accompanied by additional radiation, namely the so called mono-$X$ events. In our study we will mainly focus on mono-jet events, i.e.~emission of gluons or quarks in the initial state (while interesting correlations might be probed by exploring mono-Higgs signals \cite{Alanne:2017oqj}).\footnote{The bi-quadratic portal $(y_S^{(2)})^2$ can also give rise to interesting di-jet+DM signatures; while being out of the reach of current LHC runs they can potentially be probed in upcoming collider experiments \cite{Goertz:2019vht}.} Mono-$Z$ events originate from dimension six operators with an additional momentum dependence (and cutoff suppression), which are beyond the truncation of our EFT approach.
These mono-$X$ processes are actively studied by the LHC collaborations, see~\cite{Aaboud:2017phn,Sirunyan:2017jix} for the most recent results from ATLAS and CMS, and~\cite{Aaboud:2017yqz,Sirunyan:2018gdw,Sirunyan:2019zav} for mono-Higgs searches. Corresponding limits are, however, customarily interpreted in terms of the simplified models~\cite{Abdallah:2014hon,Abdallah:2015ter,Boveia:2016mrp,Albert:2017onk}.

\begin{figure}[t]
 \begin{center}
 \includegraphics[width=.45\textwidth]{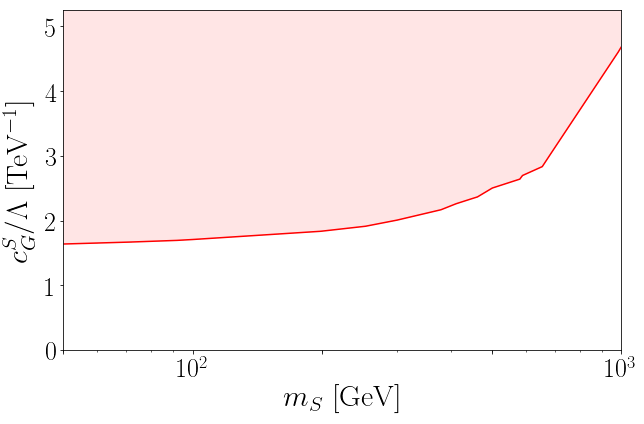}\quad
 \includegraphics[width=.45\textwidth]{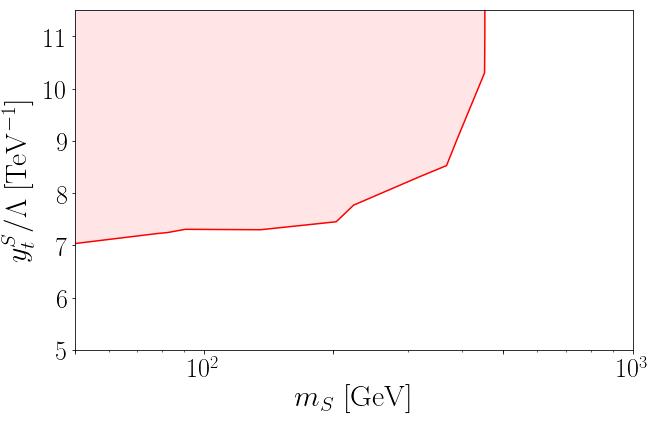}
 \end{center}
 \caption{\footnotesize Exclusion limits from the ATLAS mono-jet search \cite{Aaboud:2017phn} in the $m_S -c_{G}^{S}$ plane (left) and $m_S -y_t^{S}$ plane (right), normalized to $\Lambda$. In both cases $y_{S}=1$, $m_\chi=10$ GeV, and all other couplings are equal to zero.}
 \label{fig:monojscalar}
\end{figure}

In order to obtain bounds for the scenario under considerations, we thus needed to recast the bounds from mono-jet searches. To this end, we implemented the \eDMEFT\ setup in the event generator MadGraph5\_aMC@NLO $2.6.3$~\cite{Alwall:2014hca}.\footnote{We only simulated events with the emission of one hard jet.} The simulated events have been processed through CheckMATE $2.0.26$~\cite{Dercks:2016npn,Cacciari:2011ma,Cacciari:2005hq,Cacciari:2008gp,Read:2002hq}, linked with PYTHIA 8.1~\cite{Sjostrand:2006za,Sjostrand:2007gs} for the parton showering, and Delphes 3~\cite{deFavereau:2013} for a fast detector simulation employing the mono-jet search performed by ATLAS in 2017~\cite{Aaboud:2017phn}. In this search, twenty signal regions binned in terms of the missing transverse energy (MET) are defined. To ensure the validity of our EFT approach only the `exclusive signal regions’ (EM) with MET$<500$~GeV are used. It turns out that for most of the considered values of $m_S$, EM4 with MET~$=(400-500)$~GeV gives the strongest constraints.

As an illustration, we show in fig.~\ref{fig:monojscalar} the exclusion limits from mono-jet searches as a function of the mediator mass for a fixed value of $m_\chi=10$ GeV,\footnote{This choice does not harm the generality of the results, since we have verified that the experimental sensitivity is basically independent of the DM mass as long as it is not close to the $m_S \sim 2 m_\chi$ threshold. For heavier DM, no robust constraints can be placed since the inferred couplings are not perturbative.} assuming that the mediator exclusively couples to top quarks via the $y^S_t$ Yukawa portal (right panel) or to gluons through the effective $c_{G}^S$ portal (left panel).
Thus, the exclusion limits can be easily expressed in terms of the dimensional ratios $c_{G}^S/\Lambda$ and $y_t^S/\Lambda$.  
We find that, even for $m_\chi < m_S/2$ the region of parameter space which can be probed through current mono-jet searches is rather limited\,---\,typically allowing only $\mathcal{O}(1)$ couplings to be tested. In addition the sensitivity to the top-coupling decreases significantly for $m_S\geq 2 m_t$  due to the growing partial width $\Gamma(S\to t\bar t)$, suppressing the Br$(S\to\chi\chi)$, as shown in the right panel of fig.~\ref{fig:monojscalar}.

If the coupling $y_H^{(2)}$ is non-zero or there is sizable mass mixing between $S$ and the Higgs, DM pairs can also be produced from interactions with the Higgs bosons. Collider limits have been determined through searches of invisible decays of the Higgs, hence effective only for $m_\chi \leq m_h/2$ (see ref.~\cite{Arcadi:2019lka} for a discussion of possible prospects for $m_\chi > m_h/2$). The most recent analysis~\cite{Aaboud:2019rtt}, combining the 7--8 TeV and 13 TeV data and different signal topologies, leads to $\mbox{Br}(h \rightarrow \chi \chi)< 0.26$ (for recent reviews on searches on invisible Higgs see e.g. refs~\cite{Dawson:2018dcd,Arcadi:2019lka}).

Concerning category (ii), we note that under the assumption of a SM-Yukawa like structure for the dimension-five couplings of the scalar mediator with the SM fermions, we do not expect sizable signals at colliders from decays of the new states into SM fermions. 
 In particular, the current limits on $t\bar{t}$ resonances \cite{Aaboud:2018mjh} are too weak to constrain the range of couplings considered here.
The prospects for direct searches of decays of $S$ into SM fermion pairs are similarly poor in the case of mixing between $h$ and $S$, once compatibility with measurements of the Higgs signal strengths is required. 

Moving to category (iii), the most promising searches are, as we will see further below, those for EW gauge boson pairs. For our study we have applied the latest results from searches for $WW/ZZ$~\cite{Aaboud:2017itg,Aaboud:2017fgj,Aaboud:2017gsl} and diphoton~\cite{Aaboud:2017yyg,Sirunyan:2018wnk} resonances.\footnote{In principle searches for $Z\gamma$ resonances should be considered as well. Most recent analyses~\cite{Sirunyan:2017hsb,Aaboud:2018fgi}, however, consider masses of the scalar resonance above 1 TeV. As will be clarified in the following, these high values are not part of our analysis.} We remark that while a sizable diphoton signal relies basically on the presence of $D=5$ couplings of the scalar mediator with gauge bosons, a detectable $WW/ZZ$ signal can also be generated in presence of non-negligible $h-S$ mixing. As pointed out in e.g.~\cite{Falkowski:2015iwa}, $WW/ZZ$ searches provide the strongest constraints on the mixing angle, $\theta$, for $m_S > m_h$. In addition to EW gauge boson pair signatures we also consider limits from dijet signals~\cite{Sirunyan:2018xlo}, possibly originating from the decay of the resonance into gluon pairs. All these constraints will be implemented when exploring the \eDMEFT\ parameter space below.

Finally, the last category of signals from $S$ decay, i.e. $hh$ final states, arises for sizable values of the $\lambda_{HS}$ or $\lambda'_{HS}$ couplings, and thus, in particular, in the presence of significant $h-S$ mixing. We consequently include in our analysis limits on di-Higgs production, considering the 4b~\cite{Aaboud:2018knk,Sirunyan:2018tki}, $bbWW$~\cite{Aaboud:2018zhh}, $bb\tau \tau$~\cite{Aaboud:2018sfw}, $\gamma \gamma WW$~\cite{Aaboud:2018ewm} and $\gamma \gamma bb$~\cite{Aaboud:2018ftw} final states. A combination of the individual constraints has been given in~\cite{Sirunyan:2018two,Aad:2019uzh}. 

The collider searches just illustrated are sensitive mostly to heavy masses of the mediator, namely above the mass of the SM Higgs. Since we also consider the case of a light mediator, we do include bounds from searches for a low mass Higgs at LEP \cite{Barate:2003sz, Abreu:1990bq} as well as constraints from b-physics \cite{Aaij:2012vr,Wei:2009zv}. We have, furthermore, imposed that the sum of branching ratios of $h \rightarrow SS$, if kinematically allowed, and of the eventual $h\rightarrow \chi \chi$ decay does not exceed the constraint from the invisible width of the Higgs.

\subsection{Combined results}
\label{sec:ResS}

The general Lagrangian, eq.~\eqref{eq:lagrangian_scalar}, includes three new mass scales, i.e $m_\chi, m_S$ and $\Lambda$, and several new couplings. In order to avoid an excessively high dimensionality of the parameter space, unless differently stated, we have adopted the following simplifying assumptions. First of all we have set the scalar couplings $\lambda_S,\lambda_S^\prime, \lambda_{HS}$ (and the $D=5$ potential terms), which are expected to have a negligible impact on our analysis, to zero.
Moreover, as already pointed out we will mostly adopt a flavor-diagonal ansatz for the $D=5$ couplings of $S$ to the SM fermions following $y_f^S=c_S y_f$. Similarly we have typically assumed a single free parameter $c_{G}^S=c_{B}^S=c_{W}^S=c_{V}^S$ describing the couplings of the scalar mediator with the gauge bosons. Finally, we will neglect the effective Higgs-DM interaction \mbox{$\sim y_H^{(2)}$} for most of the analysis, since it is very well studied and stongly constrained by DD \cite{LopezHonorez:2012kv, Balazs:2017ple,Arcadi:2019lka}.
In summary the parameter space of the theory is spanned by the set $(m_\chi,\, m_S,\, y_S,\, \lambda^\prime_{HS},\, c_S,\, c_{V}^S,\, y_S^{(2)})$.

Our analysis will go through three steps of different degree of refinement. First, we will consider four basic portals which can be obtained from the \eDMEFT\ by setting all other  couplings to zero. Since a lot of results for these portals are already present in the literature, the following subsection should be seen as a brief review. The next step will consist of studying in more detail some benchmarks for a Higgs-mixing portal scenario augmented with the presence of $D=5$ couplings of $S$ with the gauge bosons. This will represent a first illustration of the strength of the \eDMEFT\ and provide some insights into the interplay of the different operators. Finally, we will present a systematic analysis of the full parameter space of the model. In order to assess the robustness of our main results we will relax some of the assumptions mentioned in the beginning of this section and comment on their impact.

\subsubsection{Four basic portals in isolation}

Four basic portals between the DM and the SM are embedded in the \eDMEFT. 
These are:
\begin{enumerate}
\item the mixing portal: $\mathcal{L} \subset - y_S \bar{\chi} \chi S  - \lambda_{H S}^\prime v |H|^2 S  $
\item the effective Yukawa portal: $\mathcal{L} \subset - y_S \bar{\chi} \chi S - \frac{y_f^S}{\Lambda} v\, \bar{f} f S$
\item the effective gauge portal: $\mathcal{L} \subset - y_S \bar{\chi} \chi S - \frac{C_{VV}^S}{\Lambda}V^{\mu \nu} V_{\mu \nu} S$ \,,\,\, $V=G,W,B$
\item the effective Higgs portal: $\mathcal{L} \subset - \frac{y_H^{(2)}}{\Lambda}\bar \chi \chi H^\dagger H$
\end{enumerate}
and correspond to subsets of operators of the Lagrangian in \eqref{eq:lagrangian_scalar}, setting the Higgs field to its vacuum expectation value, $v$, for the Yukawa portal.
These portals in isolation have been considered in the literature and received substantial attention in recent years. In the following, we will briefly summarize their main properties. 

\begin{figure}[t]
\begin{flushleft}
\vspace*{-0.6cm}\hspace*{-1.2cm}
\subfloat{\includegraphics[width=0.38\linewidth]{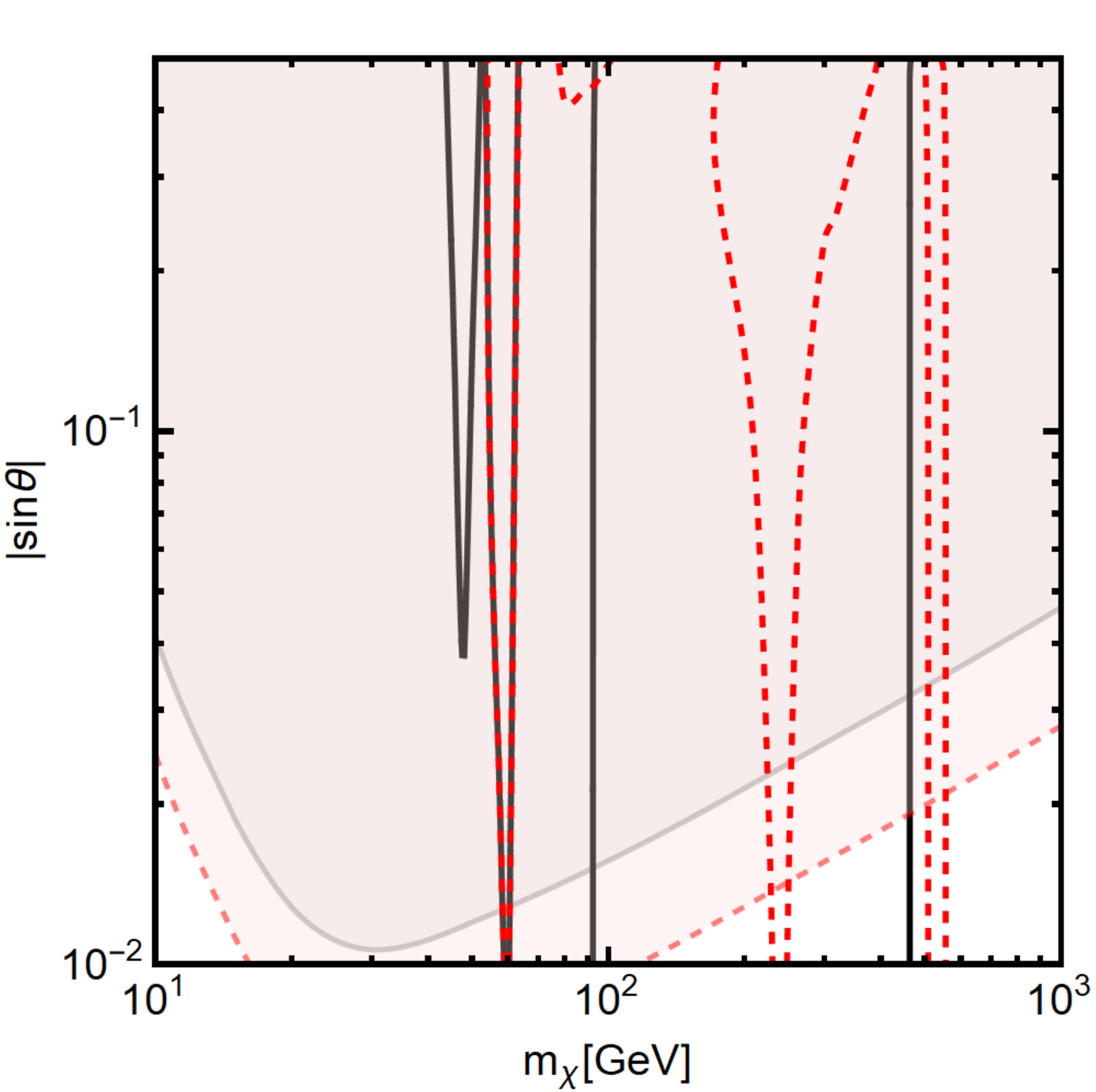}} 
\subfloat{\includegraphics[width=0.38\linewidth]{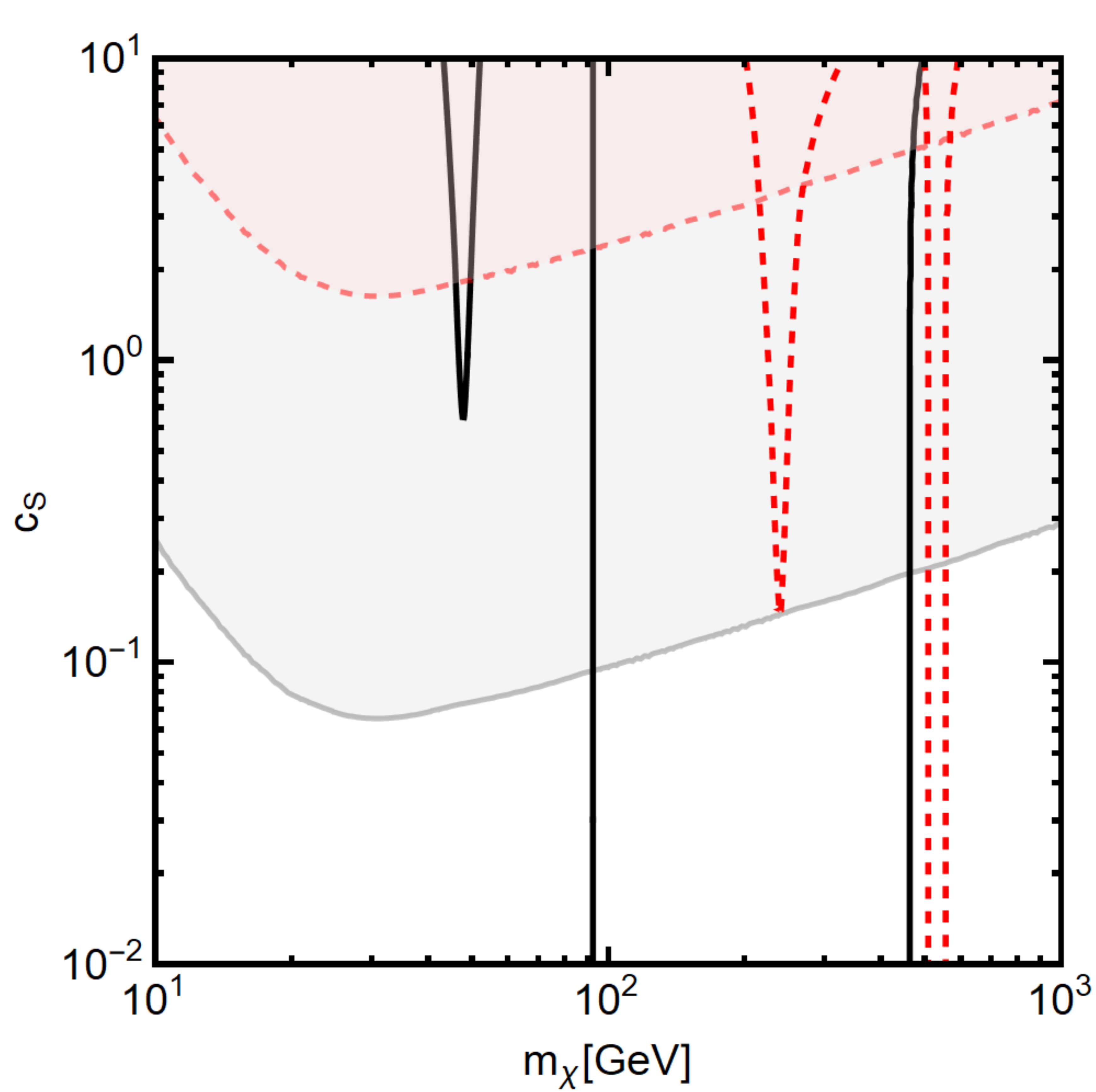}}
\subfloat{\includegraphics[width=0.38\linewidth]{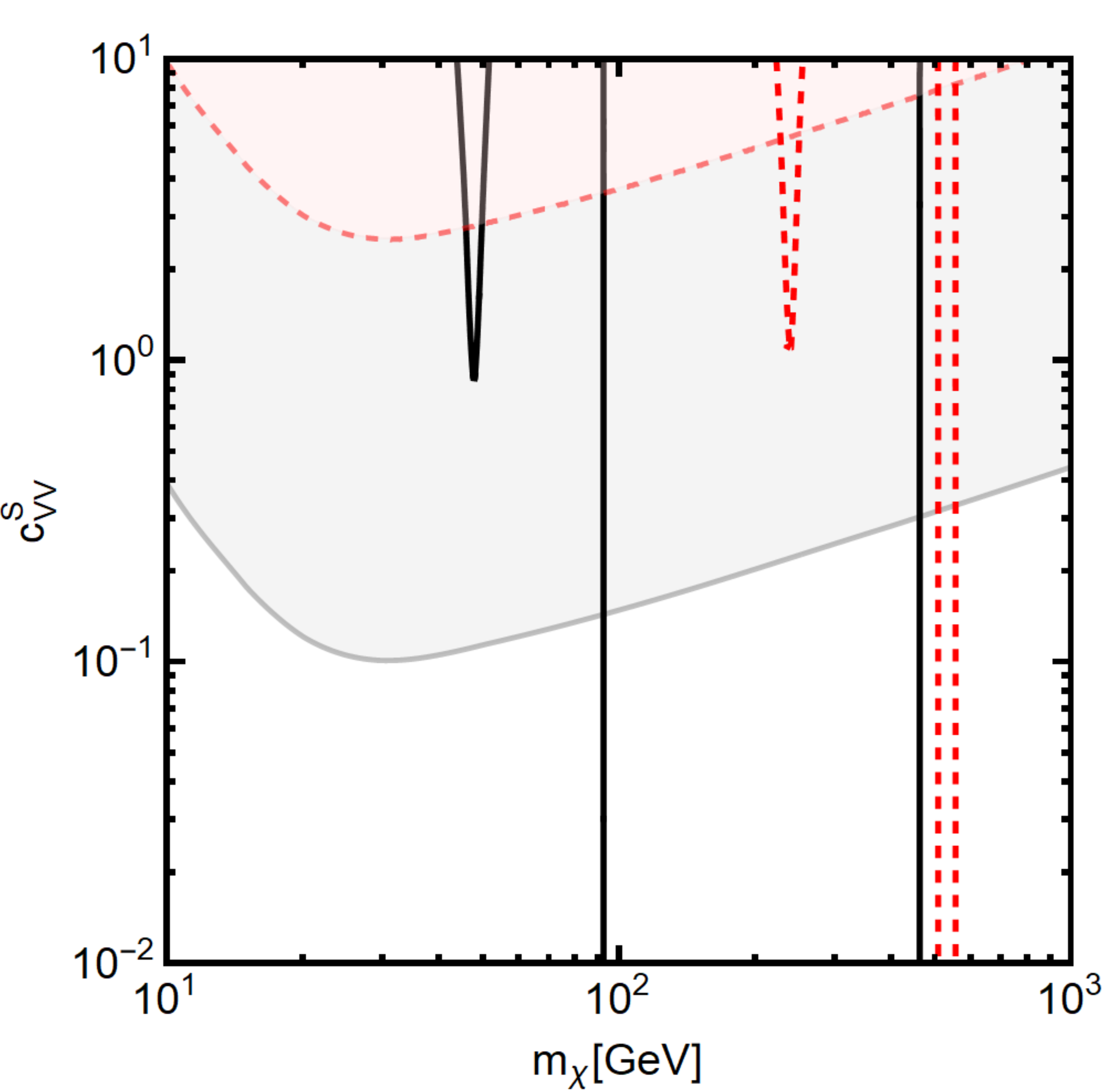}}\\
\vspace*{-0.4cm}\hspace*{-1.2cm}
\subfloat{\includegraphics[width=0.38\linewidth]{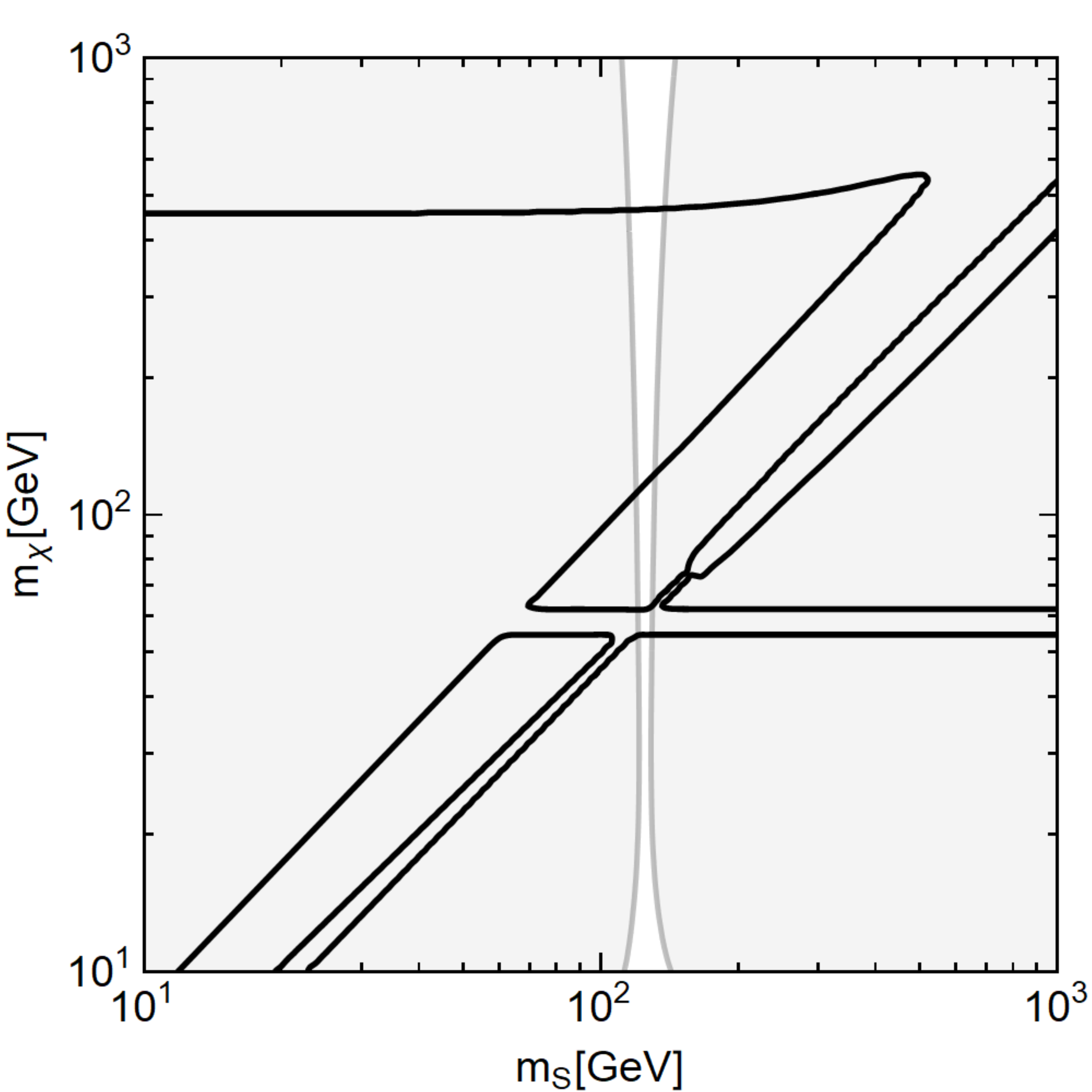}}
\subfloat{\includegraphics[width=0.38\linewidth]{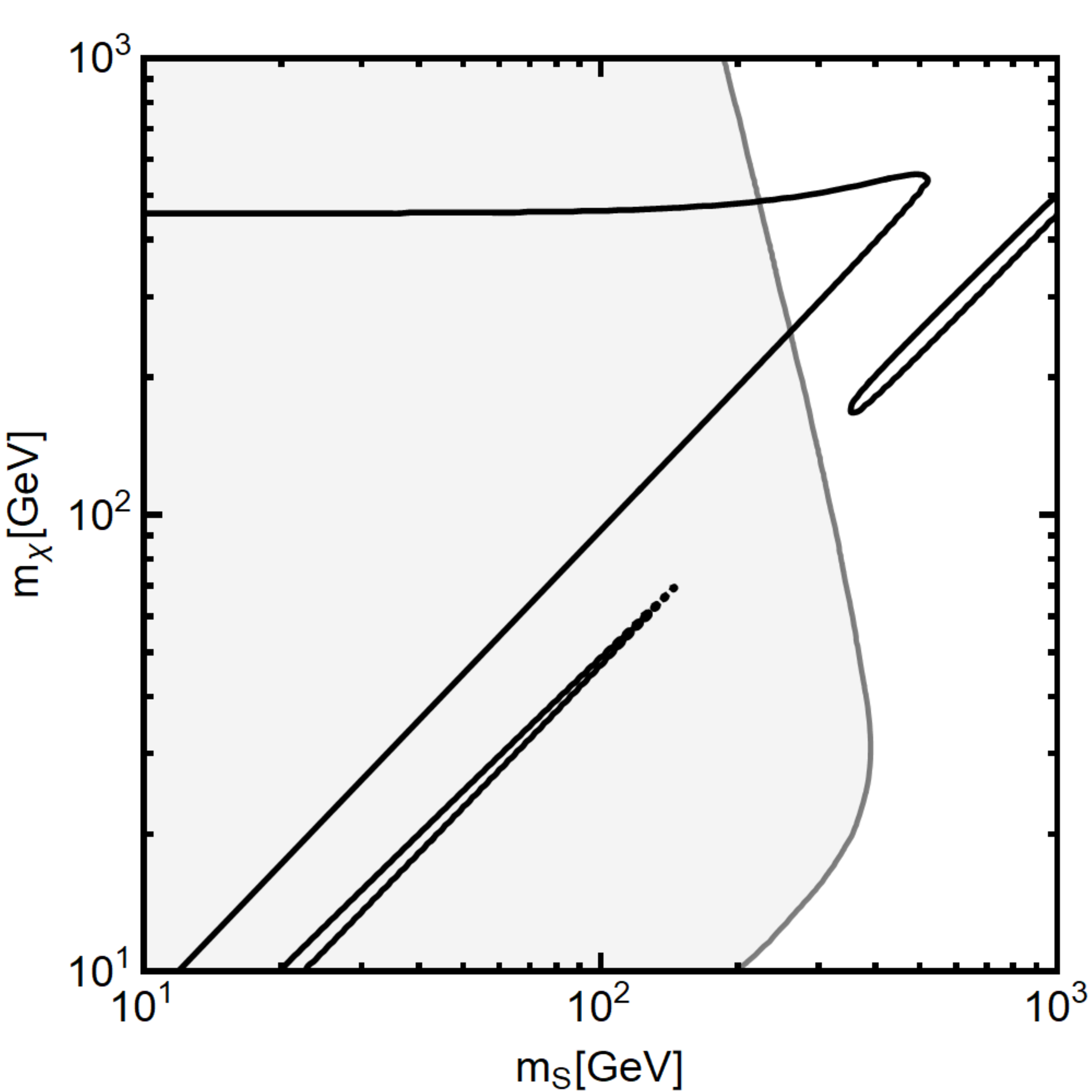}}
\subfloat{\includegraphics[width=0.38\linewidth]{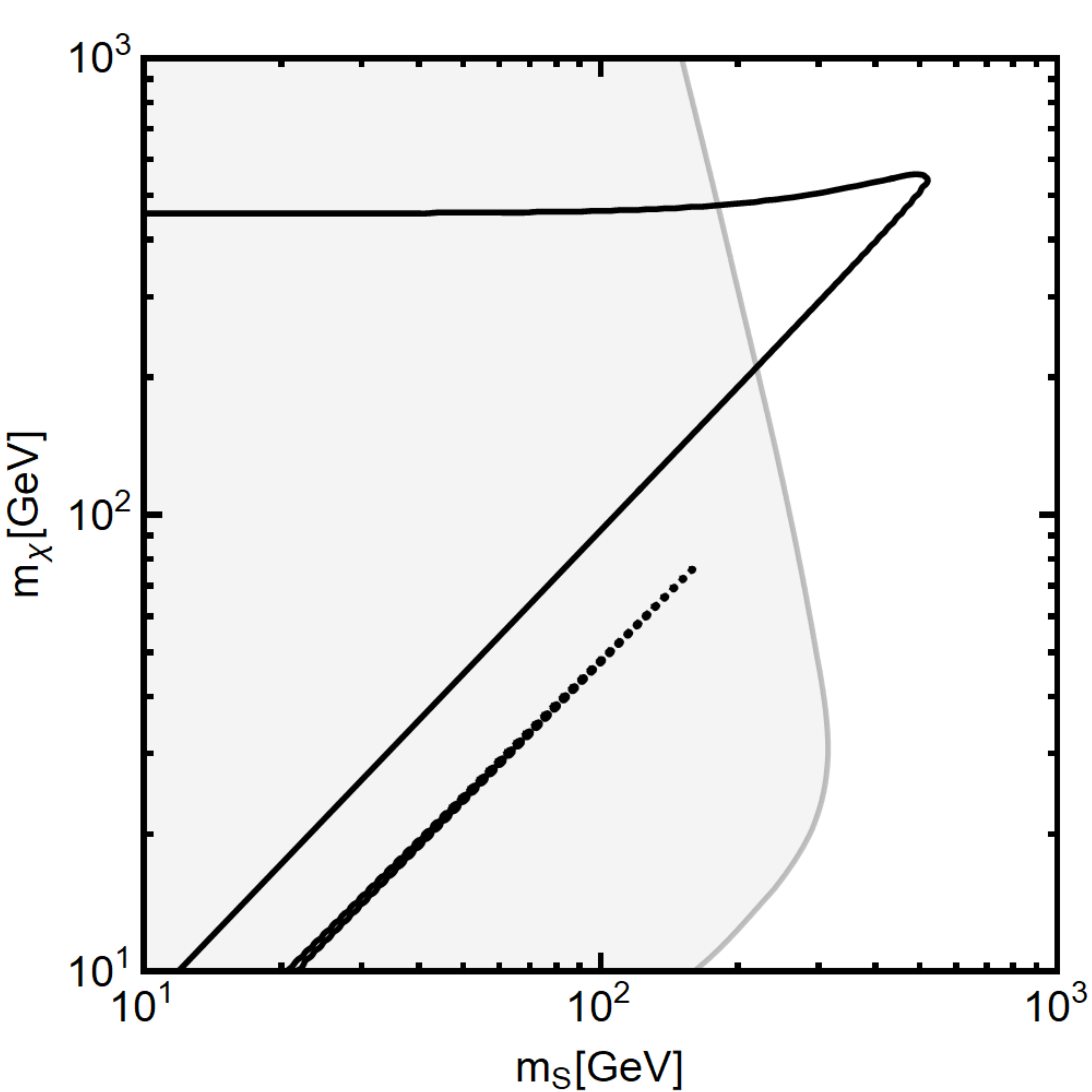}} 
\end{flushleft}
\vspace{-0.5cm}
\caption{\footnotesize{Isocontours of correct relic abundance for the mixing~(left), Yukawa (center), and gauge (right) portals for $y_S=1$. The upper panels display the $m_\chi-\mbox{coupling}$ plane for $m_S=100$\,GeV (black, solid) and $m_S=500$\,GeV (red, dashed). In the case of the mixing portal $|\!\sin \theta|$ has been put on the y-axis instead. The lower panels show the $m_S-m_\chi$ plane for $\sin\theta=0.1$, $c_S=1$, $c_{V}^S=1$ for the mixing, Yukawa and gauge portals, respectively. DD limits from XENON1T are given as light red (gray) shaded regions for the case of a light (heavy) mediator $S$. In all cases we have set $\Lambda=3\,\mbox{TeV}$.}} 
\label{fig:threeportals}
\end{figure}
 \begin{figure}[b]
\begin{center}
    \includegraphics[width=0.38\linewidth]{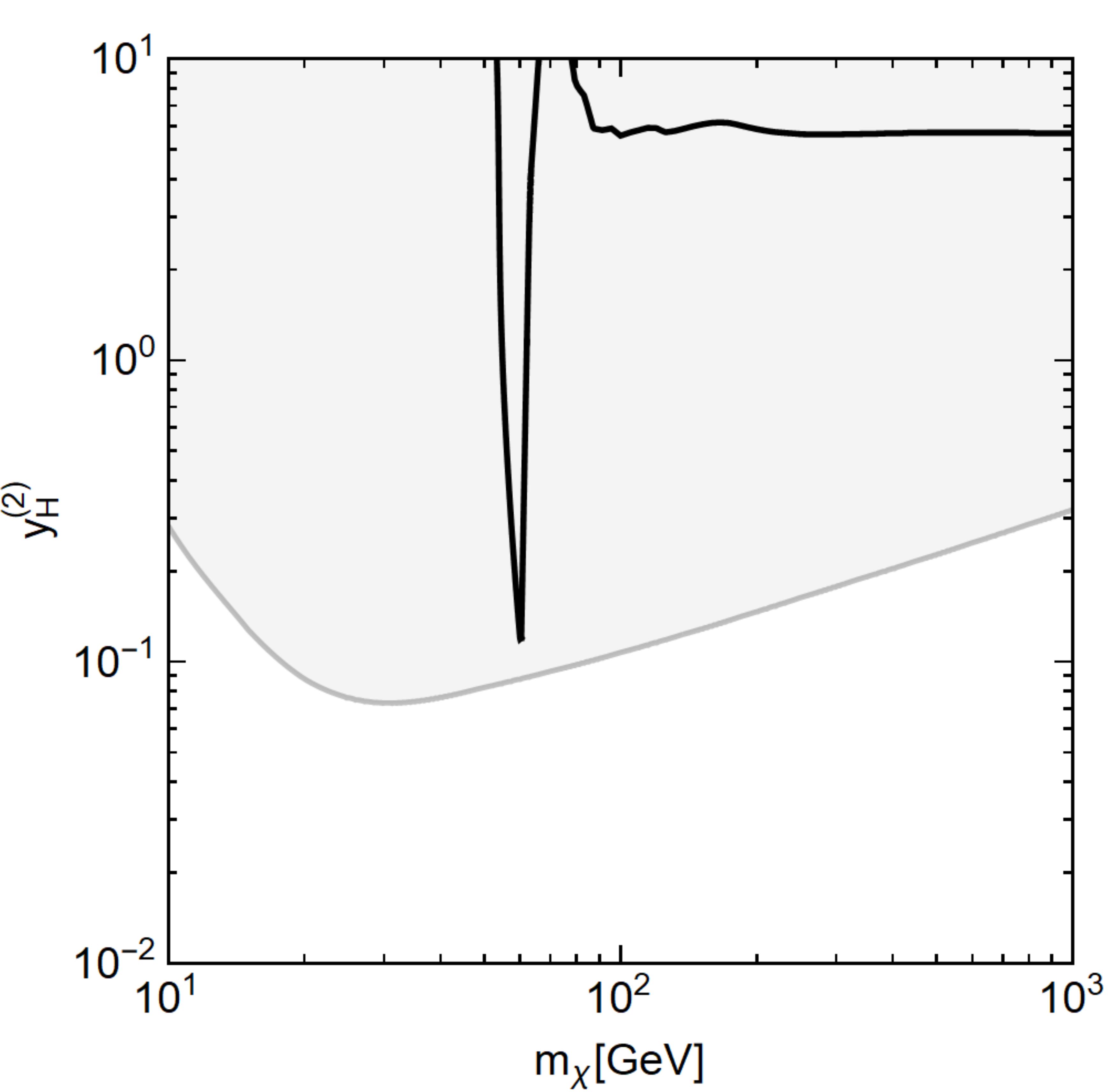}
\end{center}   
\vspace{-0.5cm}
\caption{\footnotesize{Isocontours of correct relic abundance and DD limits for the effective Higgs portal with the same conventions as in fig.~\ref{fig:threeportals}.}} 
\label{fig:effectiveHiggs}
\end{figure}

Some of the features of the mixing portal have already been discussed in sec.~\ref{sec:DMS}. In this setup the dark Yukawa operator $\bar{\chi} \chi S$ is combined with the mixing between $S$ and $h$ induced by the scalar potential. It is noteworthy that this connection between the DM and the SM is realized by renormalizable interactions. Therefore, the strength of this potential is not controlled by the scale of the higher-dimensional operators and could potentially be rather large. However, the mixing is limited by Higgs measurement as discussed in the previous section. Very roughly the bound can be approximated as $s_\theta \leq 0.2$ in substantial parts of the parameter space. This is comparable with the generic suppression of our higher-dimensional operators that feature Higgs fields, $\frac{v}{\Lambda}$, for $\Lambda \sim 1$\,TeV.  We refrain from linearizing the effects of mixing and always take the full diagonalization of the fields into account. This portal has been discussed for example in~\cite{LopezHonorez:2012kv,Arcadi:2019lka}.

The effective Yukawa portal combines the renormalizable $\bar{\chi} \chi S$ interaction with an effective coupling of the scalar mediator to SM quarks pairs. It corresponds to a realization of the popular, not gauge invariant, `simplified model' for a scalar mediator coupled to fermionic DM~\cite{Harris:2014hga,Buckley:2014fba,Abdallah:2014hon,Abdallah:2015ter,Boveia:2016mrp}.
Along an analogous reasoning the effective gauge portal connects the renormalizable DM--scalar vertex via an effective interaction to SM gauge fields. This kind of interaction is actually present in the simplified Yukawa portal, where it arises at the one-loop level from couplings of the scalar mediator with (mostly) two top quarks and plays a relevant role in its collider phenomenology \cite{Buckley:2014fba,Harris:2014hga}. Alternatively, the $G_{\mu \nu } G^{\mu \nu} S$  vertex can be generated in models with heavy vector-like fermions; models of this type were studied extensively as an explanations of the $750$ GeV diphoton excess \cite{Mambrini:2015wyu,Angelescu:2016mhl,Backovic:2015fnp,Falkowski:2015swt,DEramo:2016aee}. Here, we remain agnostic regarding its origin.
Finally, the effective Higgs portal is distinct since it does not involve the new scalar mediator at all and has only two free parameters, namely $y_H^{(2)}/\Lambda$ and $m_\chi$, see e.g.~\cite{Arcadi:2019lka} for a recent review.

Before reviewing the DM phenomenology of these portals, we remind that besides being agnostic about the origin of operators, the \eDMEFT\ uplifts them to a complete $D=5$ field theory, including for example the bi-quadratic ${S}^2 \bar{\chi}_L \chi_R$ term \cite{Alanne:2017oqj,Goertz:2019vht}. 
It thereby allows to capture a large class of DM scenarios as well as new cancellation patterns in DD emerging non-trivially in the full EFT.

The well known features of the first three basic portals are visualized in fig.~\ref{fig:threeportals}, which displays isocontours corresponding to the observed relic density as well as DD constraints in the couplings vs.\ DM-mass plane (upper panels) for fixed $m_S=100$\,GeV (black) and $m_S=500$\,GeV (red), as well as in the $m_S-m_\chi$ plane (lower panels) for fixed couplings of $\sin\theta=0.1$, $c_S=1$, $c_{V}^S=1$, respectively. 

As can be seen the correct relic density is typically achieved only in special kinematic configurations. 
A first prominent configuration is represented by the resonance at $m_{\chi} \approx \frac{1}{2}m_S$, corresponding to a strong dip in the relic density curve in the upper panels of fig.~\ref{fig:threeportals}. The correct relic density can be then achieved at the opening threshold of the of the annihilation channel $\chi \chi \rightarrow S S $, i.e., $m_\chi \approx m_S$. In the latter case, the relic density depends only very weakly on the coupling between $S$ and the SM, and it is therefore characterized by the almost vertical line right next to the threshold in the upper panels. The annihilation into $SS$ can account for the correct relic density also when $m_\chi$ is sensitively higher than $m_S$. This explains the second vertical line present in the upper row of fig.~\ref{fig:threeportals}.
Finally, the annihilation cross sections in models which lead to a direct coupling between the DM and $H$, such as the Higgs-mixing portal and the effective Higgs portal, also receive an enhancement at $m_\chi \approx\frac{1}{2} m_H$.
 
Limits from DD play the most important role in determining whether a model with a real scalar mediator is viable or not. The scattering of the DM with nuclei is induced, at the microscopic level, by three different types of interactions with the SM in the various portals. It is then worth considering the interplay of these interactions, taken individually, with the relic density. As can be seen, the DD constraints resulting from XENON1T (given by the shaded regions) are most constraining for low mediator masses and relax somewhat for $m_S\geq 200 $ GeV in the effective Yukawa and gauge portals. In the model with Higgs mixing the softening of the constraints for large scalar masses is less pronounced since the pure Higgs contribution is not directly sensitive to $m_S$. 
Note, however, that fixing a value for $\sin \theta$ while varying $m_S$ as shown in the lower left panel corresponds to changing parameters in the scalar potential. 
In general the contributions from Higgs and $S$ exchange interfere destructively  such that an unconstrained region at $m_h \approx m_S$ shows up.
In all the considered models the DM abundance relies on annihilation with velocity suppressed cross sections such that ID searches are not relevant.

Finally, it is worthwhile to take a closer look at the effective Higgs portal.  Since there is only one coupling and $S$ does not play a role, the relic density is tightly connected with the DD rate and values of the coupling $y_H^{(2)}$ required for a successful thermal freeze out are excluded by the strong constraints on the DD cross section, as can be observed in fig.~\ref{fig:effectiveHiggs}; see also refs.~\cite{LopezHonorez:2012kv,Balazs:2017ple, Arcadi:2019lka} for an in-depth discussion of the subject.

\subsubsection{Completing simplified models with EFT}

\begin{figure}[h]
\begin{flushleft}
\vspace*{-0.5cm}\hspace*{-1.8cm}\subfloat{\includegraphics[width=0.4\linewidth]{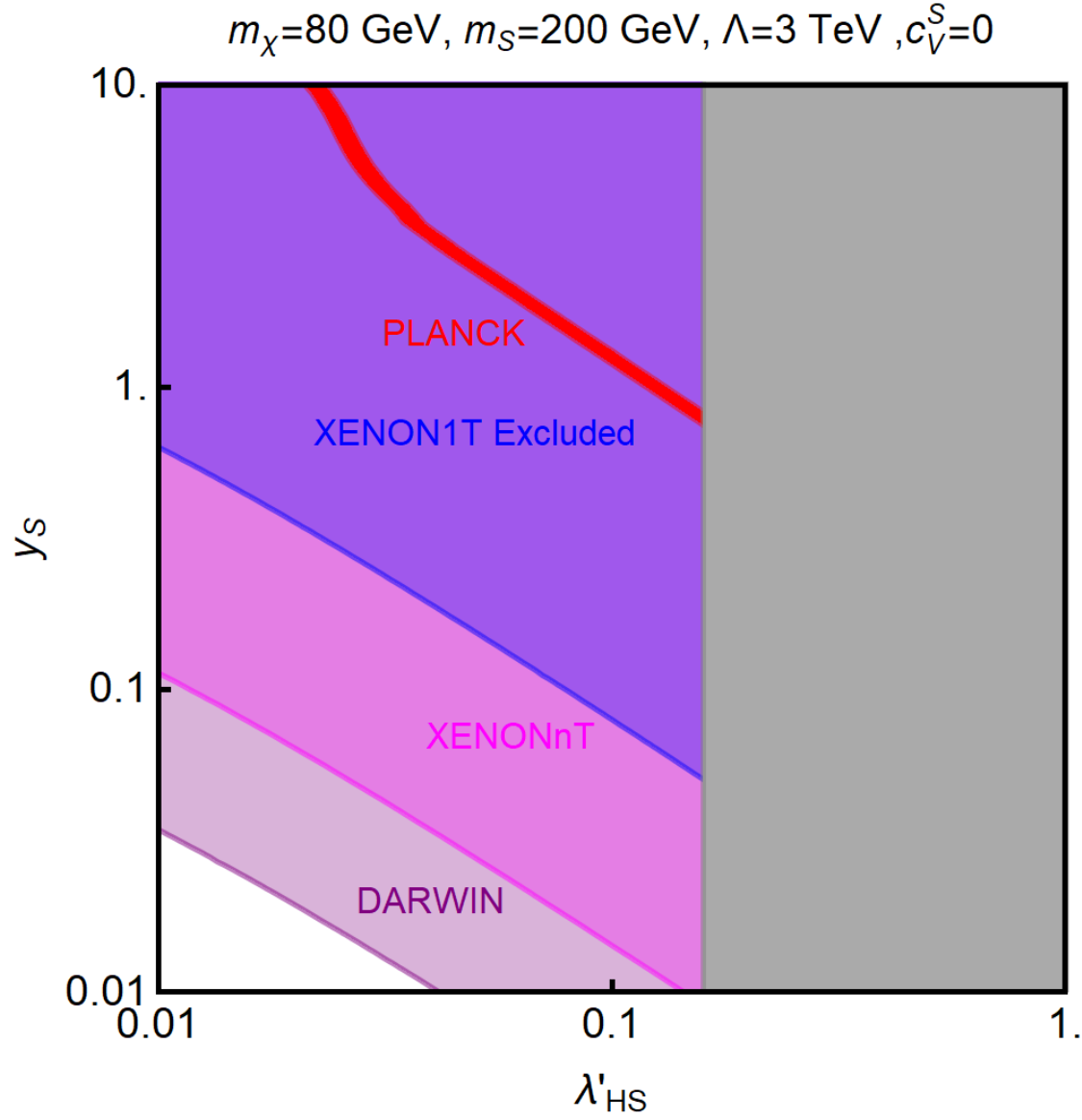}}
\subfloat{\includegraphics[width=0.4\linewidth]{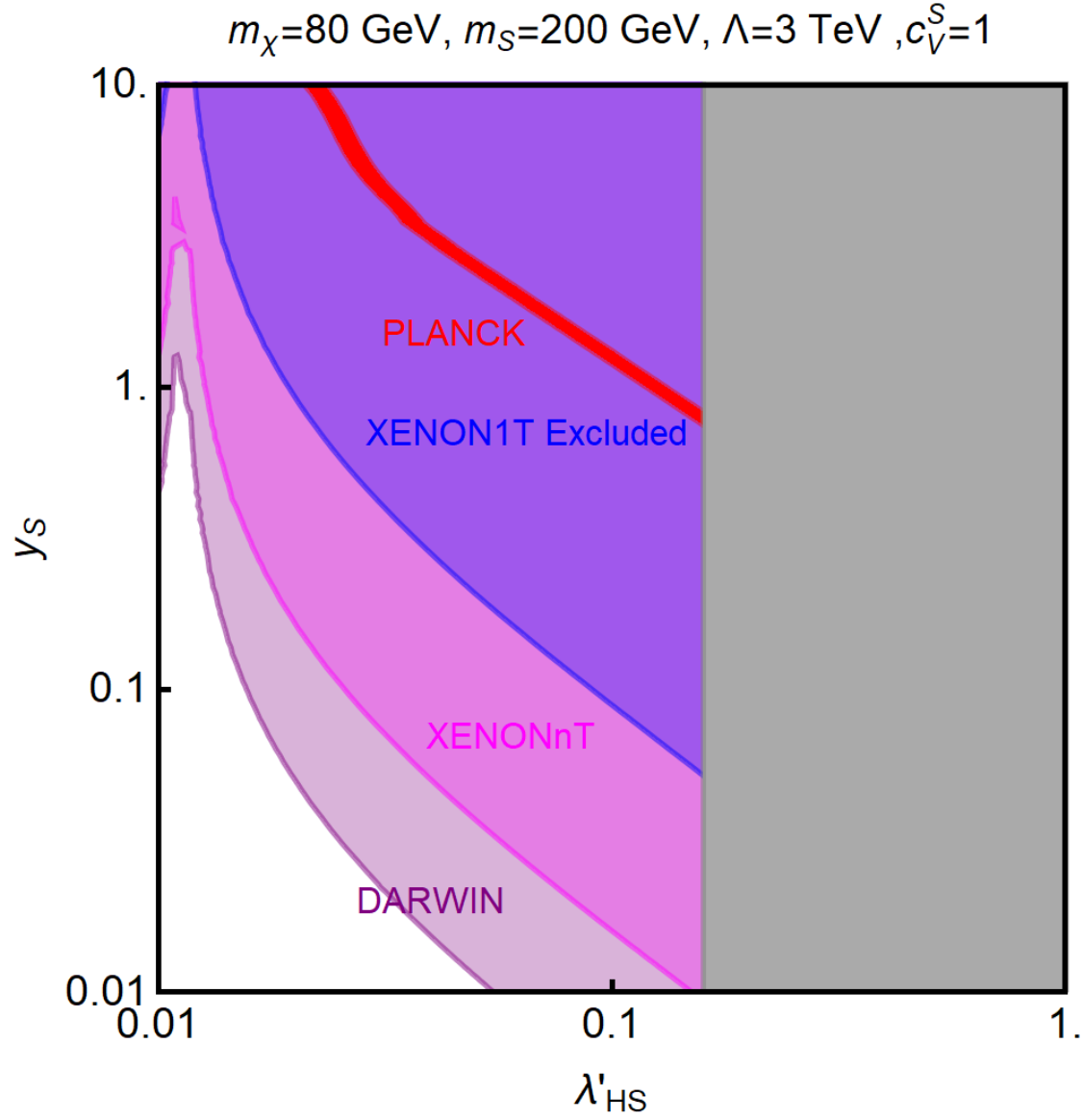}}
\subfloat{\includegraphics[width=0.4\linewidth]{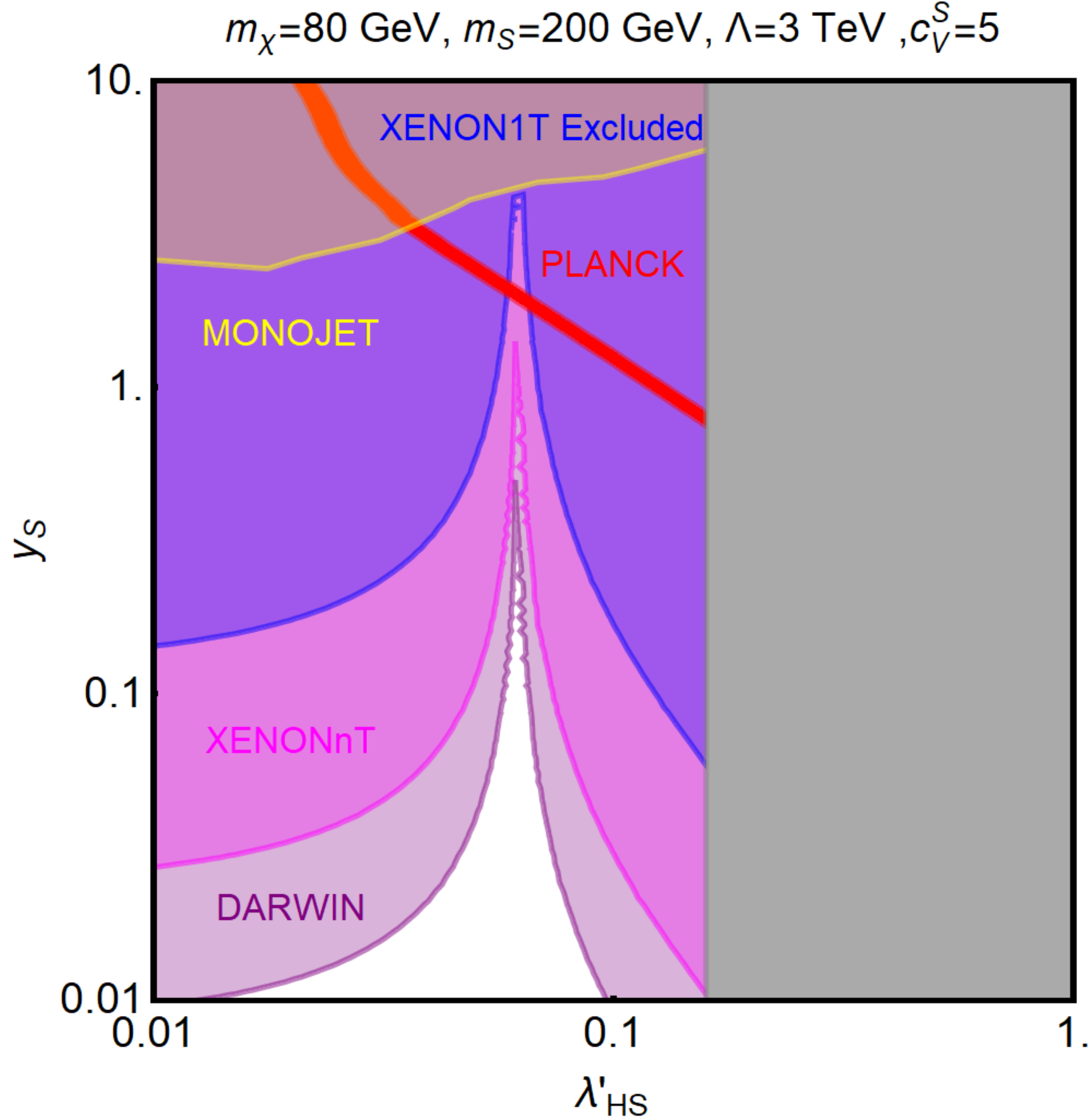}}\\
\hspace*{-1.8cm}\subfloat{\includegraphics[width=0.4\linewidth]{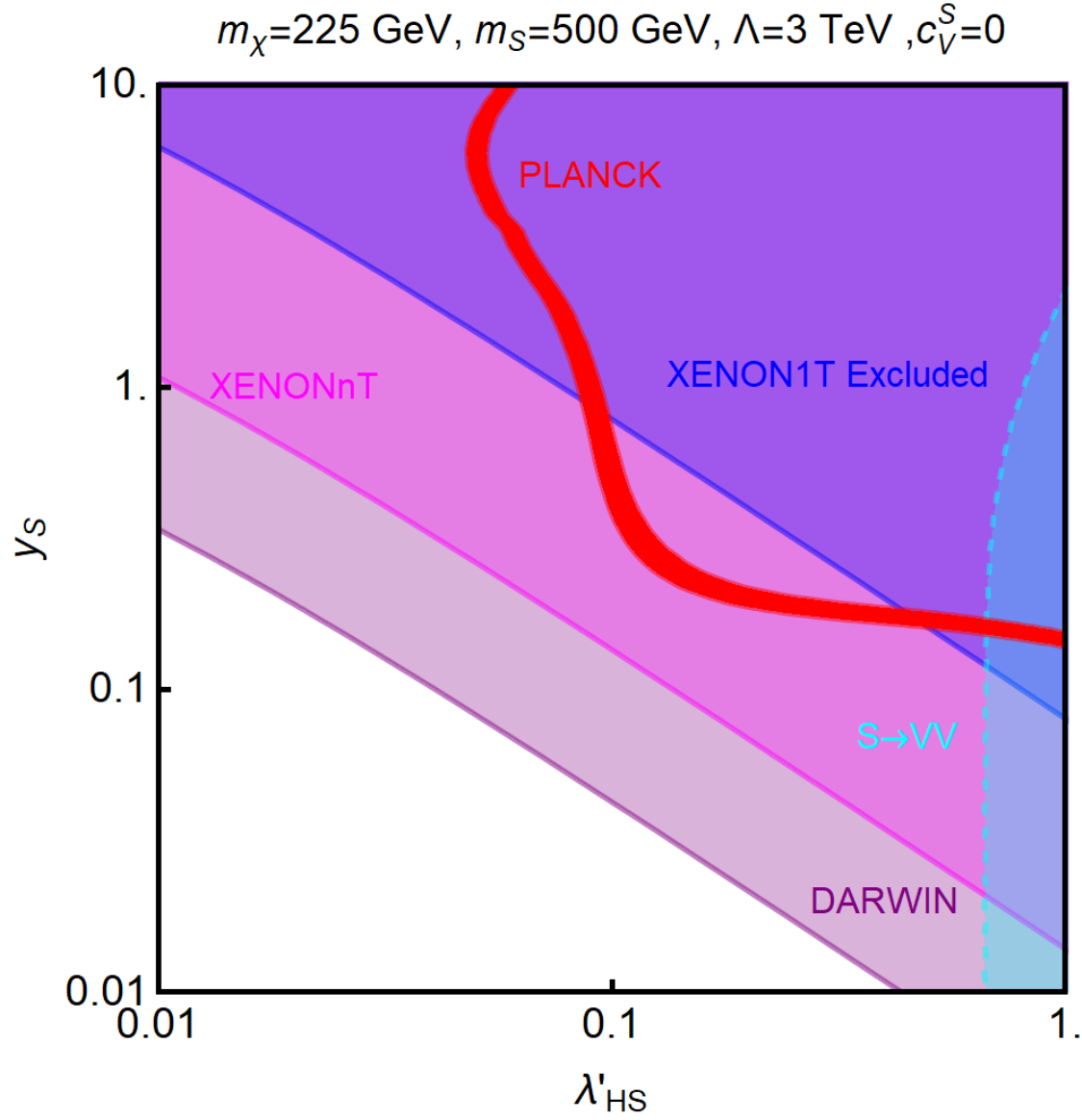}}
\subfloat{\includegraphics[width=0.4\linewidth]{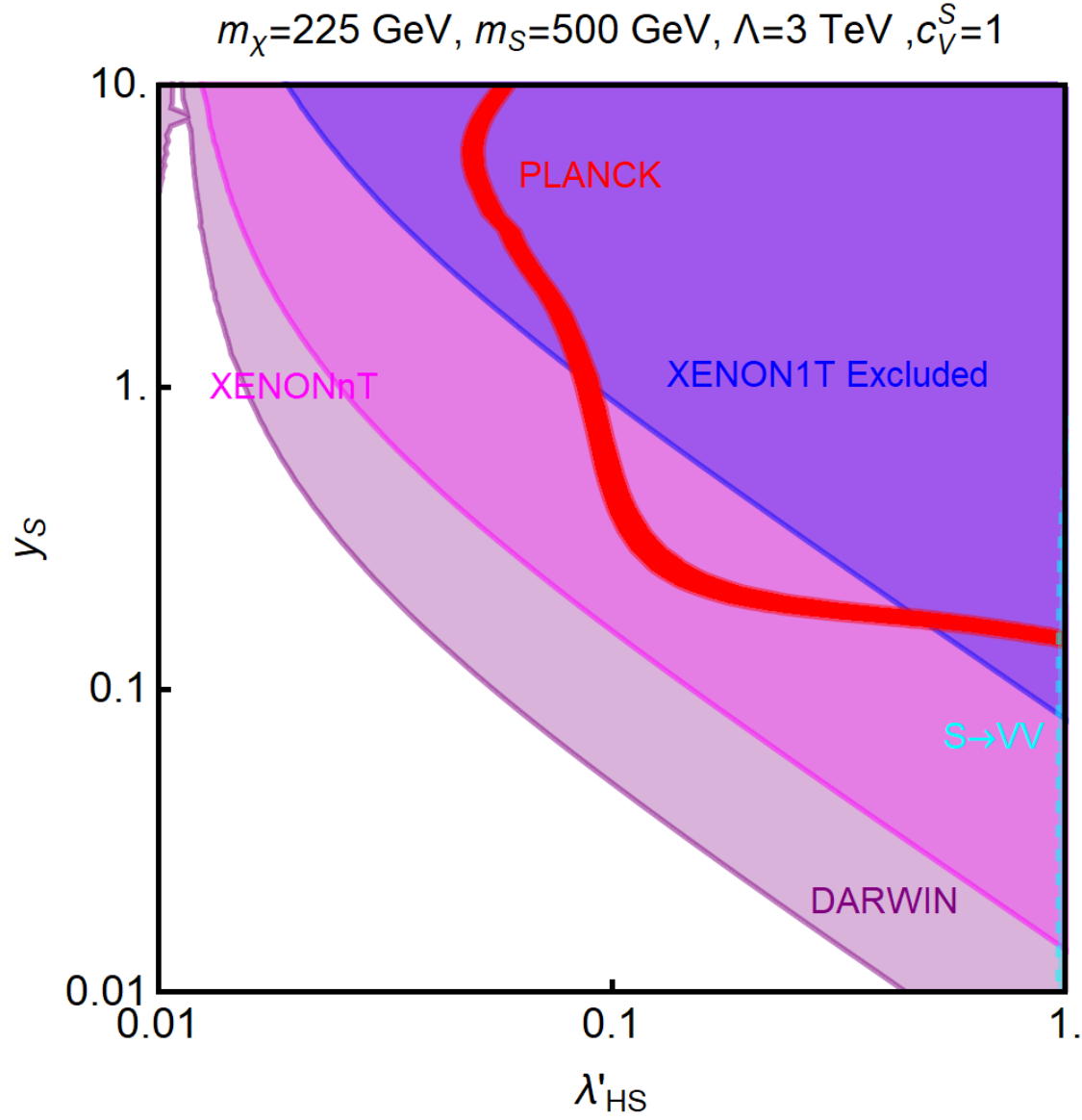}}
\subfloat{\includegraphics[width=0.4\linewidth]{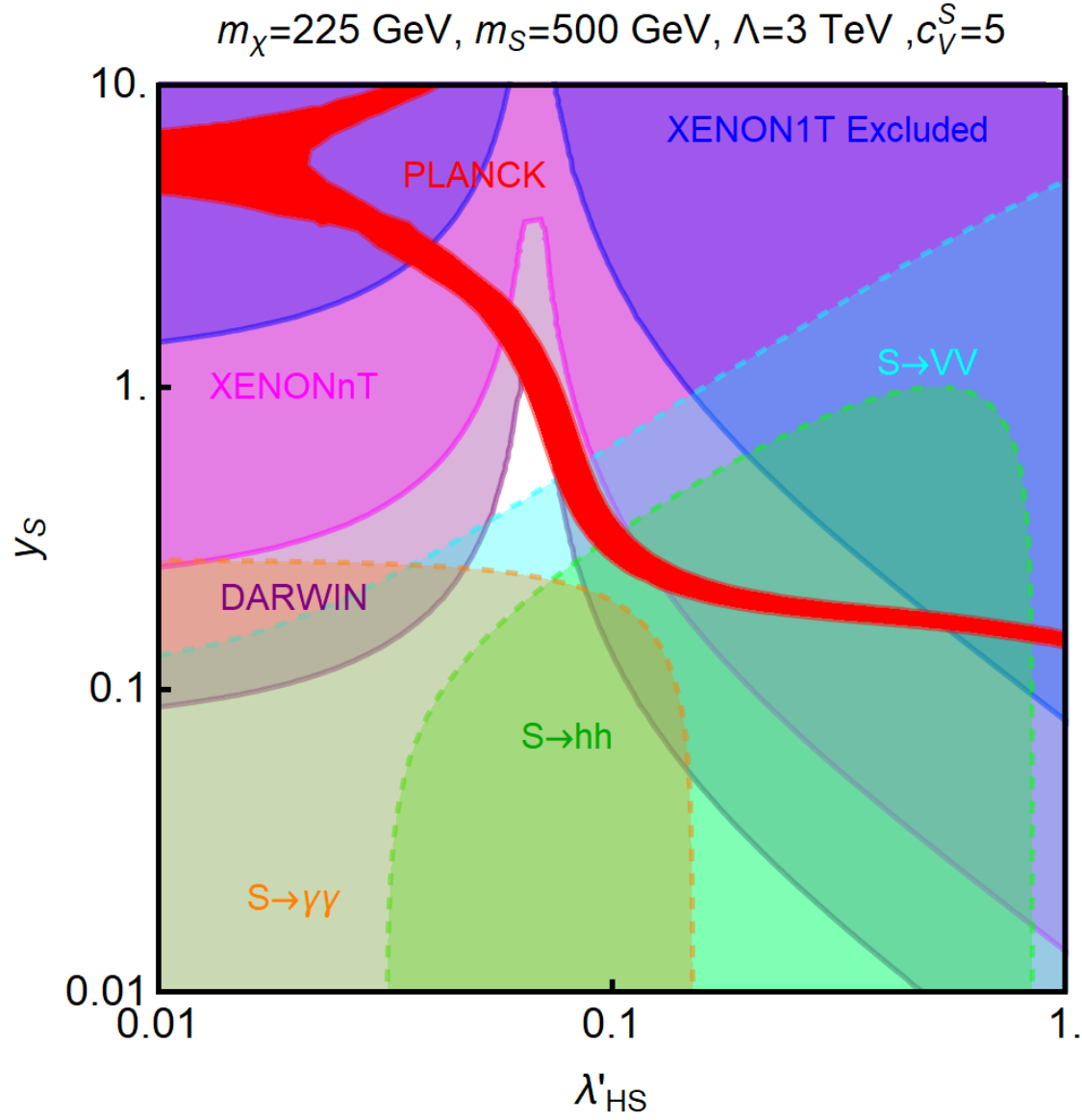}}\\
\hspace*{-1.8cm}\subfloat{\includegraphics[width=0.4\linewidth]{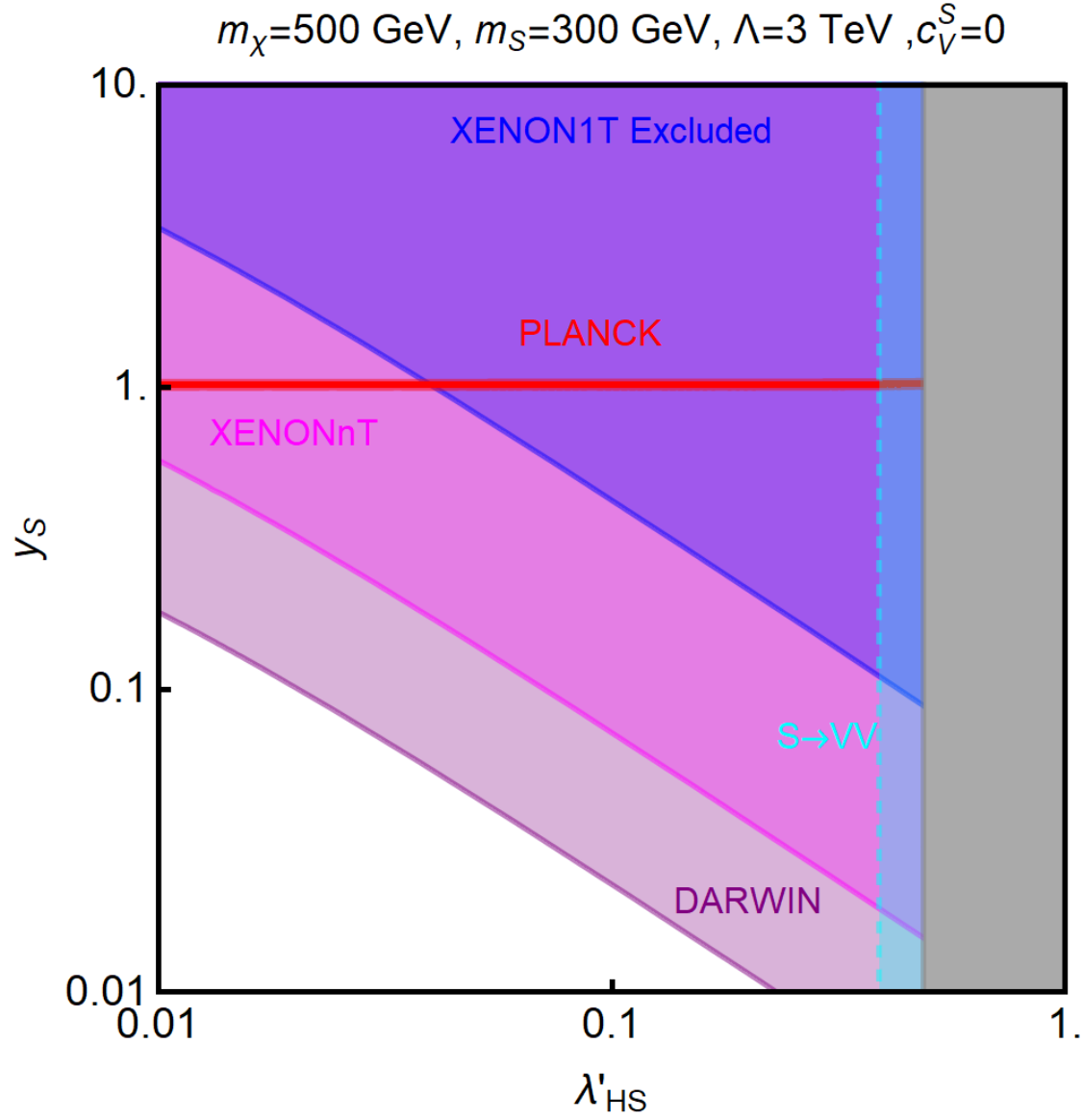}}
\subfloat{\includegraphics[width=0.4\linewidth]{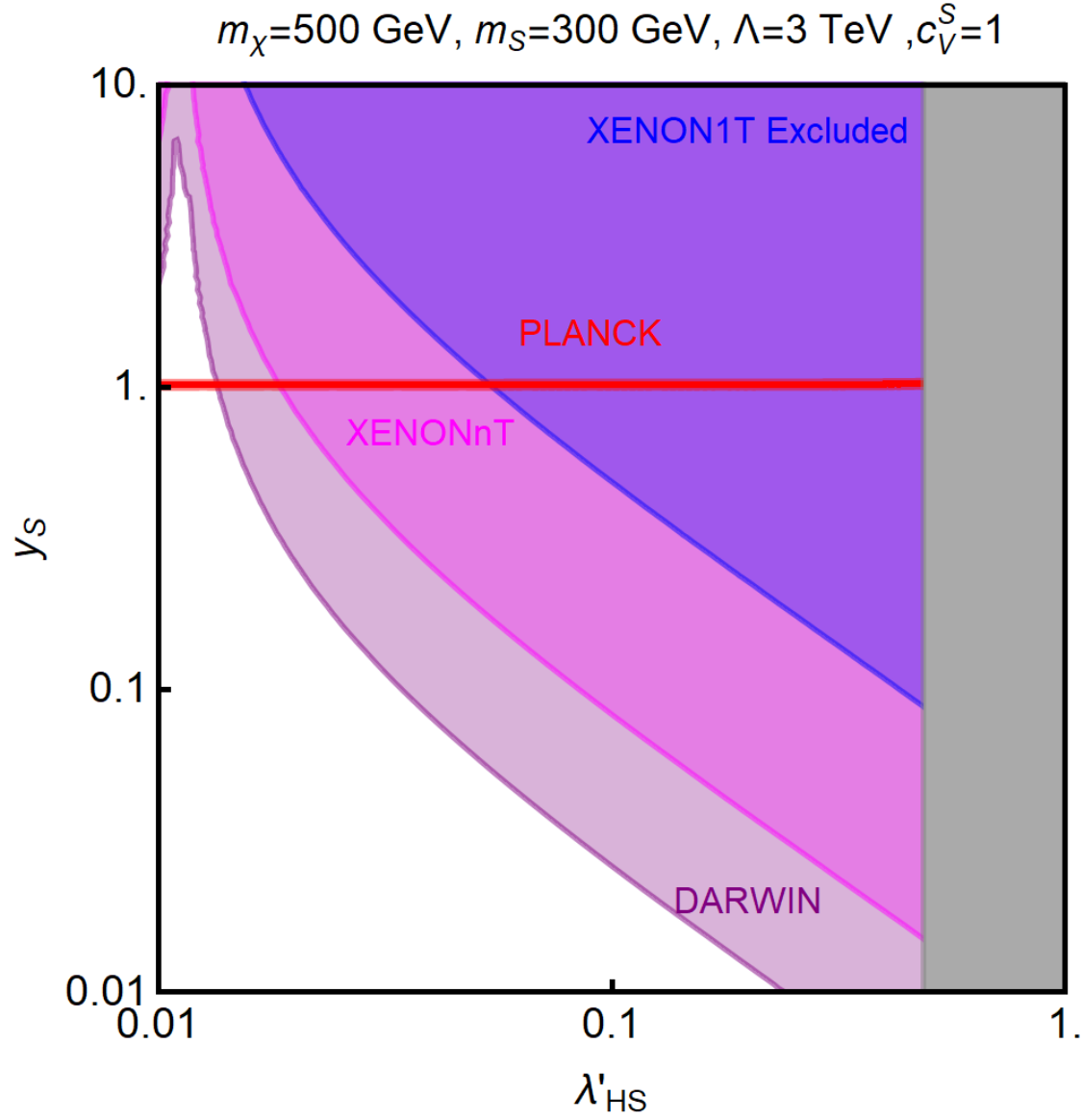}}
\subfloat{\includegraphics[width=0.4\linewidth]{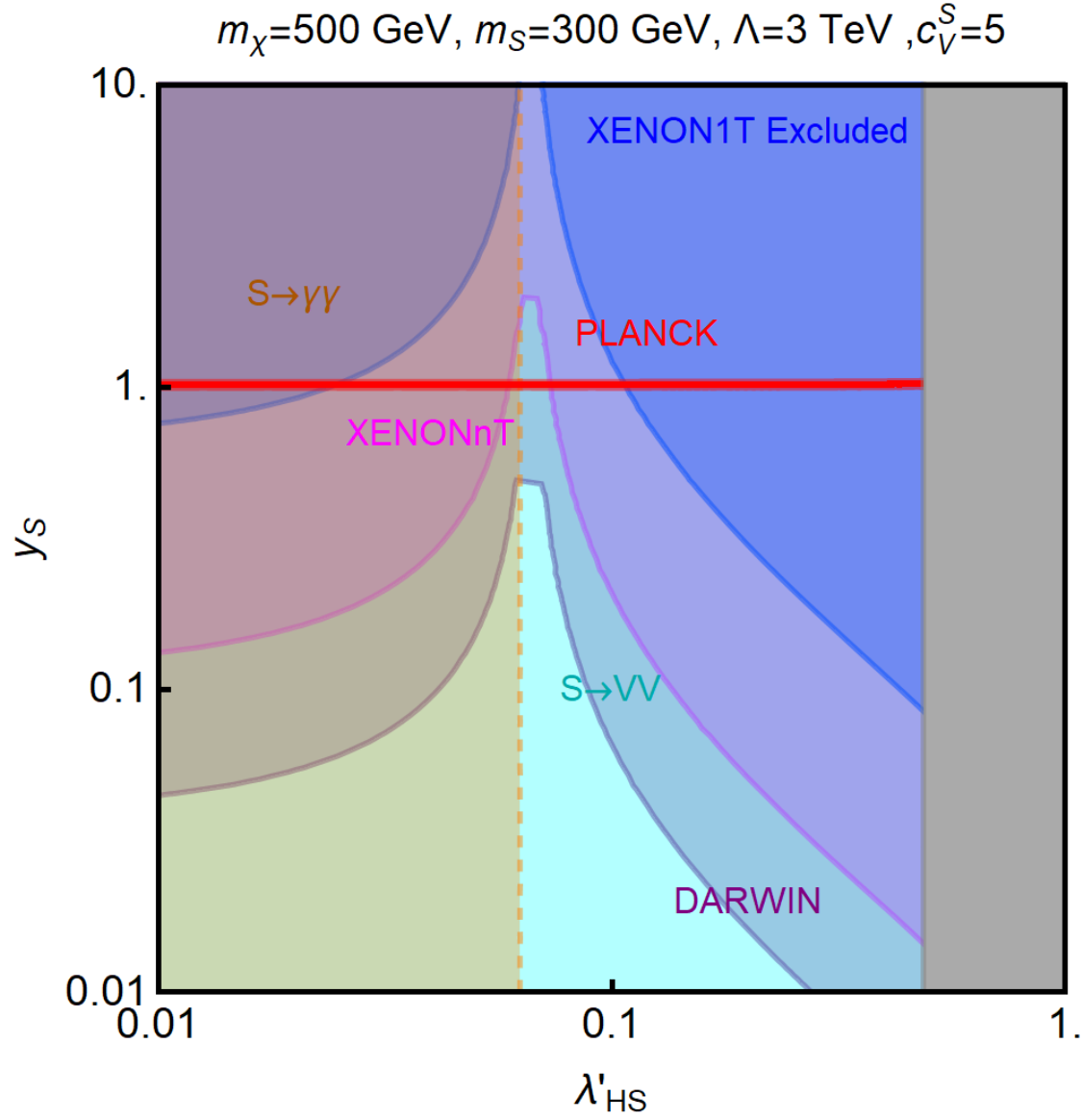}}
\end{flushleft}
\caption{\footnotesize{Summary of constraints for the mixing portal completed with $D=5$ interactions between the scalar mediator and the gauge bosons. The results are shown in the $\lambda_{HS}^{'}-y_S$ plane for $c_{V}^S=0,1,5$ and three different mass assignations as indicated in the individual panels. The red contours correspond to the correct relic density while the blue regions are excluded by limits from DD, as given by XENON1T, and the magenta and purple regions represent the projected sensitivities of XENONnT and DARWIN. The excluded regions from Higgs signal strengths are depicted in gray. Green, cyan, and orange regions are excluded by collider searches for resonances decaying into SM Higgs pairs, massive gauge bosons, and photons, respectively. The latter bounds rely on the assumption of a common parameter $c_V^S$ for the couplings with gauge bosons and could be lifted by setting $c_{B}^S,c_{W}^S\ll c_G^S$. The yellow region in the top right panel is excluded by monojet searches. }}
\label{fig:completingsimplifiedmodels}
\end{figure}

Having reviewed the simplest portals, we now move to  more complex scenarios, unfolding the strength of the \eDMEFT. To this end, we will start with the case of the Higgs mixing portal completed with the presence of effective couplings of $S$ with gluons. As already discussed, this setup may allow for the presence of \emph{blind spots} in DD and, consequently, potentially relax the strong bounds found in the simple portals. For the discussion of this phenomenon we will focus on selected benchmark scenarios. These will be identified by a specific assignation of the $(m_\chi,m_S)$ pair. For each of the considered scenarios, we will compare the different DM constraints and, where relevant, collider constraints in the $(\lambda^{'}_{HS},y_S)$ plane for three assignations of $c_{V}^S=0,\,1$ and $5$.

The first benchmark is characterized by relatively low values of $m_\chi=80\,$GeV and $m_S=200$\,GeV and is displayed in the upper row of fig.~\ref{fig:completingsimplifiedmodels}. Being $m_\chi> m_h/2$, searches for invisible Higgs decays have no impact on this benchmark.
Furthermore, current LHC searches of resonances decaying into visible products are not effective for this low value of $m_S$. On the contrary, for $c_V^S=5$ a portion of the viable region for the DM relic density is excluded by monojet searches. Another notable feature is that the presence of the effective coupling of $S$ with gauge bosons has only a modest impact on the relic density such that the corresponding isocontour is quite similar for the three different assignations of the $c_{V}^S$ parameter. The relic density is driven by annihilation processes into SM fermions, mostly $\bar b b$, with a modest $s$-channel enhancement since the DM mass is not too close to the $m_S/2$ pole. This requires rather high values of $y_S$ to comply with the correct relic density. For $c_{V}^S=0,1$, the considered benchmark is ruled out by current bounds from XENON1T. For $c_{V}^S=5$ a window remains open, where a DD blind spot can occur along the Planck constraint. This already rather small stripe will however soon be reduced substantially if DM signals at the next generation of DD experiments remain absent---or might allow for a potential discovery.

Our next benchmark is explored in the second row of fig.~\ref{fig:completingsimplifiedmodels}. It features $m_\chi=225\,\mbox{GeV}$ and $m_S=500\,\mbox{GeV}$ and is representative for a scenario with DM at the $m_S/2$ pole. Contrary to the previous benchmark, large values of $c_{V}^S$ do have a significant impact on the relic density since it is very sensitive to the total width of the scalar mediator. Given also the occurrence of the blind spot we notice that the viable region of parameter space for $c_{V}^S=5$ is wider with respect to the $c_V^S=0,1$ cases, with a small portion of it even evading future constraints from the DARWIN experiment. On the other hand our choice of $m_S$ renders this benchmark sensitive to collider experiments. The colored regions represent the exclusions from the various searches mentioned in the previous section. As evident, searches for diboson final states are most effective (with exclusions corresponding to the cyan regions in the plots). This is because the cross section $\sigma(pp \rightarrow S \rightarrow WW/ZZ)$ can be substantial both for sizable value of $\sin\theta$ (due to $\lambda_{HS}^{'}$) and for large higher-dimensional $c_{B,W}^S$ couplings.

The last scenario studied corresponds to $m_\chi=500\,\mbox{GeV}$ and $m_S=300\,\mbox{GeV}$ and is summarized in the bottom row of fig.~\ref{fig:completingsimplifiedmodels}. 
In all there panels the correct relic density is determined by a  $\mathcal{O}(1)$ value of $y_S$. This reflects the fact that the relic density is obtained in the `secluded' regime \cite{Pospelov:2007mp,Pospelov:2008jd}, i.e. it is fixed via the $\chi \chi \rightarrow SS$ process. In consequence, it is entirely set by parameters of the new particle sector, i.e. the DM and mediator masses and their coupling $y_S$. The coupling $c_{V}^S$ affects DM phenomenology by changing the position of the blind-spot region, hence determining the range of values of $\lambda^{'}_{HS}$ for which even future DD constraints can be evaded. For large values of $c_{V}^S \sim 5$, on the other hand, this benchmark is ruled out by collider bounds with those from diboson searches being again dominant.

\begin{figure}[t]
    \centering
    \subfloat{\includegraphics[width=0.4\linewidth]{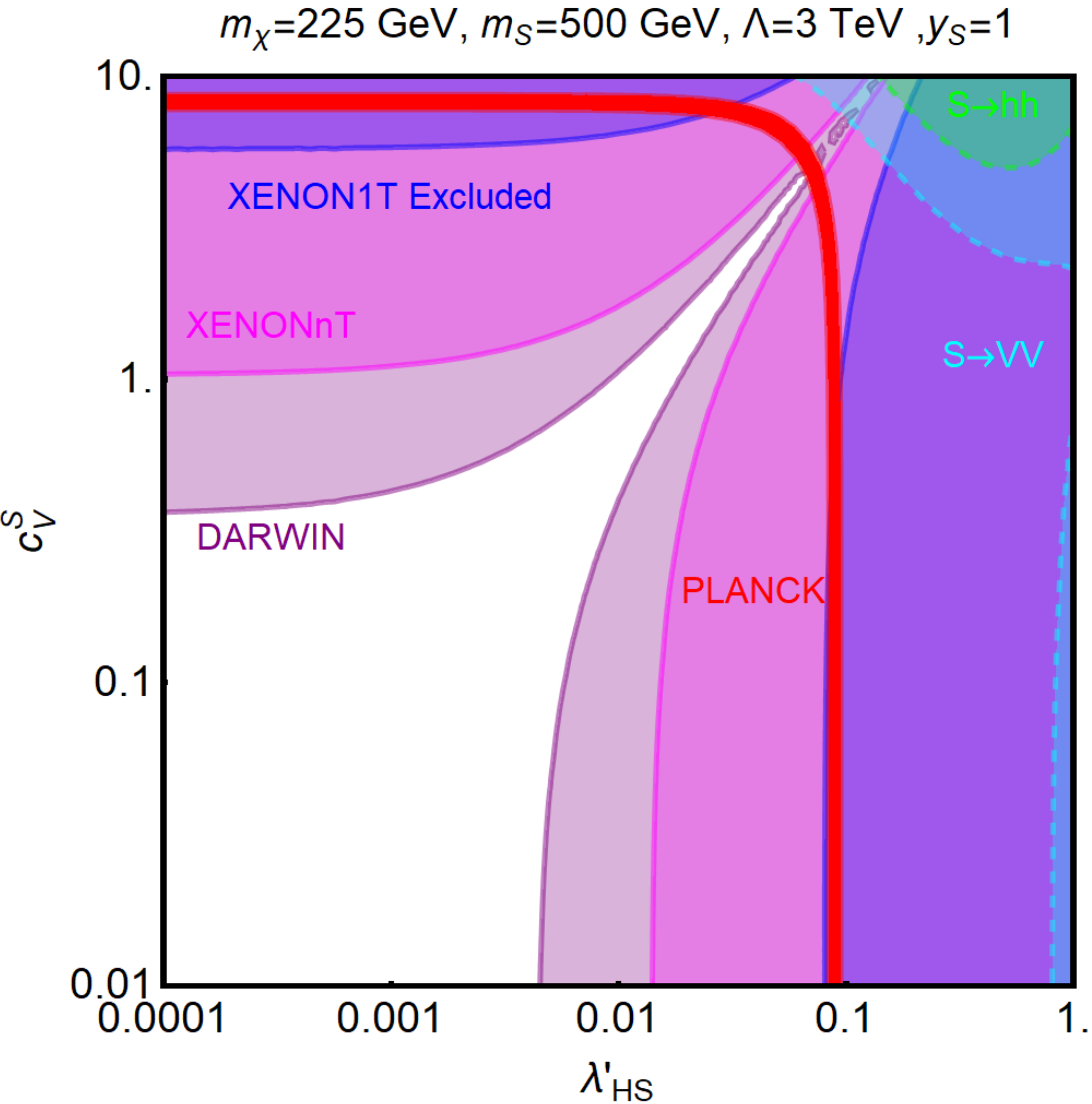}}
    \subfloat{\includegraphics[width=0.4\linewidth]{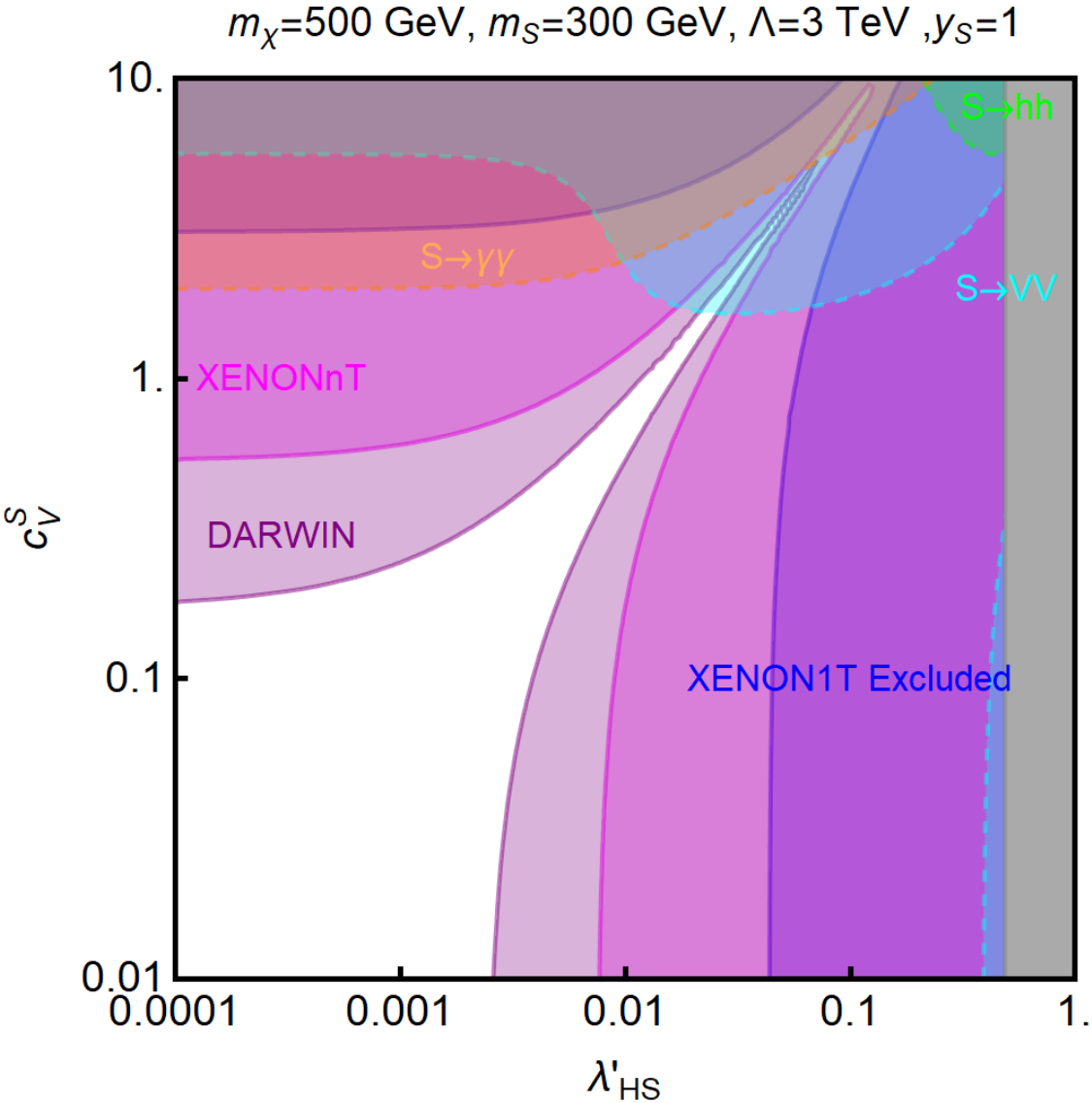}}
    \vspace{-0.2cm}
    \caption{\footnotesize{Summary of constraints in the $\lambda_{HS}^{'}-c_V^S$ plane for two of the benchmarks of fig.~\ref{fig:completingsimplifiedmodels}, employing the same color code.}}
    \label{fig:plot2D}
\end{figure}
\begin{figure}[b]
\begin{flushleft}
\hspace*{-1.8cm}\subfloat{\includegraphics[width=0.4\linewidth]{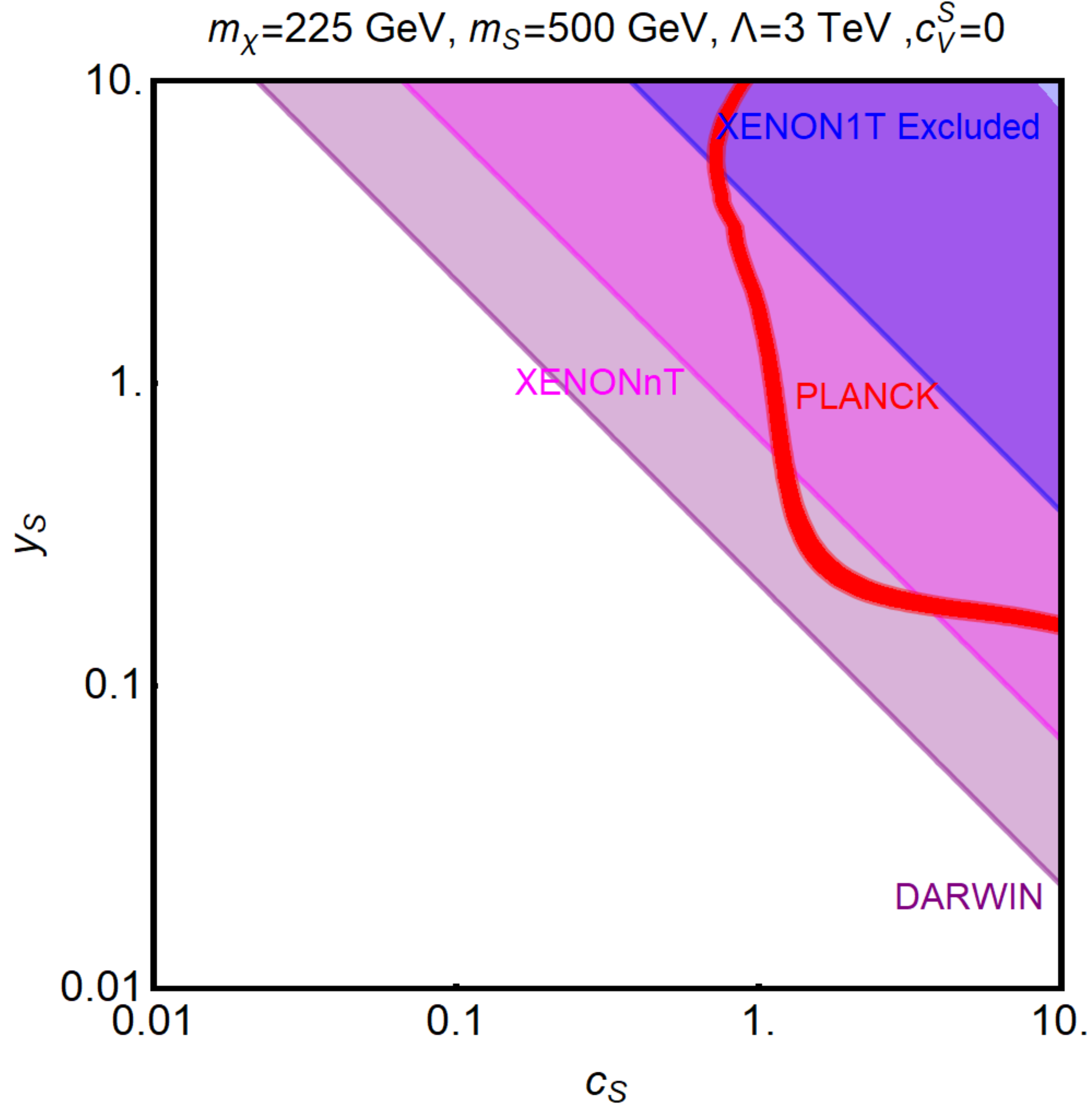}}
\subfloat{\includegraphics[width=0.4\linewidth]{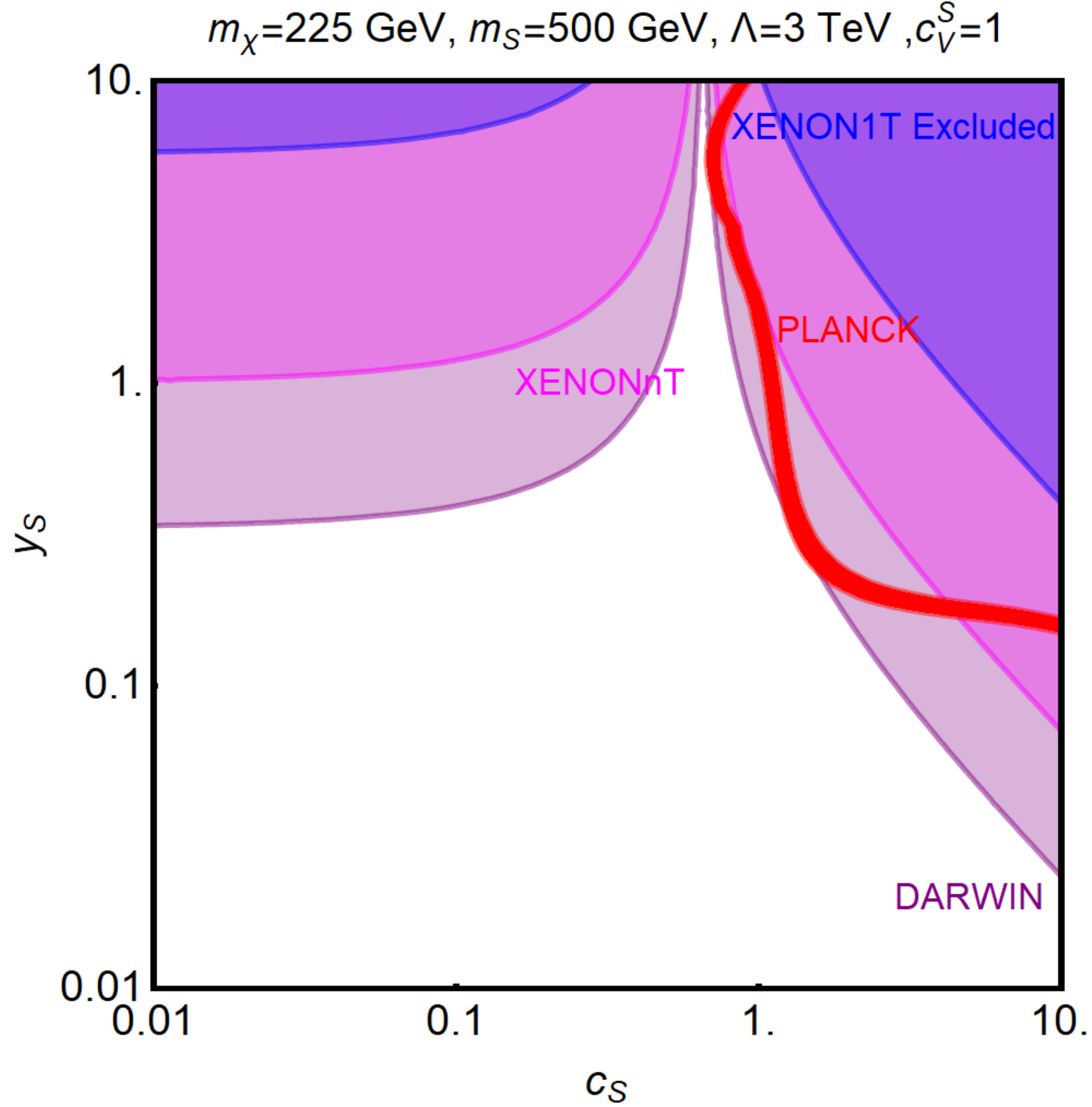}}
\subfloat{\includegraphics[width=0.4\linewidth]{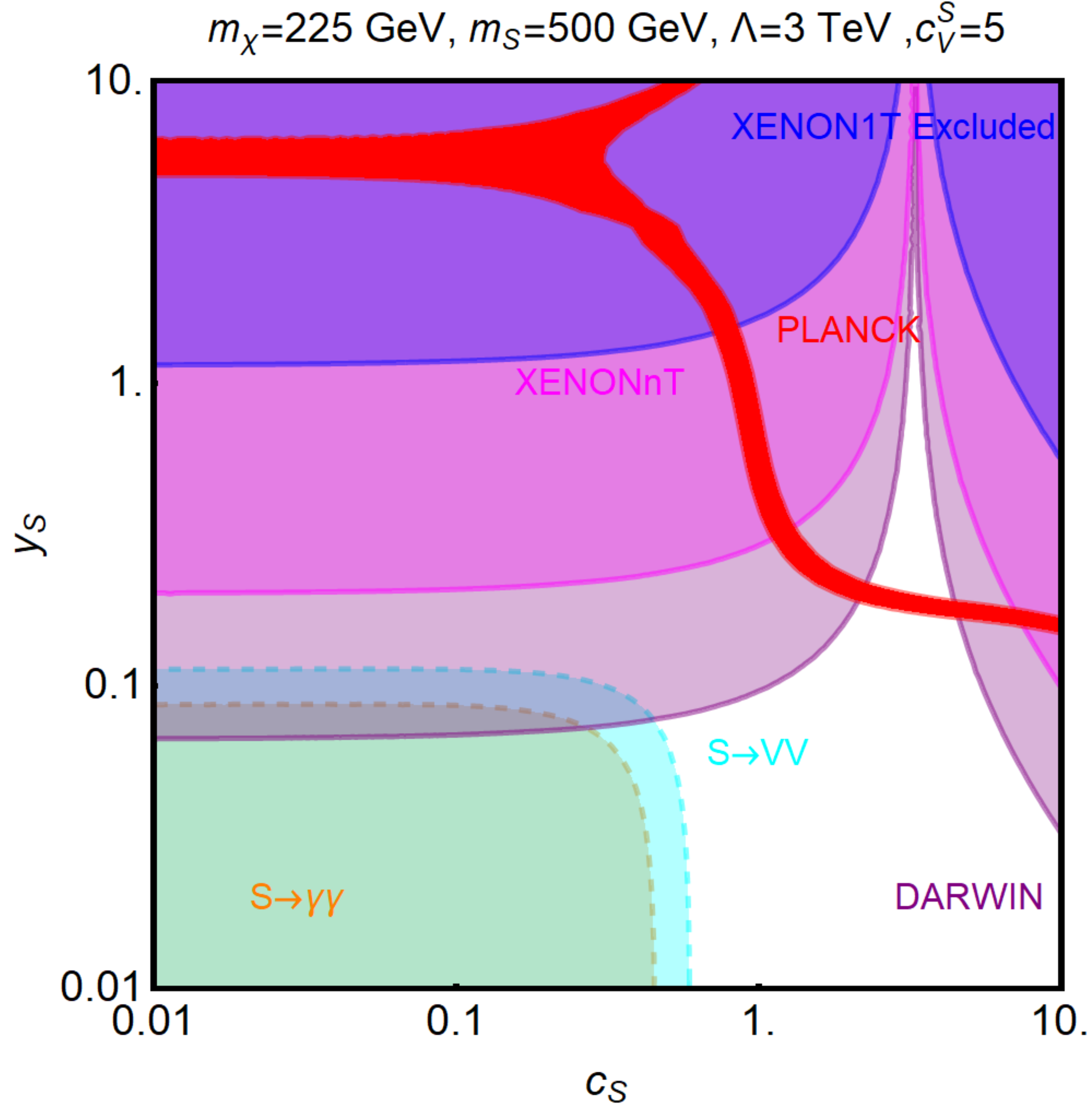}}\\
\hspace*{-1.8cm}\subfloat{\includegraphics[width=0.4\linewidth]{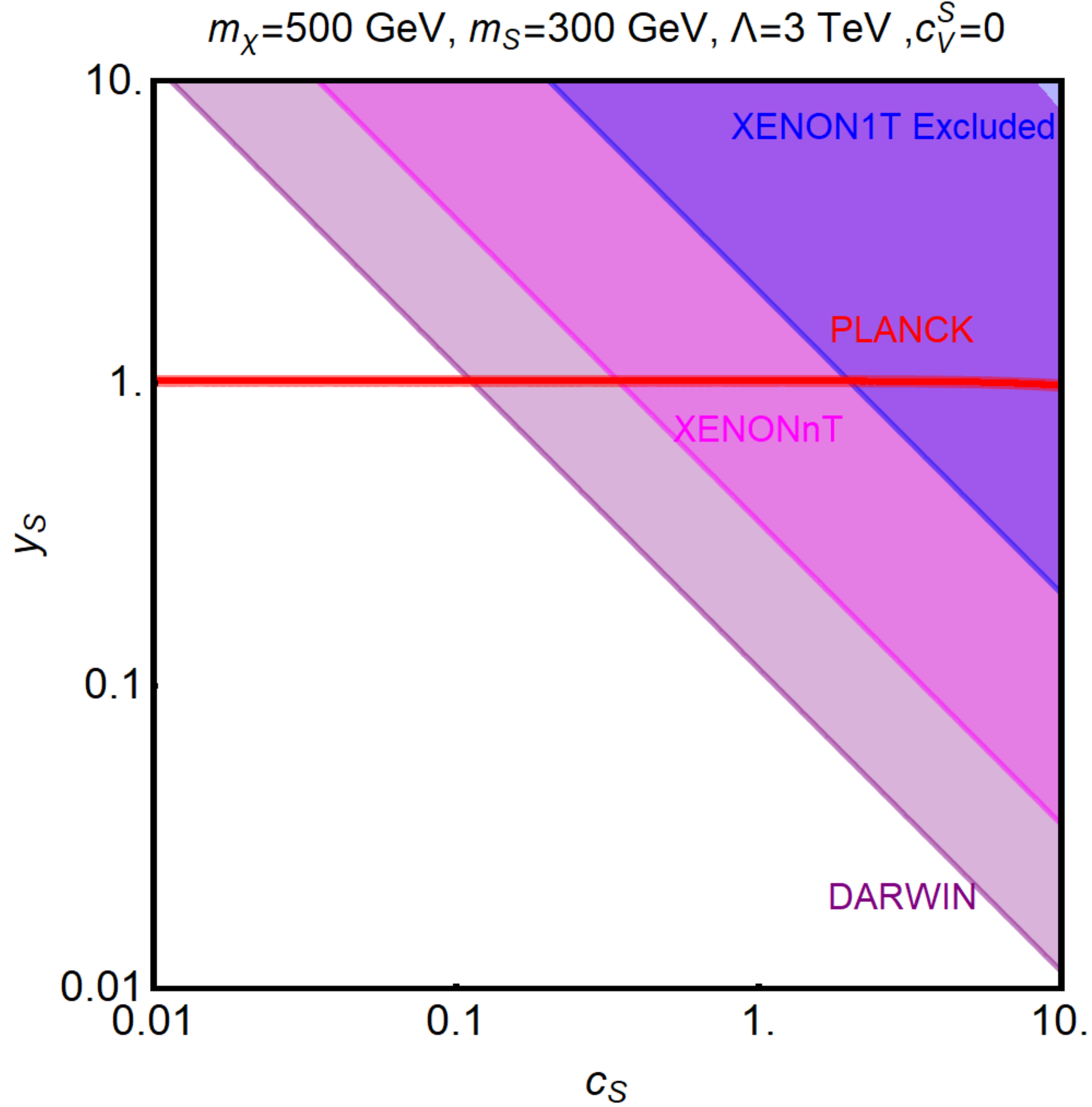}}
\subfloat{\includegraphics[width=0.4\linewidth]{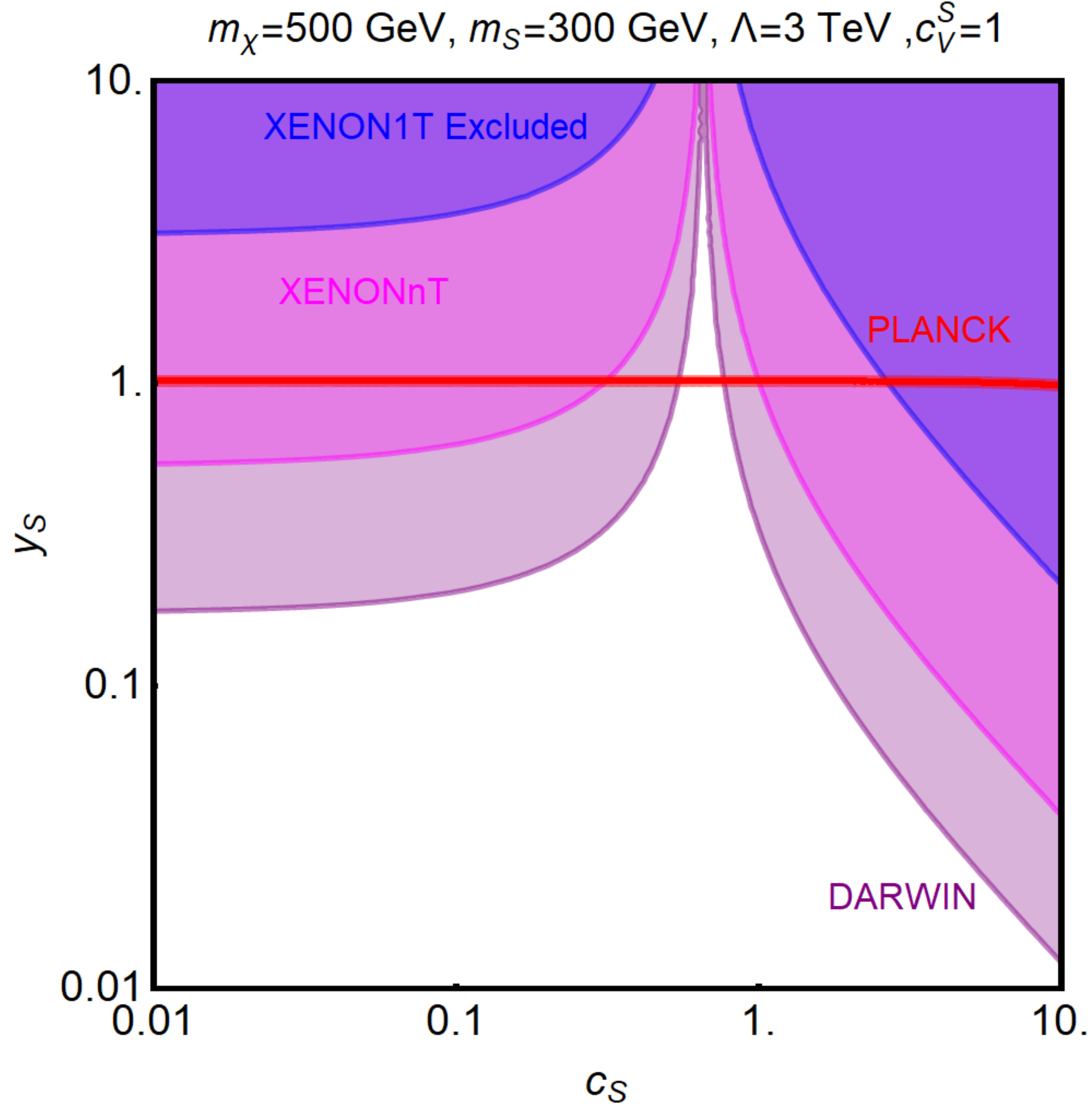}}
\subfloat{\includegraphics[width=0.4\linewidth]{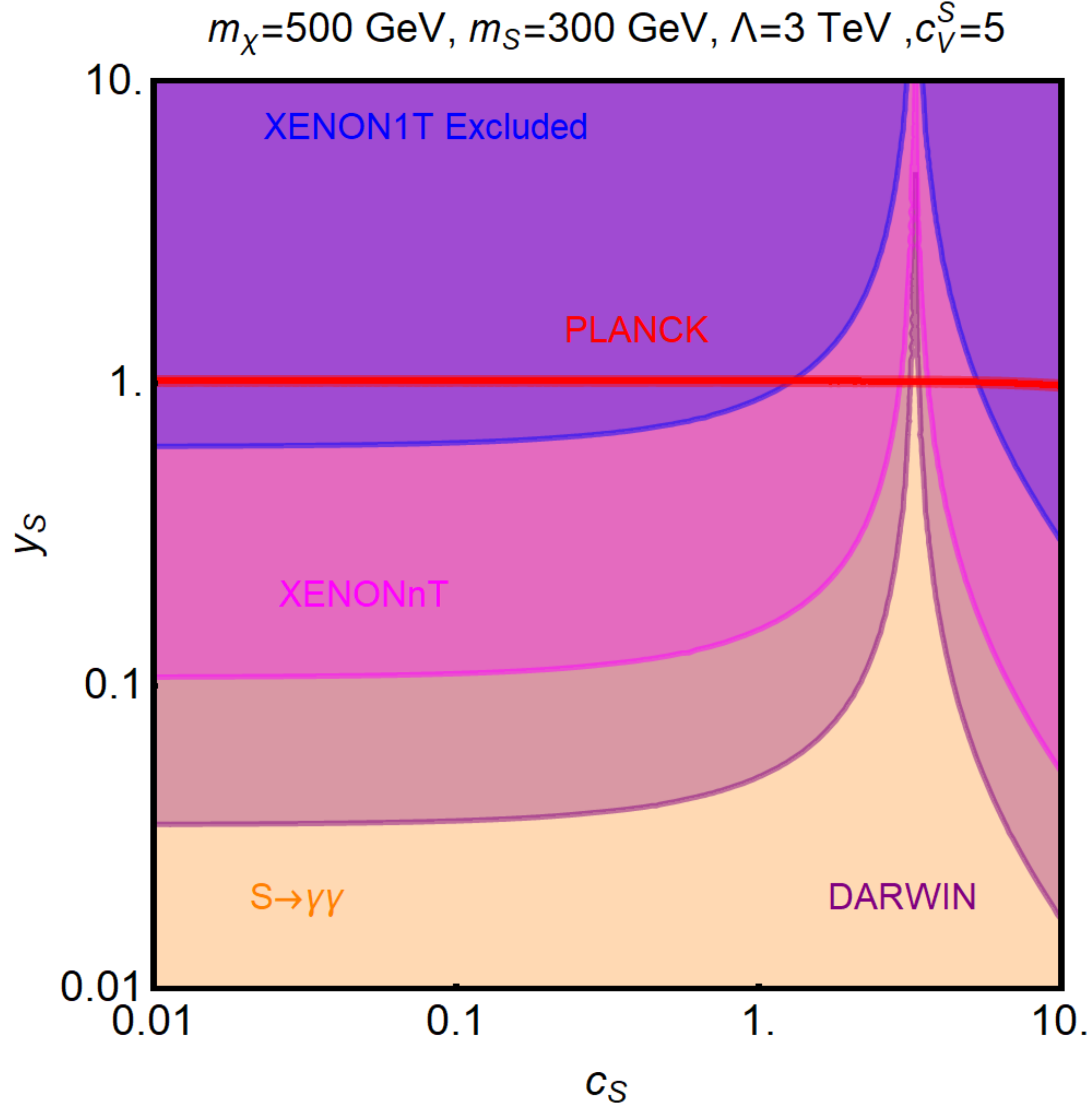}}
\end{flushleft}
\vspace{-0.5cm}
\caption{\footnotesize{Summary of constraints for the same benchmark masses as in fig.~\ref{fig:completingsimplifiedmodels} with same color code, but in the $c_S-y_S$ plane and setting the $h-S$ mixing to zero. The case $m_\chi=80\,\mbox{GeV}$, $m_S=200\,\mbox{GeV}$ is not shown, since it does not allow for the correct relic density.}}
\label{fig:completingsimplifiedmodels_bis}
\end{figure}

A remark about the results in fig.~\ref{fig:completingsimplifiedmodels} is in order. The collider constraints indeed rely strongly on the assumption of a universal coefficient $c_V^S$ and lifting this assumption would allow to evade them. Considering for example $c_{B}^S, c_{W}^S \ll c_{G}^S$ would remove the bounds from diboson searches and further open an interesting window at moderate $\lambda^{'}_{HS}$ in the secluded regime, which could evade projected XENONnT exclusions via destructive interference between different operators, but becomes testable at DARWIN.

To facilitate the understanding of the more general results discussed in the next section, in particular the cancellations in DD, 
we have re-expressed the results concerning the combined DM and collider constraints in the $(\lambda_{HS}^{'},c_V^S)$ plane for the two benchmarks with $(m_\chi,m_S)=$ $(225,500)\,{\rm GeV}$ and $(500,300)\,{\rm GeV}$. In both cases we have set $y_S=1$. The two panels in fig.~\ref{fig:plot2D} show very clearly the blind spot in the DD excluded regions due to the interplay of $\lambda_{HS}^{'}$,  responsible for the interaction of $S$ with SM fermions through mixing with the Higgs, and the $c_{V}^S$ coupling. 
Collider bounds may emerge when $c_V^S$ becomes larger than one. The second panel does not show a relic density isocontour since the DM abundance is determined entirely by the $\chi \chi \rightarrow SS$ process and is then fixed by $y_S$, irrespective of the values of $\lambda_{HS}^{'}$ and $c_V^S$. We also notice a non trivial interplay between the $\lambda'_{HS}$ and $c_V^S$ couplings regarding the shape of the excluded regions from diboson searches. In particular for both $c_V^S \sim O(1)$ and $\lambda'_{HS}\sim O(1)$ a destructive interference can be present between the different contributions to the production process of the resonance.

Before moving to the systematic survey of the parameter space, we finally consider, in fig.~\ref{fig:completingsimplifiedmodels_bis}, a different combination of portals, turning on the $D=5$ Yukawa and gauge portals with non-vanishing couplings $c_S$ and $c_V^S$, while setting the $h-S$ mixing to zero.  We stick to the same benchmark masses as in fig.~\ref{fig:completingsimplifiedmodels}, excluding the case of $(m_\chi,m_{S})=(80,200)\,\mbox{GeV}$,

\FloatBarrier
\noindent
since here the correct relic density cannot be achieved. We notice a globally weaker impact from constraints from DD, because the SM-like Higgs does no longer act as a mediator.
 Moreover, collider bounds from diboson resonances are weaker, due to the absence of $h-S$ mixing. Conversely, the lower decay branching fraction of the resonance into massive gauge bosons makes the bounds from searches for diphoton resonances stronger.

\subsubsection{Scanning the parameter space}

We finally survey the full \eDMEFT\ considering the {\it simultaneous} presence of all parameters identified at the beginning of this section, with the mentioned restrictions.
We thus perform a scan within the ranges: 
\begin{align}
\label{eq:general_scan_ranges}
     m_\chi &\in [10,1000]\,\mbox{GeV}\nonumber\\
    m_S &\in [10,1000]\,\mbox{GeV}\nonumber\\
    \lambda_{HS}^\prime &\in [10^{-4},1] \nonumber\\
        y_S &\in [10^{-2},10] \\
    c_S &\in [10^{-2},10]\nonumber \\
    c^S_{V} &\in [10^{-2},10]\nonumber\\
    y_S^{(2)} &\in [10^{-2},10]\,,\nonumber
\end{align}
with always $\Lambda=3\,$TeV.

\begin{figure}
    \centering
    \subfloat{\includegraphics[width=0.44\linewidth]{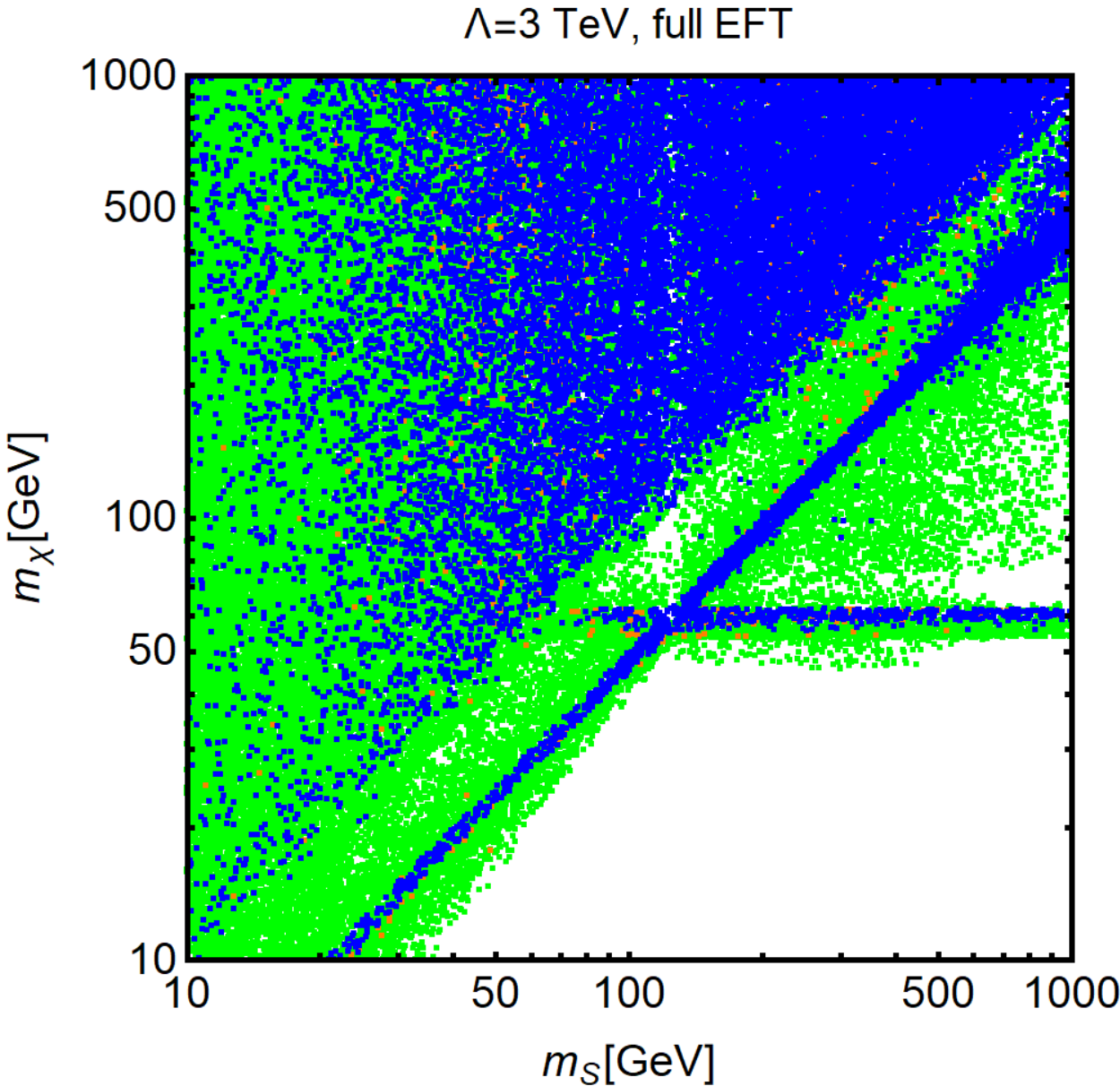}}
    \subfloat{\includegraphics[width=0.44\linewidth]{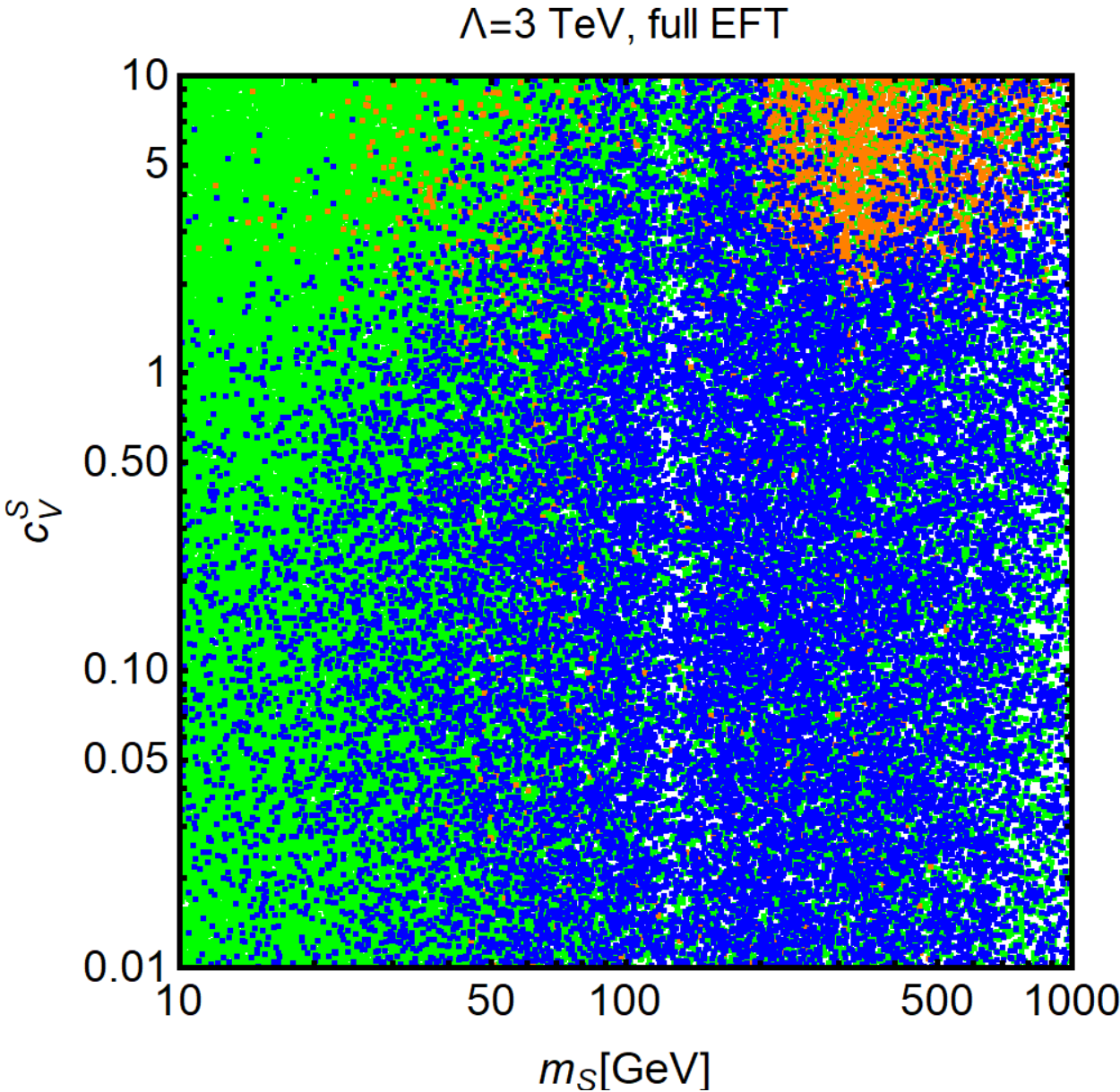}} \\
    \subfloat{\includegraphics[width=0.44\linewidth]{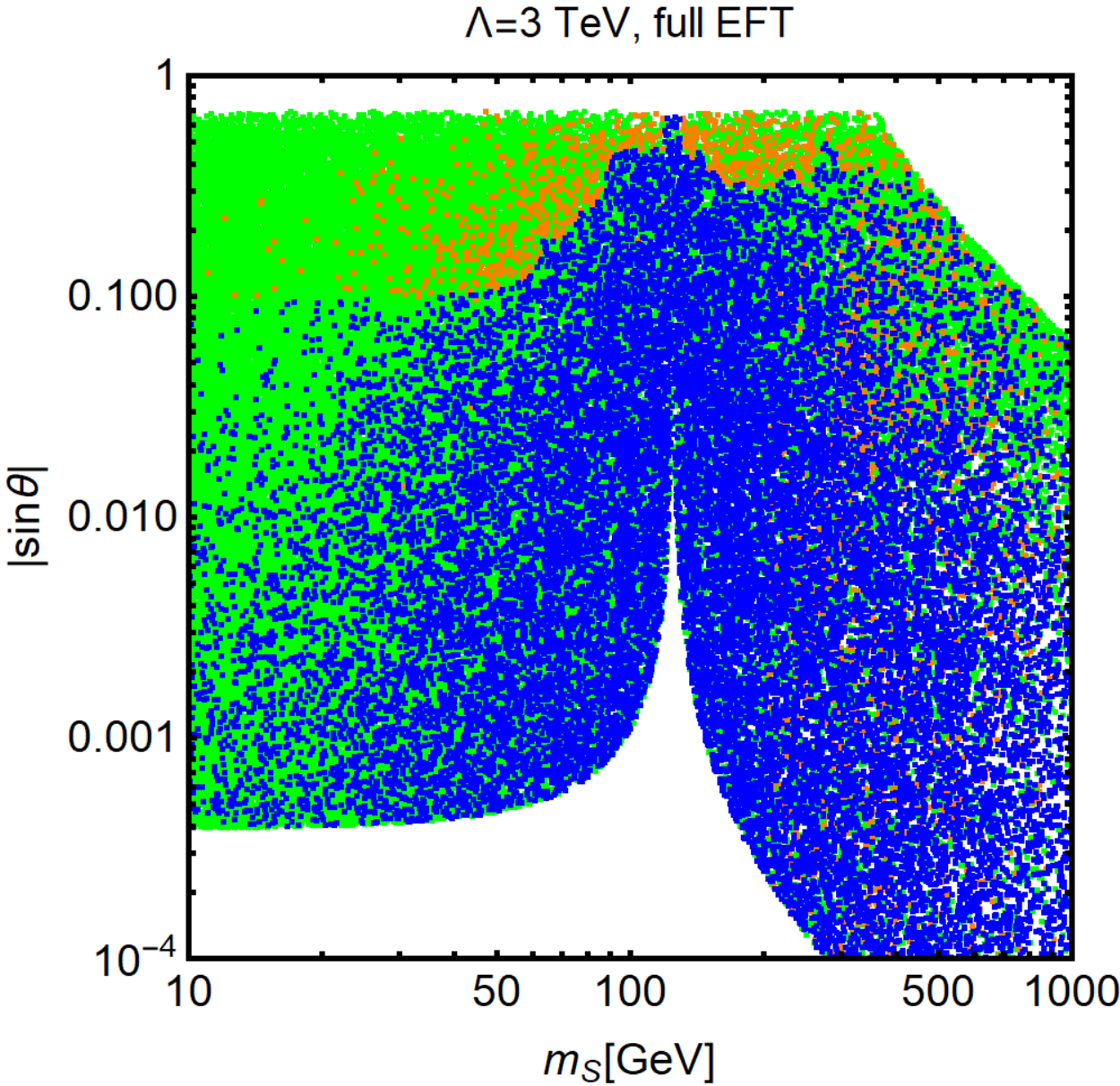}}
    \subfloat{\includegraphics[width=0.44\linewidth]{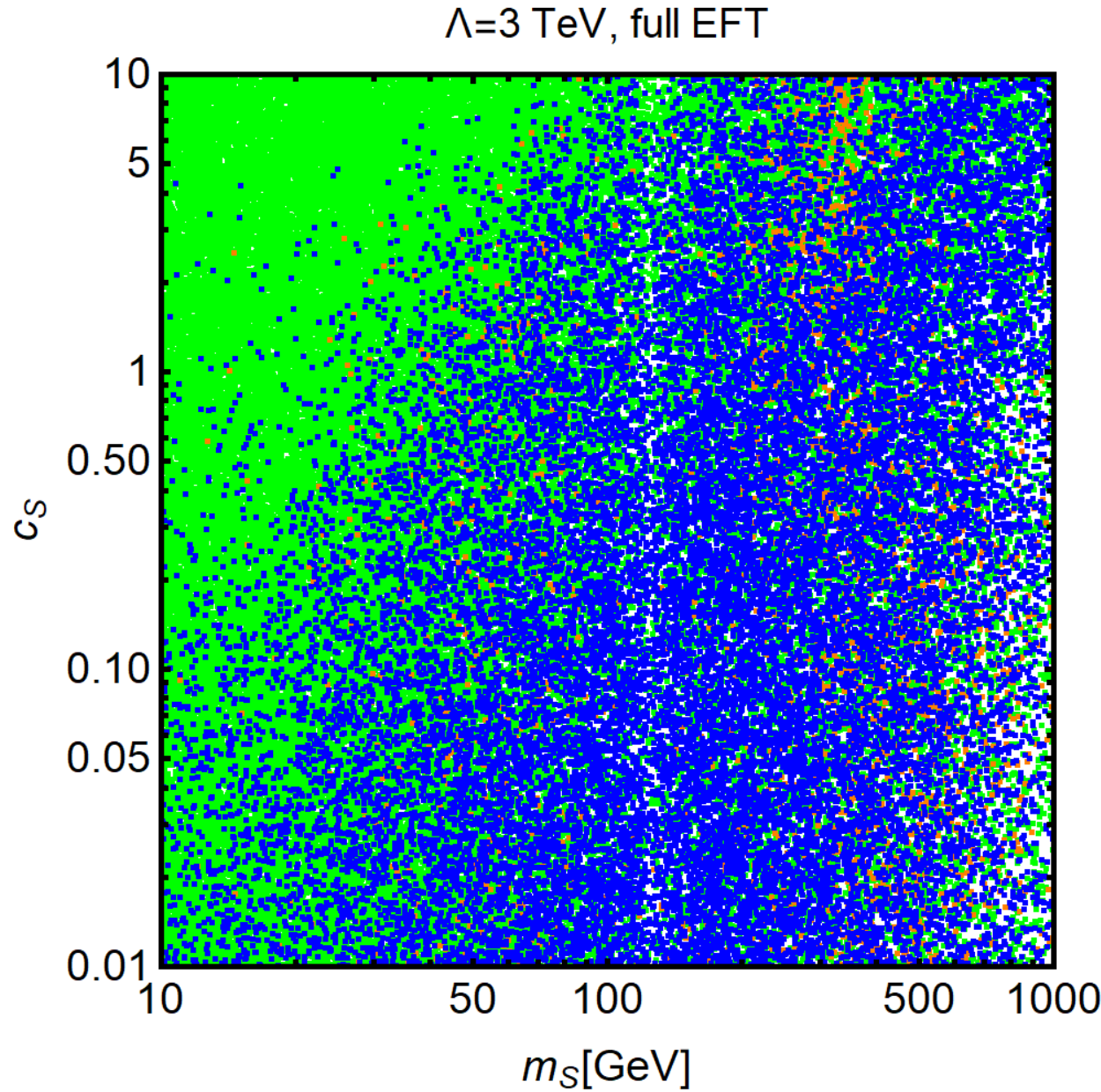}}
    \caption{\footnotesize{Results of the parameter scan for the scalar \eDMEFT\ setup in the $m_S-m_\chi$, $m_S-c_{V}^S$, $m_S-|\sin\theta|$, and $m_S-c_S$ planes. The green points provide the correct relic density according to the WIMP paradigm, while the orange points are, in addition, compatible with constraints from DD. The blue points are, finally, also compatible with collider constraints. See main text for details on the scan and the constraints accounted for.}}
    \label{fig:scan1}
\end{figure}

For each configuration obtained in this way, we compute a comprehensive set of observables. We consider relic density, the SI scattering cross section, the invisible width of the Higgs, the LHC monojet rate and the production cross sections of diboson resonances listed above. In addition, we apply the general bounds to the mixing between the Higgs and a real scalar singlet as determined e.g. in refs \cite{Falkowski:2015iwa,Huitu:2018gbc}.  In fig.~\ref{fig:scan1}  the results of our analysis are shown. We project the points found in our scan into the $m_S-m_\chi$ (upper left), $m_S-c_{V}^S$ (upper right), $m_S-|\sin\theta|$ (bottom left), and $m_S-c_S$ (bottom right) planes. 
The color code identifies three sets of model points:
\begin{itemize}
    \item {\it green}: points that account for the correct relic density but are otherwise excluded;
    \item {\it orange}: points allowed by the relic density and DD but excluded by collider constraints;
    \item {\it blue}: points which satisfy all constraints\,.
\end{itemize}

\begin{figure}
    \centering
    \subfloat{\includegraphics[width=0.44\linewidth]{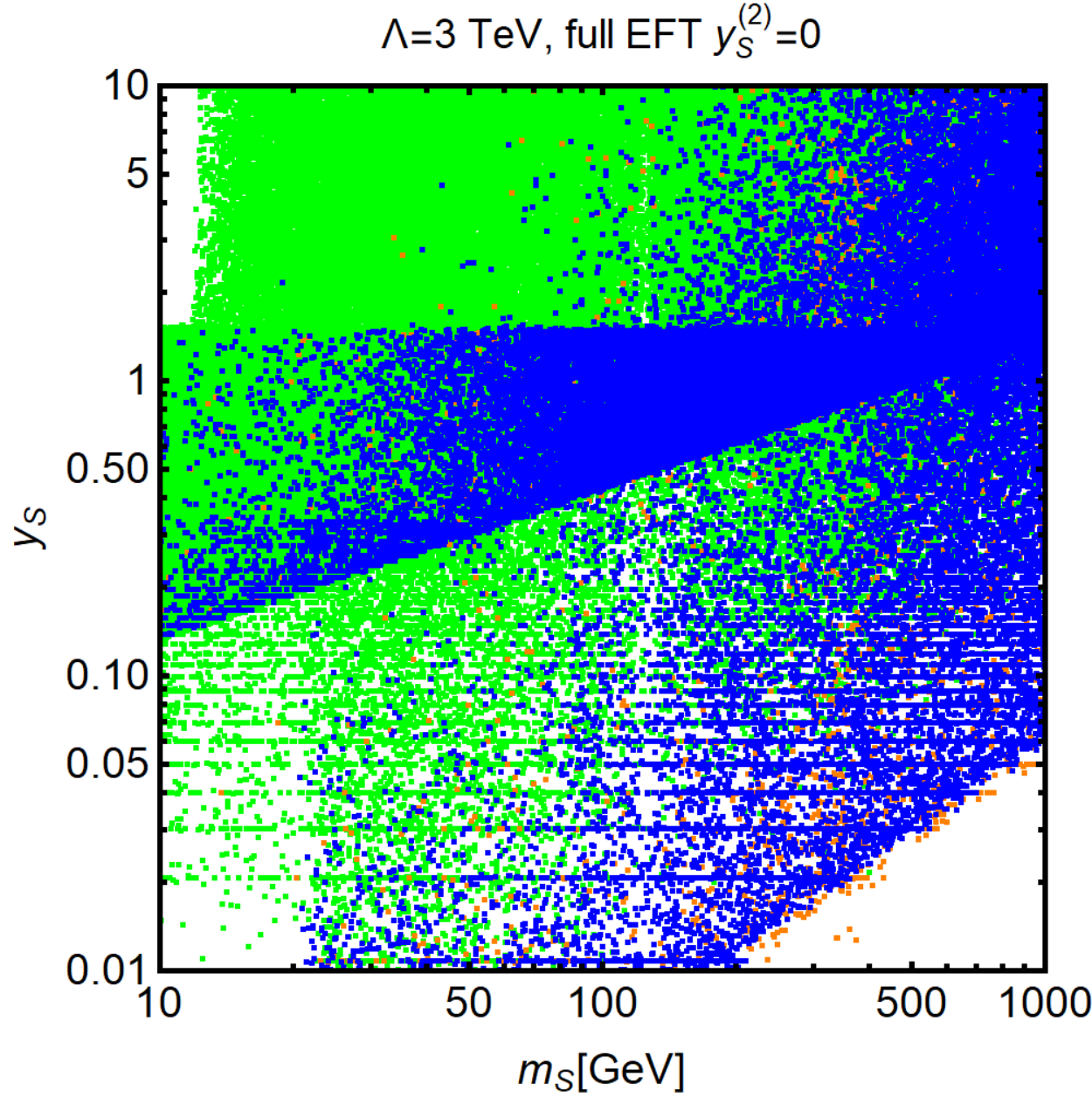}}
    \subfloat{\includegraphics[width=0.44\linewidth]{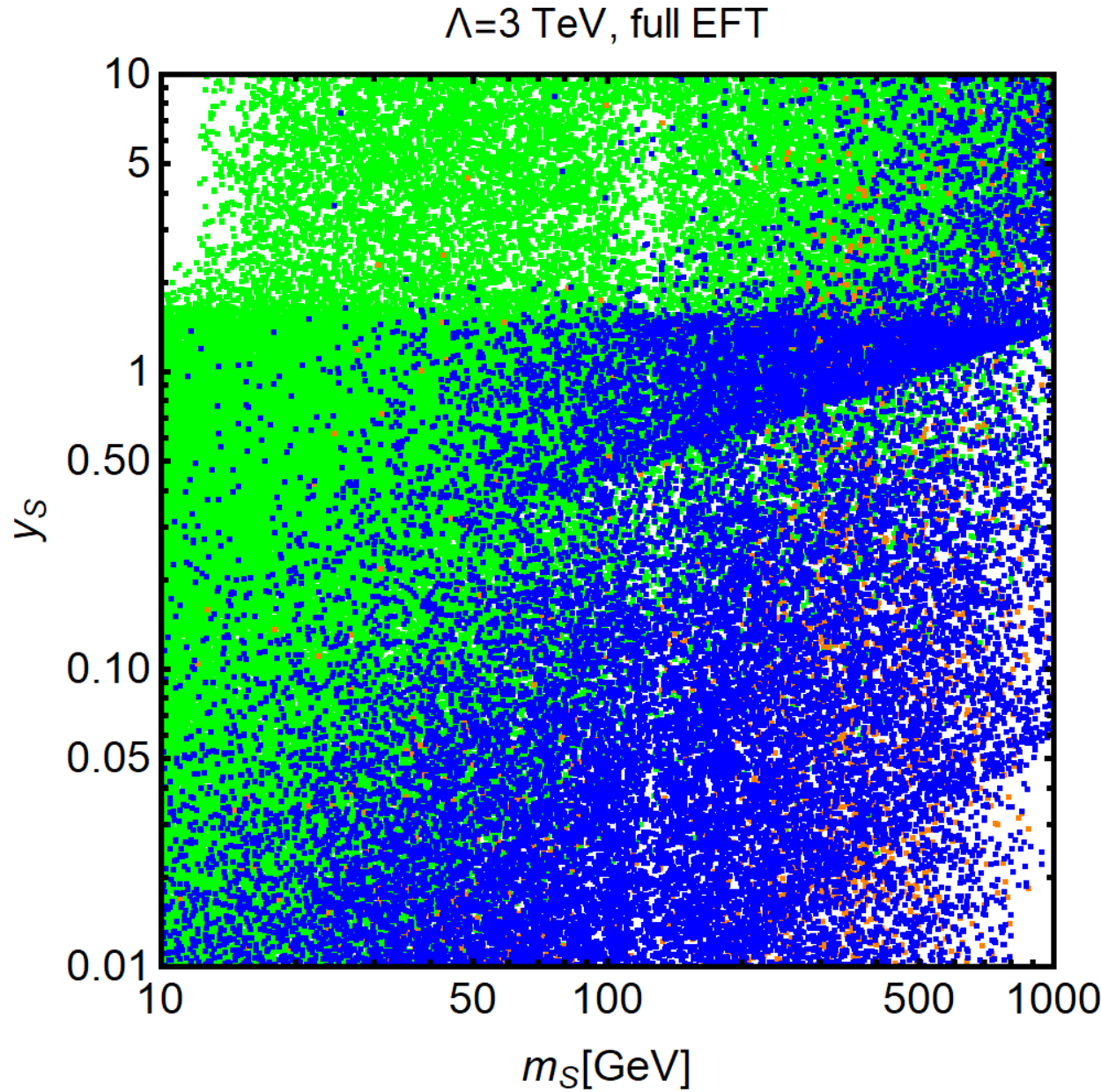}} \\
    \subfloat{\includegraphics[width=0.44\linewidth]{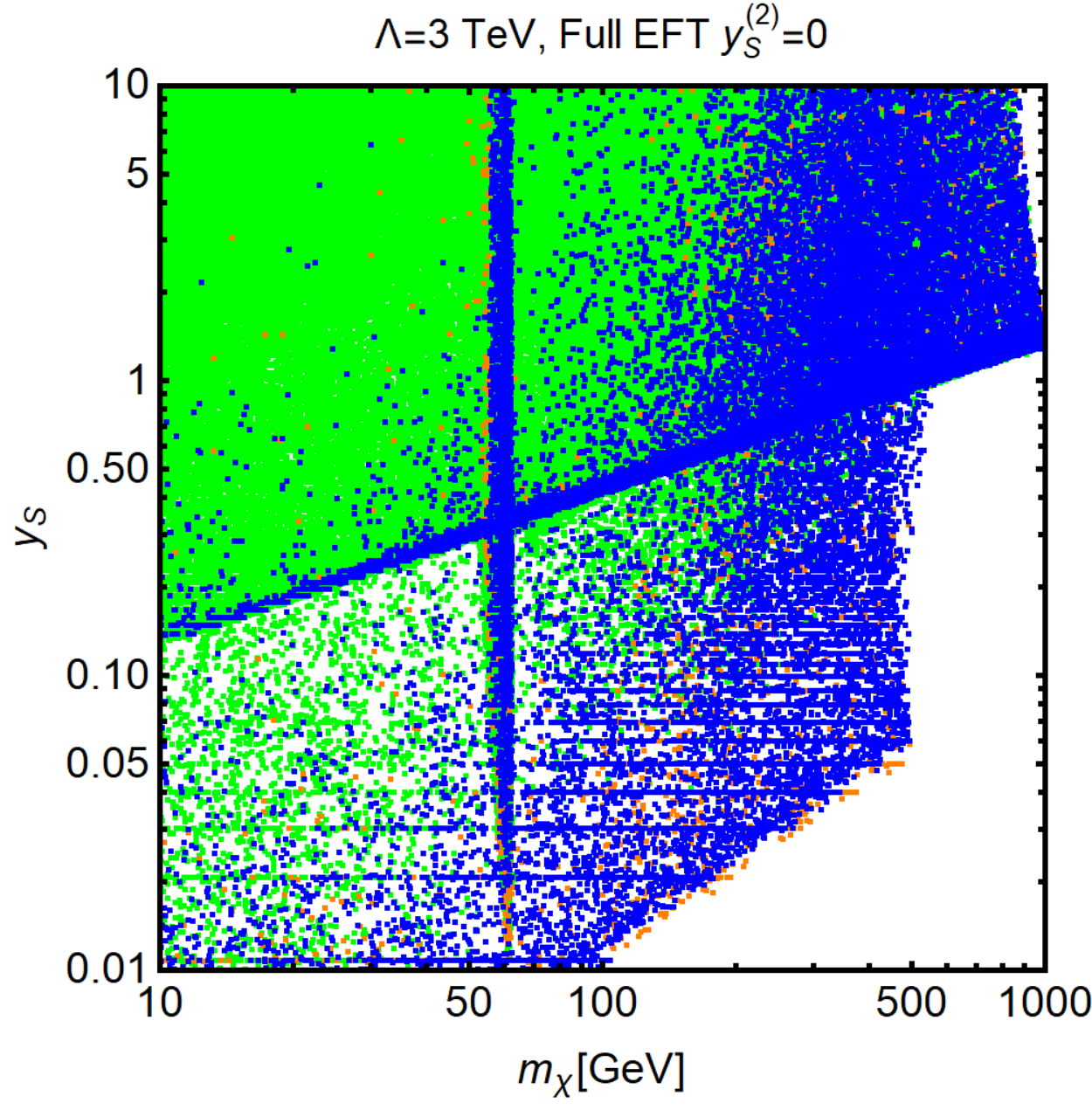}} 
    \subfloat{\includegraphics[width=0.44\linewidth]{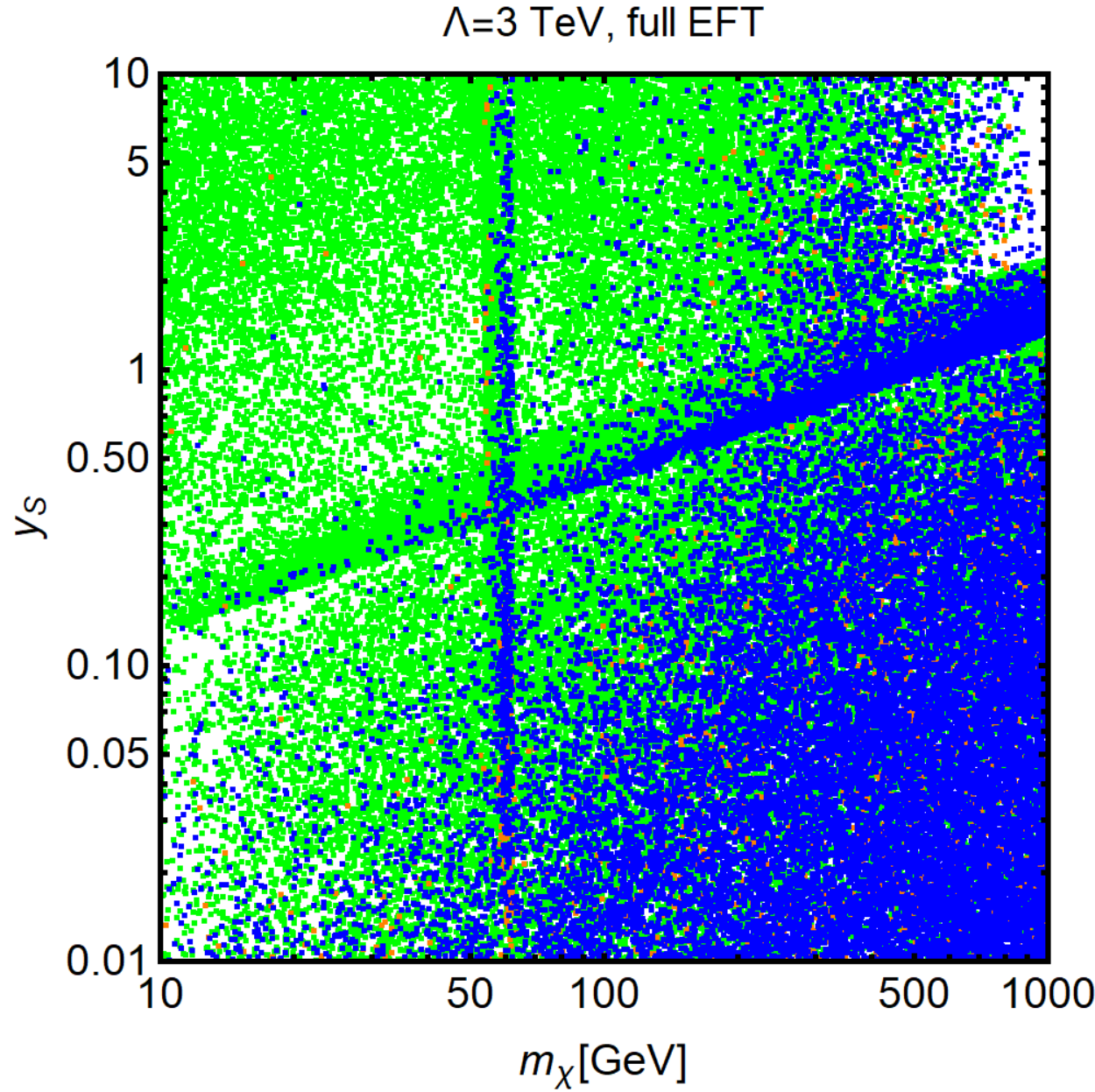}} 
    \caption{    
     \footnotesize{Results of the parameter scan for the scalar \eDMEFT\ setup in the $m_S-y_S$ and $m_\chi-y_S$ planes, following the same color code as in fig.~\ref{fig:scan1}. 
     While the right panels correspond to the full \eDMEFT, in the left panels we have set $y_S^{(2)}=0$.}
     }
    \label{fig:scan1_bis}
\end{figure}

\begin{figure}
    \centering
    \subfloat{\includegraphics[width=0.46\linewidth]{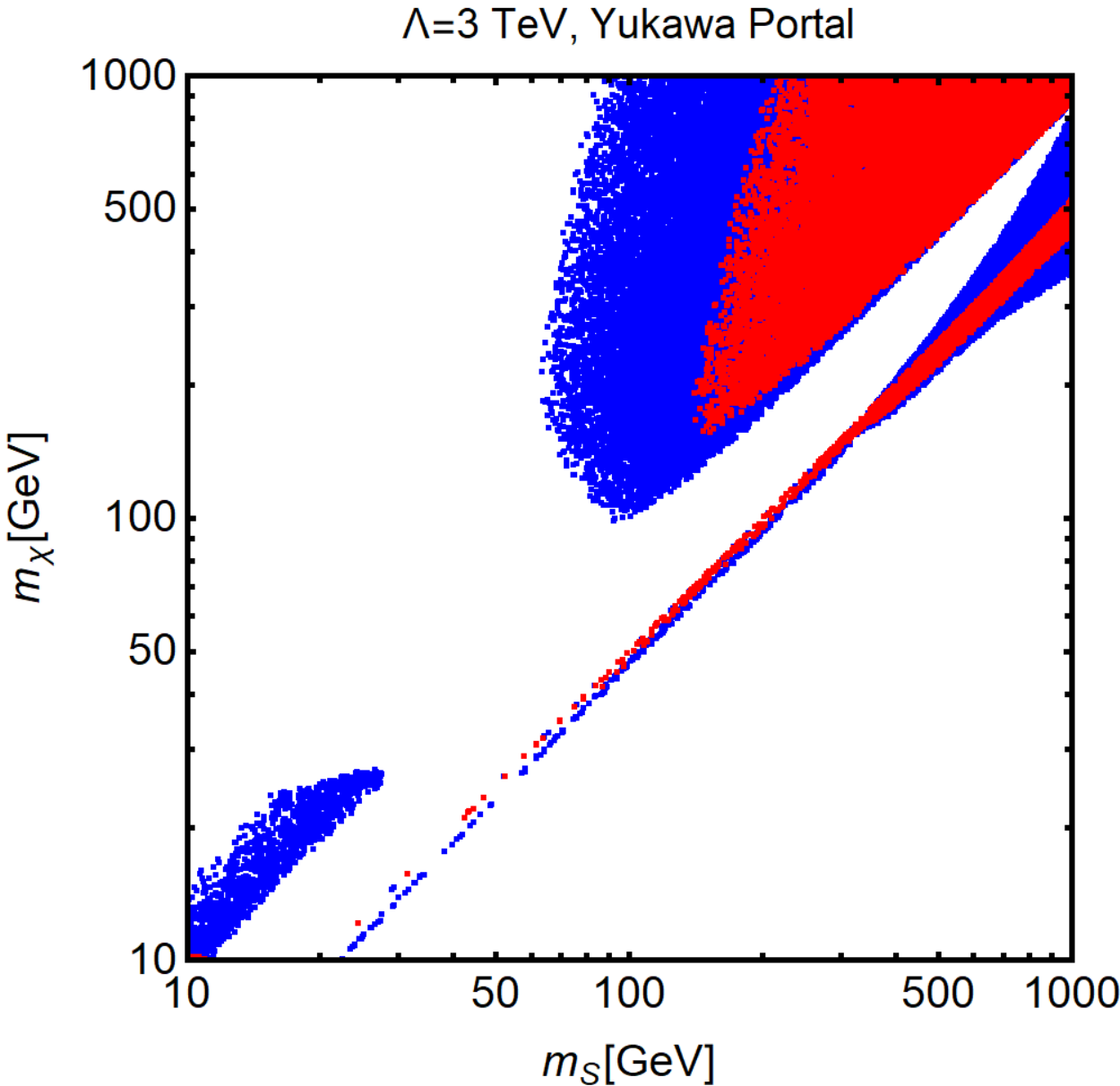}}
    \subfloat{\includegraphics[width=0.46\linewidth]{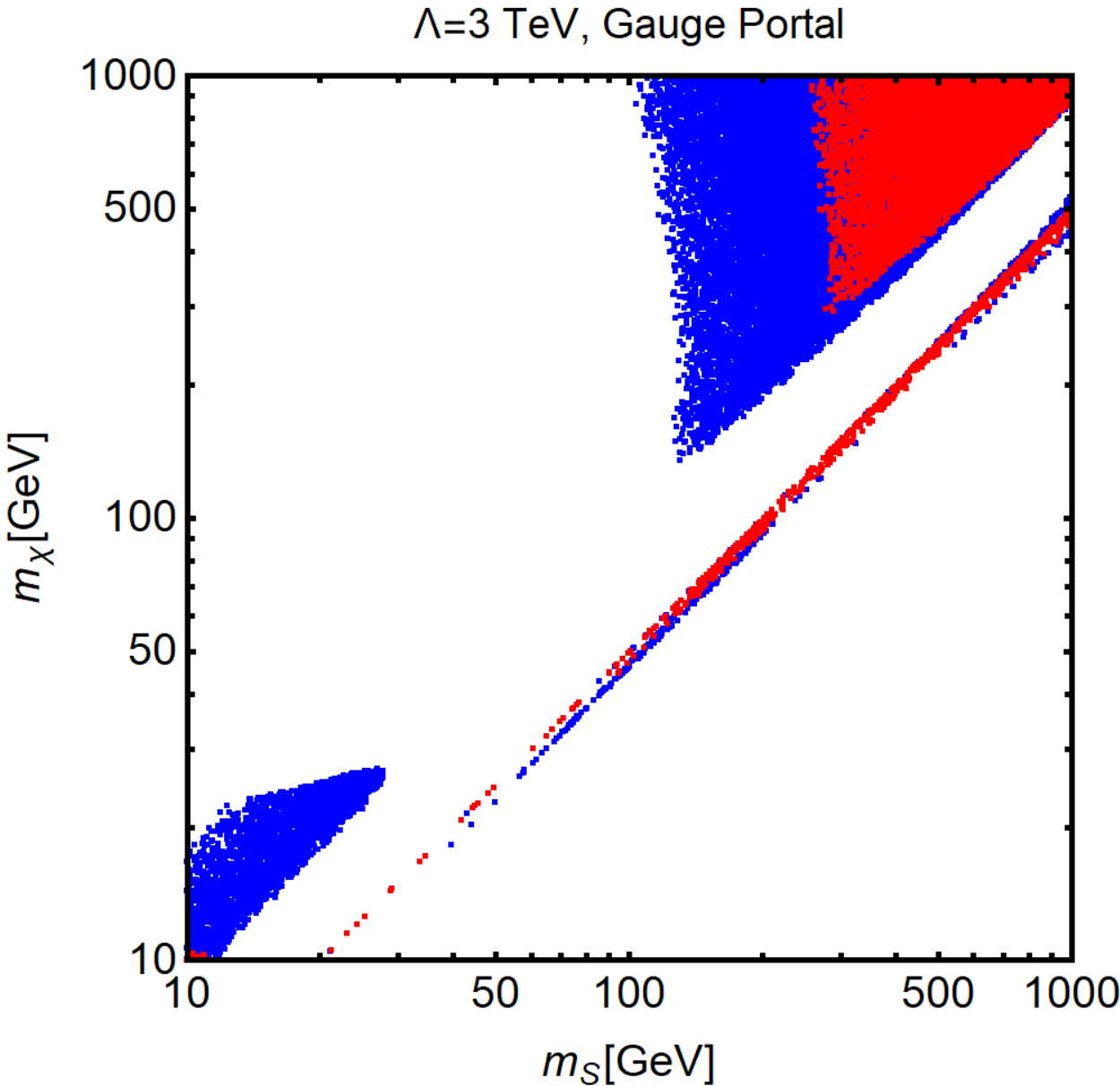}} \\
    \subfloat{\includegraphics[width=0.46\linewidth]{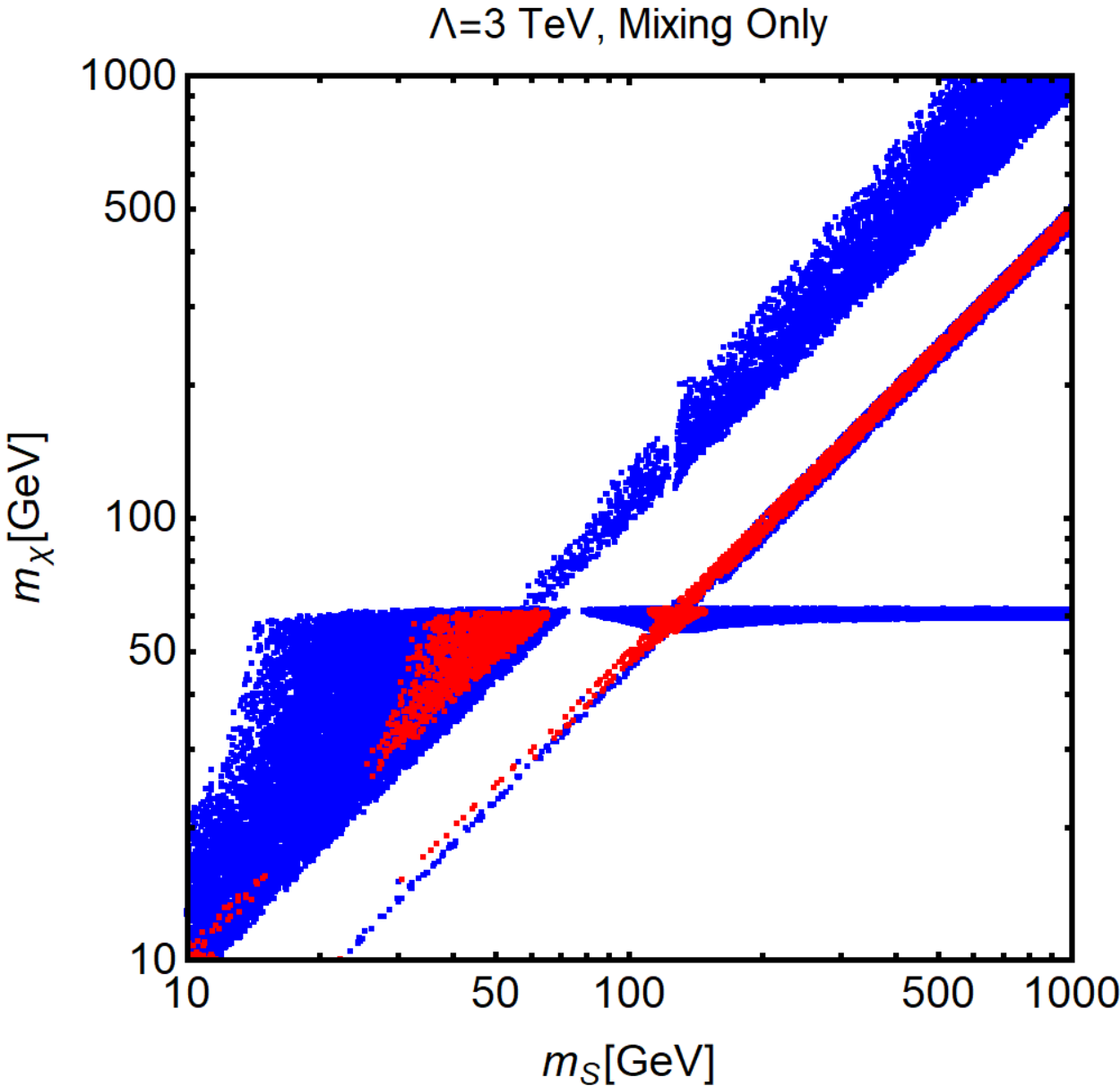}} \\ 
    \subfloat{\includegraphics[width=0.46\linewidth]{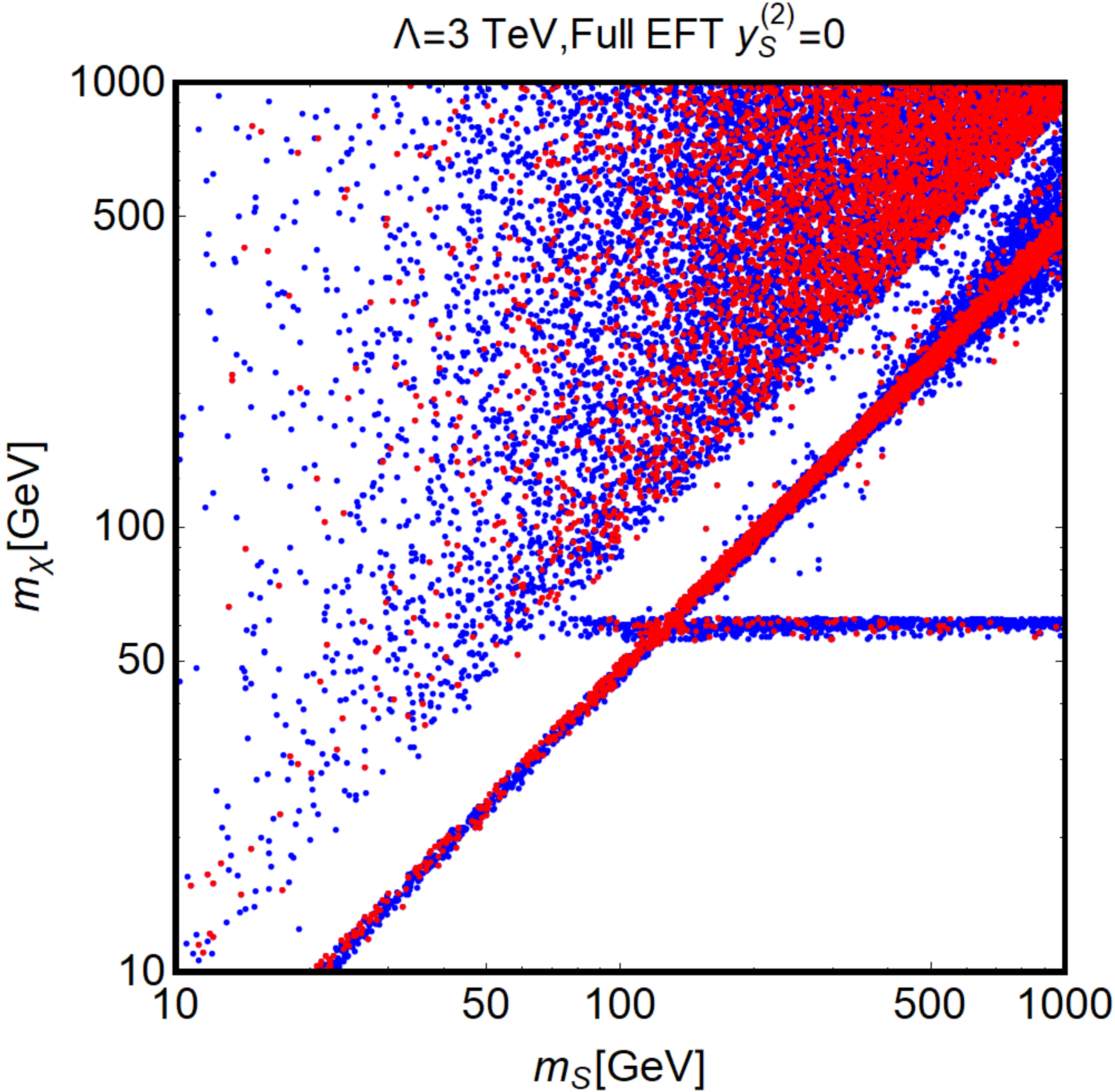}}
    \subfloat{\includegraphics[width=0.46\linewidth]{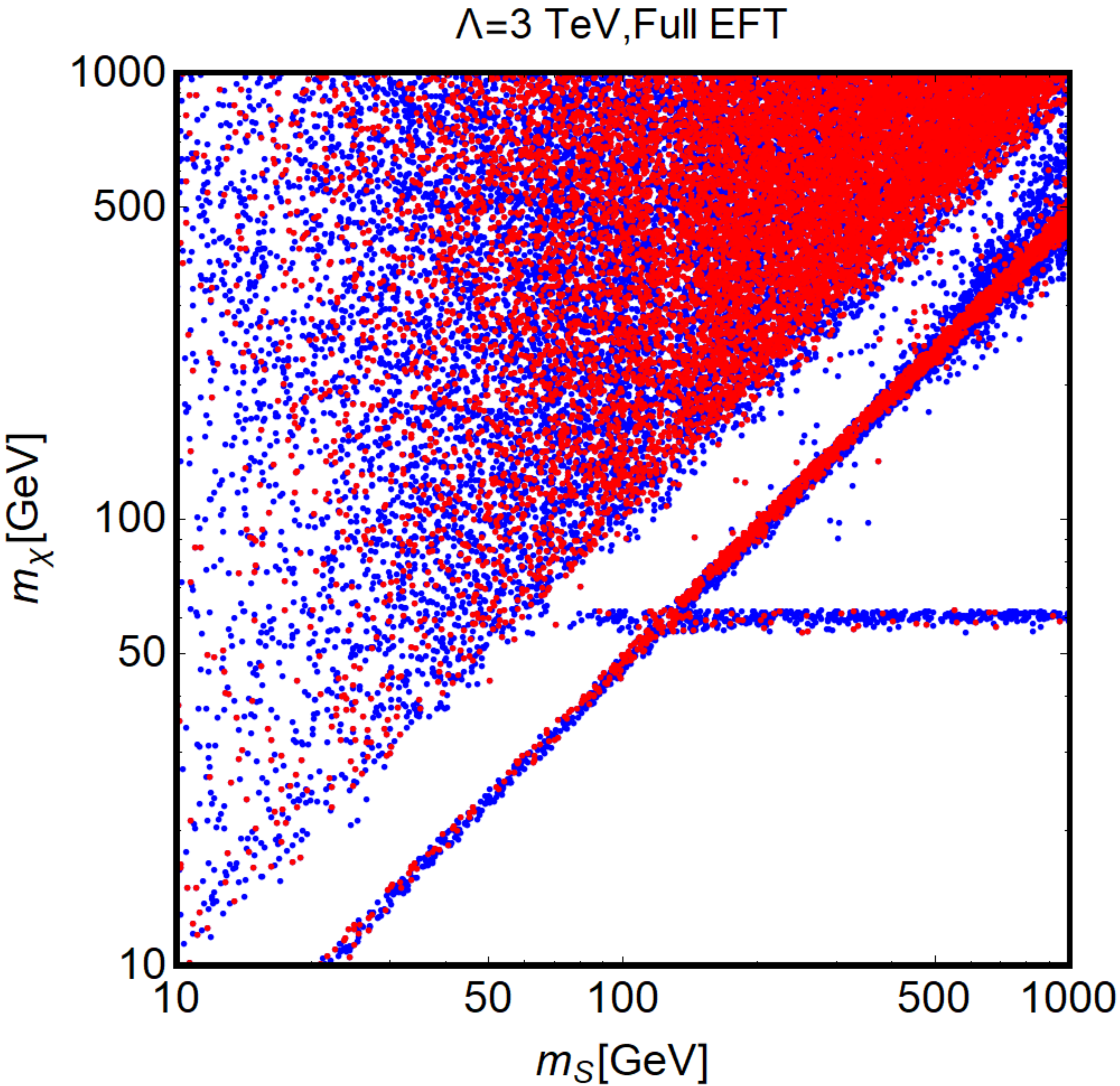}}
    \caption{\footnotesize{Viable model points for the $D=5$ flavor-diagonal Yukawa portal (upper left), $D=5$ scalar gauge portal (upper right), Higgs mixing portal (center),  and full \eDMEFT\ with (lower right) and without (lower left) considering $y_{S}^{(2)}$, in the $m_S-m_\chi$ plane. While the blue points satisfy all DM and collider bounds entertained before, the red points represent configurations evading also a projected XENONnT exclusion.}}
    \label{fig:compScan}
\end{figure}

Among the different panels of fig.~\ref{fig:scan1}, the $m_S-m_\chi$ plot in the first panel is particularly illustrative. As evident, the regions with the highest density of viable points (blue points), are the special kinematical regions already identified in the previous sections: the secluded regime $m\chi> m_S$, in particular with $m_S$ above 100 GeV in order to avoid DD constraints, and the $m_h/2$, $m_S/2$ poles. Still, as we will explore in detail below, there emerge important quantitative differences and new regions in parameter space do open up in the \eDMEFT\ allowing in particular to survive projected XENONnT constraints in significant portions of parameter space. 
Another notable feature of the first plot in fig.~\ref{fig:scan1} is the relatively small number of orange points, compared to the green and blue ones. This suggest that, at least for what concerns present bounds, the ones from DD are the most severe. As shown in the second plot of fig.~\ref{fig:scan1}, collider bounds primarily depend on the value of the $c_{V}^S$ coupling and are relevant only for values of the latter above one and for $m_S \gtrsim 200\,\mbox{GeV}$. It should be kept in mind that the impact of collider constraints depends on the hypothesis $c_G^S=c_B^S=c_W^S=c_V^S$ and results might change in case the latter is lifted.  
One can finally also inspect how larger mediator masses $m_S$ allow for bigger couplings to SM states regarding the DD constraints. The lower panel of fig.~\ref{fig:scan1} shows the impact of the bounds, applied in our study, on $\sin\theta$ and $c_{S}$. As can be seen, the latter parameter appears to be constrained only by DD and mostly for $m_S \lesssim 100\,\mbox{GeV}$. In the case of $\sin\theta$, instead, the combination of searches illustrated in the previous section forms a useful complement for $50\,\mbox{GeV} \lesssim m_S \lesssim 300\,\mbox{GeV}$.

In fig.~\ref{fig:scan1_bis} we continue to explore the parameter space, focusing now on the possible size of $y_S$ in dependence on $m_S$ (upper panel) and $m_\chi$ (lower panel). In addition we confront the full \eDMEFT\ (right panel), as defined at the beginning of sec.~\ref{sec:ResS}, with a scenario without the bi-quadratic $S^2\bar \chi \chi$ interaction (left panel). In the upper left plot, one can clearly identify the `secluded region' around $y_S \lesssim 1$, opening up for smaller $m_S$. Including the couplings $y_S^{(2)}$, as done in the upper right plot, shifts the points significantly towards smaller $y_S$, since now the bi-quadratic mediator-DM portal allows for efficient annihilation of the DM.
This relaxes DD constraints in the full \eDMEFT. A similar trend is visible in the $m_\chi - y_S$ plane, shown in the lower panel. Furthermore, the latter plane evidences a narrow strip of viable model points for $m_\chi \simeq 60\,\mbox{GeV}$, corresponding to the $m_\chi \simeq m_h/2$ resonance.

In order to improve the understanding of our results, and to further pinpoint the viable regions of parameter space that conventional simplified models miss, we now disentangle the fundamental portals. Thus, in addition to the general scan over the full set $(m_\chi,\, m_S,\, y_S,\, \lambda^\prime_{HS},\, c_S,\, c_{V}^S,\, y_S^{(2)})$, we have performed dedicated scans of the restricted set of parameters corresponding to the different portals taken individually. The corresponding intervals remain as given in eq.~\eqref{eq:general_scan_ranges}.
In fig.~\ref{fig:compScan} we compare the viable regions of parameter space, in the $m_S-m_\chi$ plane, populated by these portals, confronting them with the results of the full \eDMEFT. 
For simplicity, we consider only points that simultaneously satisfy all the constraints entertained before, and add, as complementary information, the projected XENONnT bounds, with points passing also the latter depicted in red. This makes particularly transparent which scenarios can remain open while constraints from DD get even tighter.
One can easily see that regarding current bounds the $D=5$ Yukawa and gauge portals, shown in the upper panel, tend to occupy rather similar regions of the parameter space, mostly restricted to the secluded and $m_S/2$ pole regimes. However, the former portal features a larger viable region around $m_{\chi} \approx m_S/2$ for $m_S\geq 2 m_t$.  In this regime the decay $S \rightarrow \bar tt$ is allowed which increases the total decay width of the mediator. Since a wider resonance boosts the annihilation cross section further away from the exact resonance condition the allowed region also broadens. Here, it could be interesting to also consider correlated limits from mono-Higgs final states, that necessarily emerge, but are not captured in the simplified Yukawa portal. 
The mixing portal, explored in the central panel, exhibits a significantly smaller population in the secluded region, especially for lower $m_S$, since even the modest scalar couplings considered in our scan are under pressure from direct detection in this regime. 
However, it adds a pronounced $m_\chi \approx m_h/2$ resonance region.

Once the `full' \eDMEFT\ is finally considered, as done in the lower panel, we see a significant extension of the allowed parameter space, which can be traced back to basically two effects: non-trivial interplay between the different operators, originating for example from blind spots in DD, and possible new DM annihilations via the $S^2 \bar \chi \chi$ contact term, as explained before.
To disentangle these effects, here the EFT without (with) $y_S^{(2)}\!>\!0$ is shown in the left (right) plot.
We see that even with $y_S^{(2)}=0$ the parameter space increases notably due to the operator interplay, leading to a broader region of points around the 
$m_{\chi} \approx m_S/2$ resonance and the opening of the region towards lighter mediators of $m_S \lesssim 100$\,GeV for moderate and larger DM masses. 
Including finally the coupling $y_S^{(2)}$ fully opens the light-$S-$heavy-DM quadrant, including even smaller $m_S$, due to possible annihilations via the $S^2 \bar \chi \chi$ operator which allows for more modest values of $y_S$ and thereby to evade DD limits.

Before continuing the discussion, a remark is important. In the secluded region the correct relic density depends basically just on the $y_S$ and $y_{S}^{(2)}$ parameters. One would consequently obtain viable model points, evading even future detection prospects, just by considering extremely small values for the coupling of $S$ with the SM sector, namely $c_V^S,\, \lambda'_{HS},\, c_S$. The restriction of the viable parameter regions for $m_\chi > m_S$ of the {\it individual} 

\FloatBarrier
\noindent
portals shown in fig.~\ref{fig:compScan} is basically due to the choice of the lower limits for $c_V^S,\, \lambda'_{HS},\, c_S$ given in eq.~\eqref{eq:general_scan_ranges}, which is motivated by the approach of considering non-vanishing values for each operator that is not forbidden by a symmetry.
 Moreover, we in fact cannot just set all these couplings to zero, since a small but non-zero coupling with SM states is needed to ensure that the DM was in thermal equilibrium in the early Universe, so that the standard thermal freeze-out computations are valid.
Beyond that, when we refer to an enlargement of the viable parameter space in the secluded region due to considering the full EFT (or its variant with $y_S^{(2)}=0$) rather than simpler portal models, we have in mind that in the former case there are viable model points for moderate values of the $c_V^S,\, \lambda'_{HS},\, c_S$ parameters, which would be already excluded in the corresponding portals and could be large enough to be probed by future experimental upgrades. We will further explore this last point below.

\begin{figure}[t!]
    \centering
    \subfloat{\includegraphics[width=0.435\linewidth]{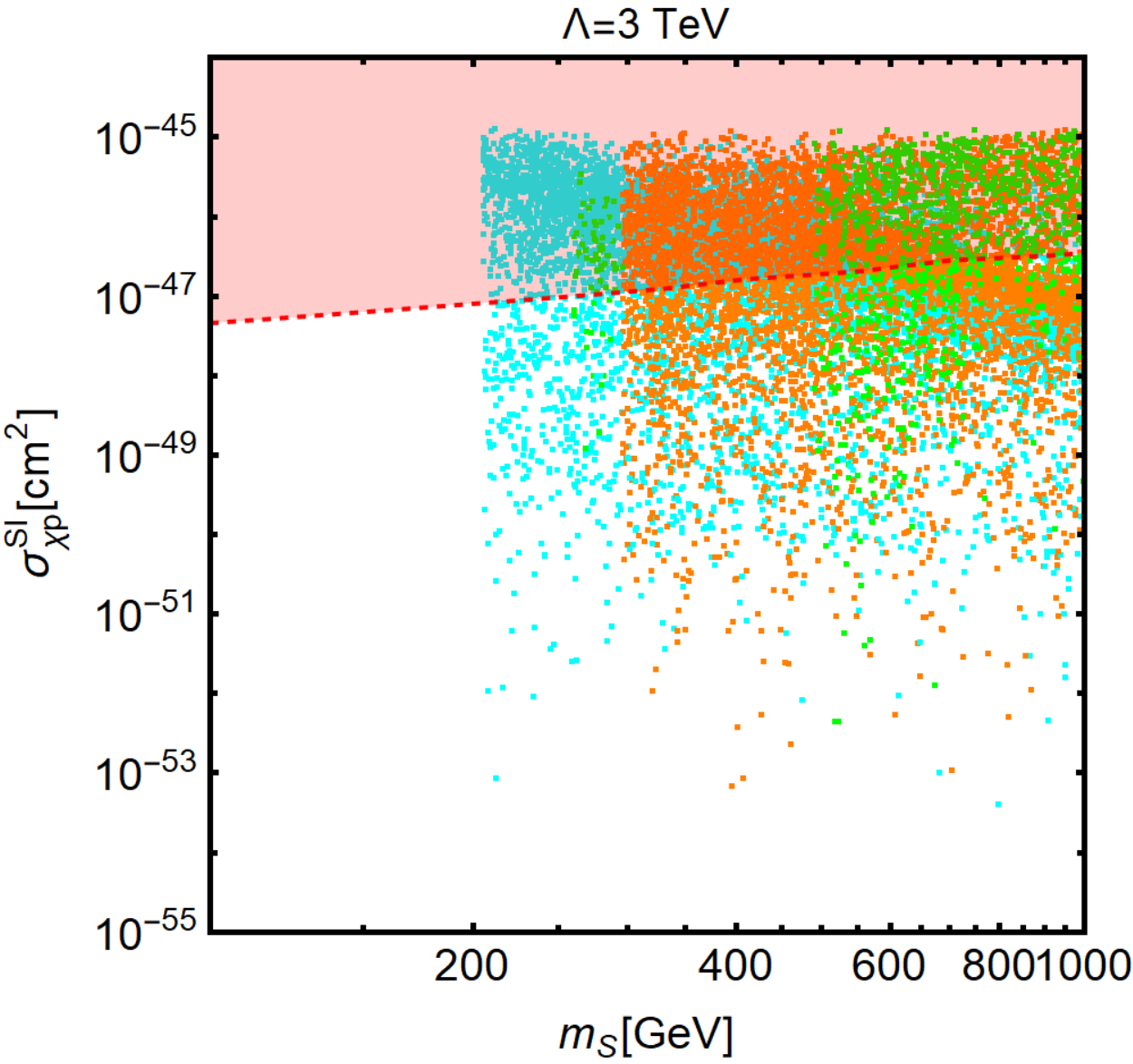}}
    \subfloat{\includegraphics[width=0.4\linewidth]{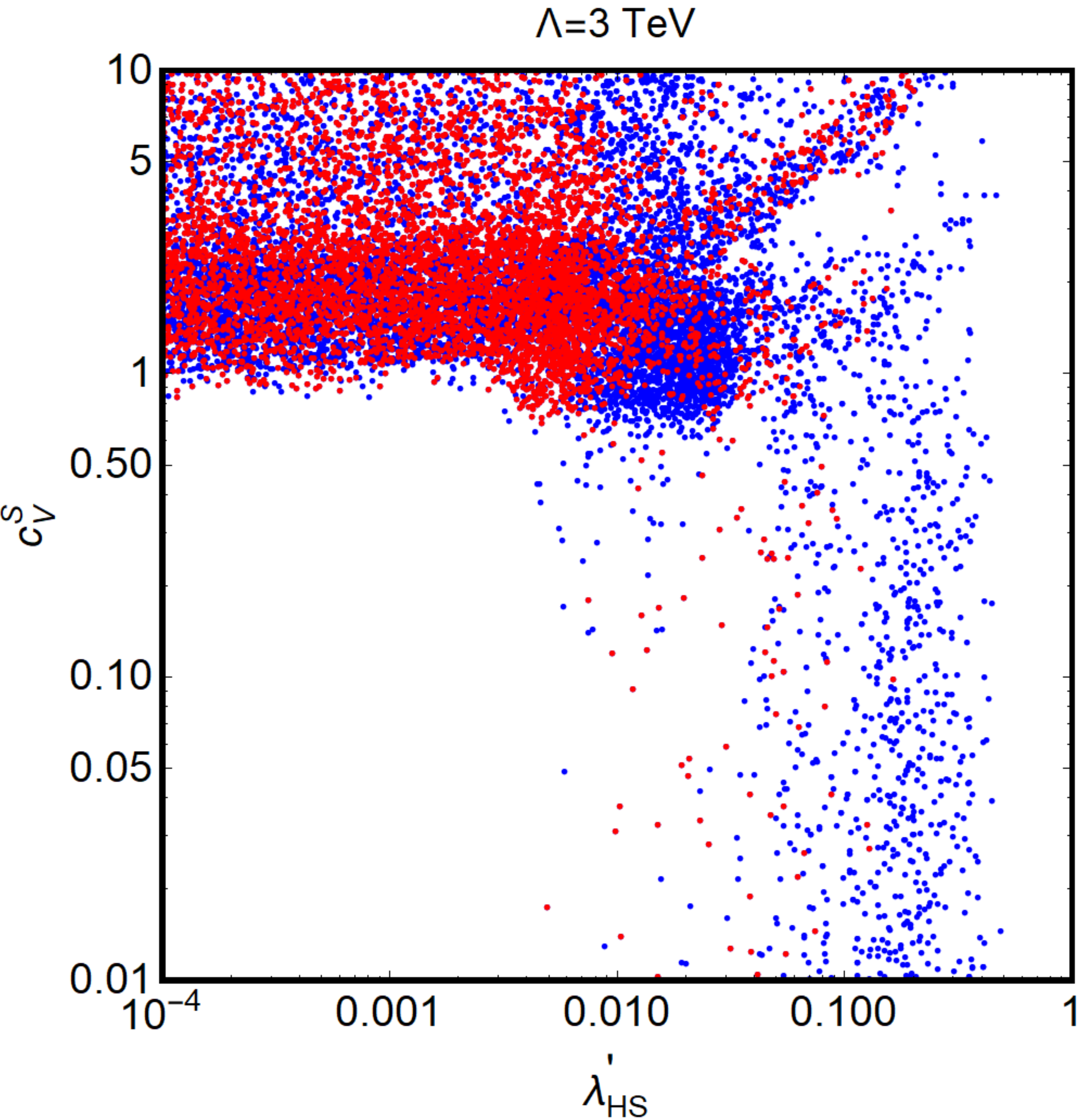}}
    \caption{\footnotesize{{\it Left Panel:} Model points in the $m_S-\sigma_{\chi p}^{\rm SI}$ plane for the full EFT, complying with all current experimental constraints but being potentially testable at colliders in the near future according to the criteria given in the main text. The different colors indicate the processes which can be used to probe the corresponding points, namely $pp \rightarrow S \rightarrow VV$ (cyan), $pp \rightarrow S \rightarrow hh$ (green) and $pp \rightarrow S \rightarrow \gamma \gamma$ (orange). The regions marked in red will be excluded in the case of no signal at XENONnT. {\it Right panel}: Same model points in the $\lambda_{HS}'-c_V^S$ plane, where blue points feature DM scattering cross-sections above the projected limit from XENONnT while red ones will also pass this upcoming constraint.}}
    \label{fig:ppLHC}
\end{figure}

Irrespectively of this, the differences between the portals and the \eDMEFT\ get even more pronounced once we finally look at projected XENONnT constraints, with points passing also the latter depicted in red. As one can observe, while with further strengthening DD constraints the viable region of the Higgs-mixing portal would be mostly bound to a rather tuned resonance band with $m_S \approx 2 m_\chi$, and also the Yukawa and gauge portals would be constrained to rather narrow regions, the full \eDMEFT\ stays vital in large areas of parameter space. This includes in particular the Higgs resonance region, which in the presence of the new operators remains viable due to cancellations in DD.

In summary, the \eDMEFT\ scenario for fermionic DM with real scalar mediator appears currently to be most constrained by DM DD experiments, while LHC can exclude limited regions of the parameter space, characterized by high values of the mixing angle $\theta$ or of the $D=5$ coupling $c_V^S$. We now explore whether this situation might change in the near future, i.e.~if upgrades of LHC results could be more competitive, at least in some regions of the parameter space of the \eDMEFT, than future constraints from DD and thus allow for a potential `discovery' of DM at colliders, within this extended framework. A naive estimate puts the potential improvement of the collider limits at $\sqrt{L_{\mbox{\tiny HL-LHC}}/L_{\mbox{\tiny curent}}}\approx 10$.
Therefore, we select from our general parameter scan all viable points that have a cross section less than one order of magnitude below the present bound in at least one of the collider processes considered here.   
The model points selected in this way are displayed in the left panel of fig.~\ref{fig:ppLHC} in the $(m_S,\sigma_{\chi p}^{\rm SI})$ plane, and compared to the expected exclusion from XENONnT (red region). As can be seen, a sizable fraction of these points is characterized by very suppressed SI cross sections, far below the future experimental reach.
To better characterize these parameter points we finally display them, in the right panel of fig.~\ref{fig:ppLHC}, in the $(\lambda_{HS}',c_V^S)$ plane. In this plot the data points with DM scattering cross sections above the projected XENONnT bound are marked in blue, while those evading it are shown in red. The distribution resembles the shape of fig.~\ref{fig:plot2D}. We notice in particular a stripe at high values of $c_V^S$ and substantial $h-S$ mixing which corresponds to the blind spot highlighted in fig.~\ref{fig:plot2D}.

\section{Pseudoscalar mediator}
\label{sec:PS}

We will now turn to the case where the SM is connected to the dark sector via a pseudoscalar mediator $\tilde{S}$. While sharing basic features with the scalar model discussed above, the pseudoscalar mediator leads to striking phenomenological differences in a number of observables. The \eDMEFT\ Lagrangian for a fermionic dark matter particle $\chi$ and a pseudoscalar mediator $\tilde{S}$ reads~\cite{Alanne:2017oqj}
\begin{align}
\label{eq:pseudo_lag}
 \mathcal{L}_{\rm eff}^{ \tilde{S} \chi} &=  \mathcal{L}_{\rm SM} +
    \frac{1}{2}\partial_\mu \tilde{S} \partial^\mu \tilde{S}
    - \frac 1 2 \mu_{\tilde{S}}^2  {\tilde{S}}^2 + \bar \chi i \slashed{\partial}\chi- m_\chi \bar \chi \chi \nonumber\\
    &- \frac{\lambda_{\tilde{S}}}{4} {\tilde{S}}^4 - \lambda_{H\tilde S} |H|^2 {\tilde{S}}^2 \nonumber\\
    &- i y_{\tilde{S}} {\tilde{S}} \bar \chi_L \chi_R -\frac{y_{\tilde{S}}^{(2)} {\tilde{S}}^2 + y_H^{(2)} |H|^2}{\Lambda}\	\bar{\chi}_L \chi_R +\mathrm{h.c.} \nonumber\\  
    &- \frac{\tilde{S}}{\Lambda} \left[i (y_d^{\tilde{S}})^{ij} \bar{Q}_{\mathrm{L}}^i H d_{\mathrm{R}}^j + i (y_u^{\tilde{S}})^{ij}\bar{Q}_{\mathrm{L}}^i\tilde{H}u_{\mathrm{R}}^j+i (y_\ell^{\tilde{S}})^{ij} \bar{L}_{\mathrm{L}}^i H \ell_{\mathrm{R}}^j +\mathrm{h.c.}\right]\nonumber\\
	&-\frac{\tilde{S}}{\Lambda}\left[C_{BB}^{\tilde{S}} \tilde{B}_{\mu\nu} B^{\mu\nu}+ C_{WW}^{\tilde{S}} \tilde{W}^{I\mu\nu} W_{\mu\nu}^I +C_{GG}^{\tilde{S}} \tilde{G}^{a\mu\nu}G_{\mu\nu}^a\right]\,.
\end{align}

The notation follows the conventions detailed in sec.~\ref{sec:S}, and we adopted a similar rescaling as in eq.~(\ref{eq:loop_coefficient}) between $c_V^{\tilde{S}}$ and $C_{VV}^{\tilde{S}}$. Due to the assumption of CP-conservation, the operator $|H|^2 \tilde{S}$ is forbidden and we assume that $\tilde{S}$ does not develop a vacuum expectation value. Therefore, there is no mixing between the SM Higgs and the pseudoscalar mediator, and Higgs precision observations are less sensitive to this model.
The effective Higgs portal operator $|H|^2 \bar{\chi_L}\chi_R$ does not depend on $\tilde{S}$ (or $S$) and can also be included in the pseudoscalar model. In order to work out the differences between the scalar and the pseudoscalar \eDMEFT{}, and since the strength of $y_H^{(2)}$ is rather constrained, we restrict our analysis to  $y_H^{(2)}=0$. For analogous reasons we will neglect the $y_{\tilde{S}}^{(2)}$ coupling since the corresponding operator does not distinguish the CP-even versus CP-odd nature of the mediator.

In the following we revisit the various observables considered in the previous section and work out the differences and similarities between the scalar and the pseudoscalar mediator before analyzing the full parameter space of the pseudoscalar \eDMEFT{}.

\subsection{DM phenomenology}

\subsubsection{Relic density}

In analogy to the case of the scalar mediator, we present the DM annihilation cross sections in the relevant channels, retaining just the leading piece. 
The thermally averaged cross section into fermions can be approximated by
\begin{align}
    \langle \sigma v\rangle_{ff} &\approx \frac{N_c}{4\pi}\, \frac{ v^2}{\Lambda^2}\, \frac{y_{\tilde{S}}^2 (y^{\tilde{S}}_f)^2 m_\chi^2}{(m_{\tilde{S}}^2-4 m_\chi^2)^2} \,\nonumber\\
    &\approx \left \{\begin{array}{cc}
    5.0 \times 10^{-2} \sigma_v^0 N_c \big(\frac{3\,{\rm TeV}}{\Lambda}\big)^2 {\big(\frac{m_\chi}{100\,{\rm GeV}}\big)}^2 {\big(\frac{500\,{\rm GeV}}{m_{\tilde{S}}}\big)}^4 y_{\tilde{S}}^2 (y_f^{\tilde{S}})^2,    & m_\chi \ll \frac{m_{\tilde{S}}}{2}\\
    1.95\, \sigma_v^0 N_c {\big(\frac{3\,{\rm TeV}}{\Lambda}\big)}^2 {\big(\frac{100\,{\rm GeV}}{m_\chi}\big)}^2 y_{\tilde{S}}^2 (y_f^{\tilde{S}})^2,    & m_\chi \gg \frac{m_{\tilde{S}}}{2} \,.
    \end{array} \right.
\end{align}

In contrast to the case of a scalar mediator this expression is of order $v_\chi^0$, i.e.~$s$-wave instead of $p$-wave. Consequently, the annihilation cross section at freeze-out is typically enhanced by a factor $1/v_\chi^2 \approx 10$, and the cosmologically preferred values of the couplings are smaller than found in the previous section.
The same effect can be observed in the cross section into gauge bosons. Taking gluons as a representative choice for $VV$ final states, we find
 \begin{align}
    \langle \sigma v\rangle_{GG} &\approx \frac{2}{\Lambda^2 \,\pi}\, \frac{(C^{\tilde{S}}_{GG})^2 y_{\tilde{S}}^2 m^4_\chi}{(4 m_\chi^2-m_{\tilde{S}}^2)^2}\,\nonumber\\
    &\approx\left \{\begin{array}{cc}
    6.6 \times 10^{-2} \sigma_v^0 {\big(\frac{3\,{\rm TeV}}{\Lambda}\big)}^2 {\big(\frac{m_\chi}{100\,{\rm GeV}}\big)}^4 {\big(\frac{500\,{\rm GeV}}{m_{\tilde{S}}}\big)}^4y_{\tilde{S}}^2 {\big(\frac{\alpha_s c_{G}^{\tilde{S}}}{4\pi}\big)}^2 \,,      & m_\chi \ll \frac{m_{\tilde{S}}}{2} \\
    2.58\, \sigma_v^0 {\big(\frac{3\,{\rm TeV}}{\Lambda}\big)}^2 y_{\tilde{S}}^2 {\big(\frac{\alpha_s c_{G}^{\tilde{S}}}{4\pi}\big)}^2\,,     &  m_\chi \gg \frac{m_{\tilde{S}}}{2}\, .
    \end{array} \right.
\end{align}

The situation is different for $\tilde{S}\tilde{S}$ final states. Here, the thermal cross section can be approximated, in the limit $m_\chi \gg m_{\tilde{S}}$, as 
\begin{align}
  \langle \sigma v\rangle_{\tilde{S}\tilde{S}} &\approx
  \frac{y_{\tilde{S}}^4\, v^2_\chi}{192 \pi m_\chi^2} \approx 10\, \sigma_v^0 {\left(\frac{100\,{\rm GeV}}{m_\chi}\right)}^2 y_{\tilde{S}}^4 \,.
\end{align}

As can be seen, the $p$-wave suppression is not lifted by switching the CP-properties of the mediator in this case and the leading contribution arises at $\mathcal{O}(v_\chi^2)$. 
However, it should be kept in mind that this annihilation channel can be realized without resorting to higher dimensional operators and lacks the $1/\Lambda^2$ suppression that characterizes the annihilation channels discussed above. Unless the masses of the new states are rather close to the scale of the effective interaction this can compensate for the velocity suppression and makes $\tilde{S}\tilde{S}$ one of the most important channels for setting the relic density in the secluded regime.
The $\chi\bar{\chi}\tilde{S}^2$ operator can also contribute to $\tilde{S}\tilde{S}$ final states. The annihilation cross section is identical to the scalar case, and eq.~$\eqref{eq:xsec_XXSS}$ can be used to estimate its importance.

Interestingly, the mixed annihilation channel into $\tilde{S}h$ final states is both $s$-wave and can be realized with dimension-4 terms only. The leading contribution is given, taking for simplicity the limit $m_\chi \gg \frac{m_{\tilde{S}}+m_h}{2}$, by
\begin{align}
  \langle \sigma v\rangle_{\tilde{S}h} &\approx
\frac{y_{\tilde{S}}^2\lambda^2_{\tilde{S}H}v^2}{256\pi m_{\chi}^4} \approx 4.4 \times 10^2 \sigma_v^0 {\left(\frac{100\,{\rm GeV}}{m_\chi}\right)}^4 y_{\tilde{S}}^2\lambda^2_{\tilde{S}H}\,.
\end{align}
In particular for $m_\chi$ around the electroweak scale this annihilation rate can naturally become large and should be expected to contribute significantly to the relic density.

Besides shifting the parameter space for a successful thermal freeze-out towards lower couplings, the presence of $s$-wave cross sections also makes ID effective. We will comment more on this later.

\subsubsection{Direct detection}

\begin{figure}
\centering
\hspace{10mm}
\begin{minipage}[c]{.47\textwidth}
  \begin{tikzpicture}[node distance=1.5cm]
    \coordinate[label=above:$\chi$] (v1);
    \coordinate[label=above:$q$,below = 2.57cm of v1] (v2);
    \coordinate[right = of v1] (v3);
    \coordinate[below right = of v3] (v4);
    \coordinate[above right = of v4] (v5);
    \coordinate[label=above:$\chi$,right = of v5] (v6);
    \coordinate[below = of v4] (v7);
    \coordinate[label=above:$q$, below = 2.57cm of v6] (v8);
    \draw[fermion] (v1) -- (v3);
    \draw[fermion] (v3) -- (v5);
    \draw[fermion] (v5) -- (v6);
    \draw[rscalar] (v3) -- node[label=left:$\tilde{S}$] {} (v4);
    \draw[rscalar] (v4) -- node[label=right:$\tilde{S}$] {} (v5);
    \draw[rscalar] (v4) -- node[label=left:$h$] {} (v7);
    \draw[fermion] (v2) -- (v7);
    \draw[fermion] (v7) -- (v8);
  \end{tikzpicture}
\end{minipage}
  \hspace{-5mm}
\begin{minipage}[c]{.47\textwidth}
  \begin{tikzpicture}[node distance=1.5cm]
    \coordinate[label=above:$\chi$] (v1);
    \coordinate[label=above:$q$,below = 2.57cm of v1] (v2);
    \coordinate[right = of v1] (v3);
    \coordinate[below right = of v3] (v4);
    \coordinate[above right = of v4] (v5);
    \coordinate[label=above:$\chi$,right = of v5] (v6);
    \coordinate[below = 2.57cm of v3] (v7);
    \coordinate[label=above:$q$, below = 2.57cm of v6] (v9);
    \coordinate[below = 2.57cm of v5] (v8);
    \draw[fermion] (v1) -- (v3);
    \draw[fermion] (v3) -- (v5);
    \draw[fermion] (v5) -- (v6);
    \draw[rscalar] (v3) -- node[label=left:$\tilde{S}$] {} (v7);
    \draw[rscalar] (v5) -- node[label=right:$\tilde{S}$] {} (v8);
    \draw[fermion] (v2) -- (v7);
    \draw[fermion] (v7) -- (v8);
    \draw[fermion] (v8) -- (v9);
  \end{tikzpicture}
\end{minipage}
\caption{\label{fig:feynloop} \footnotesize Representative loop diagrams contributing to the scattering cross-section on nucleons in the case of a pseudoscalar mediator.}
\end{figure}
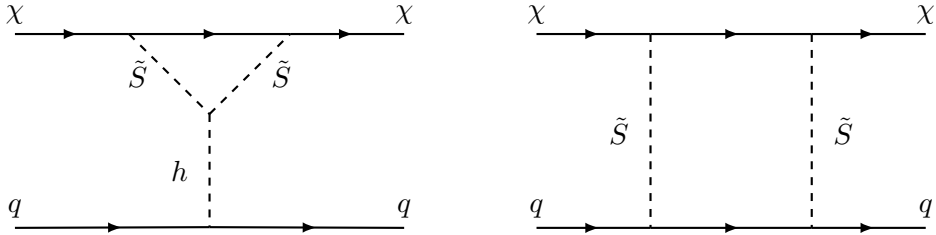

In the limit in which the direct coupling of DM to the Higgs boson
through the effective interaction $y_H^{(2)}$ is absent or suppressed, the DD phenomenology is crucially different from the previous case. The coupling of the DM with quarks via a pseudoscalar field leads to a spin-dependent cross section which is also suppressed by $q^4/(m_\chi^2 m_p^2)$, with $q$ the (small) momentum transfer. The scattering rate induced by this interaction is far from the experimental sensitivity unless the mass of the mediator is significantly below the nuclear scale~\cite{Arina:2014yna}. The effective coupling of $\tilde{S}$ with gluons similarly leads to a tiny momentum-suppressed cross section \cite{DelNobile:2013sia}. Therefore, the most relevant interactions with nuclei arise at higher order and are induced at the one-loop level  \cite{Arcadi:2017wqi,Sanderson:2018lmj,Abe:2018emu,Ertas:2019dew}. For illustration we report a set of representative  diagrams in fig.~\ref{fig:feynloop}. From these it is straightforward to notice that the contribution to the amplitude from box-shaped diagrams is suppressed by a factor $1/\Lambda^2$ while, on the contrary, the triangle shaped diagrams contain no coupling depending on $\Lambda$. To good approximation, we can then compute the DM scattering cross section retaining only the latter and, hence, write the cross section as:
\begin{equation}
    \sigma_{\chi p}^{\rm SI\,loop}=\frac{\mu_{\chi p}^2}{\pi}\left \vert \sum_{q=u,d,s}m_p f_q \frac{\lambda_{H\tilde{S}}}{m_h^2}C_{\tilde{S}}^{\rm triangle}+\frac{6}{27} m_p f_T  \frac{\lambda_{H\tilde{S}}}{m_h^2}C_{\tilde{S}}^{\rm triangle} \right \vert^2 \,,
\end{equation}
where
\begin{equation}
    C_{\tilde{S}}^{\rm triangle}=\frac{y_{\tilde{S}}^2}{(4\pi)^2} m_\chi  C_2 (m_\chi^2,m_{\tilde{S}}^2,m_\chi^2) \,
\end{equation}
with
\begin{align}
    C_2 (m_\chi^2,m_{\tilde{S}}^2,m_\chi^2) =&-\frac{1}{m_\chi^2}+\frac{(-m_{\tilde{S}}^4+3 m_{\tilde{S}}^2 m_\chi^2)\sqrt{m_{\tilde{S}}^2-4 m_\chi^2}}{ m_{\tilde{S}} m_\chi^4 {\left(m_{\tilde{S}}^2-4 m_\chi^2\right)}}\, \log\left[\frac{m_{\tilde{S}}^2+\sqrt{m_{\tilde{S}}^2 (m_{\tilde{S}}^2-4 m_\chi^2)}}{2 m_\chi m_{\tilde{S}}}\right]\nonumber\\
   & +\frac{m_{\tilde{S}}^2-m_\chi^2}{2 m_\chi^4}\, \log\left[\frac{m_{\tilde{S}}^2}{m_\chi^2}\right] \,.
\end{align}

Even though the bounds will turn out to be less strong than in the case of a scalar mediator, those from Xenon-based experiments are not negligible. We will illustrate them in more detail below.

\subsubsection{Indirect detection}

The annihilation cross sections into SM fermions and gauge boson pairs, as well as the one into the $h\tilde{S}$ final state, are $s$-wave dominated; their values at thermal freeze-out and at present times are comparable. Therefore, ID experiments have the potential to test thermally produced dark matter in this setup. There are various signatures that can be used to search for DM annihilation happening in our local environment today.  In our analysis we will include two of the cleanest bounds on a DM annihilation signal:
i) the limits on the continuum $\gamma$-ray flux produced by DM annihilations in dwarf galaxies from  Fermi-LAT data~\cite{Ackermann:2015zua}, and 
ii) the limits on gamma-ray lines from $\gamma \gamma$ and $Z\gamma$ final states derived by the Fermi collaboration in~\cite{Ackermann:2015lka}. Concerning the $h\tilde{S}$ final state, it will mostly lead to a $\bar b b \bar b b$ signature which could be again probed through $\gamma$-ray signatures. To our knowledge there are, however, no dedicated studies for this kind of signature.

\subsection{Collider signals}

The collider phenomenology of the pseudoscalar model is very similar to the scalar case. In order to gain some insight into the relevance of new physics searches it is instructive to consider first resonant mediator production. As before, the cross section can be estimated using eq.~\eqref{eq:res_production}, and only the width needs to be reevaluated. For the gauge-portal interaction the square matrix elements of the $\tilde{S}\rightarrow g g $ and the $\tilde{S}\rightarrow \gamma \gamma $ processes are identical to the scalar case, while the width into massive gauge bosons tends to the same value for $m_{\tilde{S}} \gg m_{W/Z}$. Therefore, the bounds from searches for the  
 visible decays of the pseudoscalar mediator are essentially the same as in the scalar model. 
 For the loop-induced production from gluons due to the Yukawa-like operator a minor modification of the width given in eq.~\eqref{eq:loop_width} is necessary; the loop function has to be replaced by the one for a pseudoscalar mediator 
\begin{equation}
    F_S(x) \rightarrow F_{\tilde{S}}(x)=x^2 \left| \arctan^2\frac{1}{\sqrt{x-1}} \right|^2 \,.
\end{equation}
Since $F_{\tilde{S}}(x)\geq F_{S}(x)$ for all $x$, the production rate of pseudoscalars is always bigger than the one of a scalar of the same mass and the bounds are stronger by an $\mathcal{O}(1)$ factor. Our practical implementation of the limits on a beyond-the-SM resonances decaying to visible final states is analogous to the case of a scalar mediator. 

\begin{figure}[t]
\begin{center}
 \includegraphics[width=.45\textwidth]{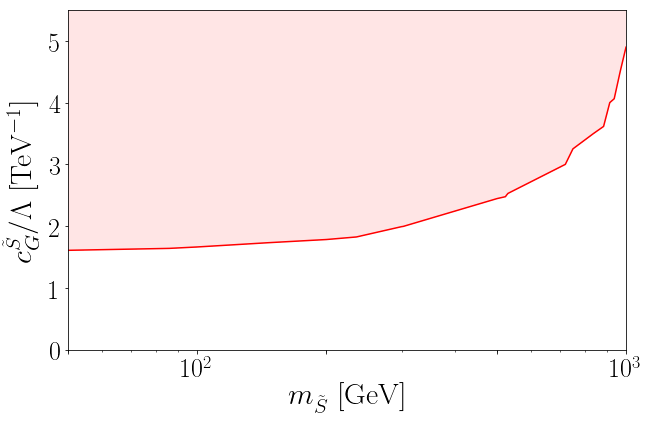}\quad
 \includegraphics[width=.45\textwidth]{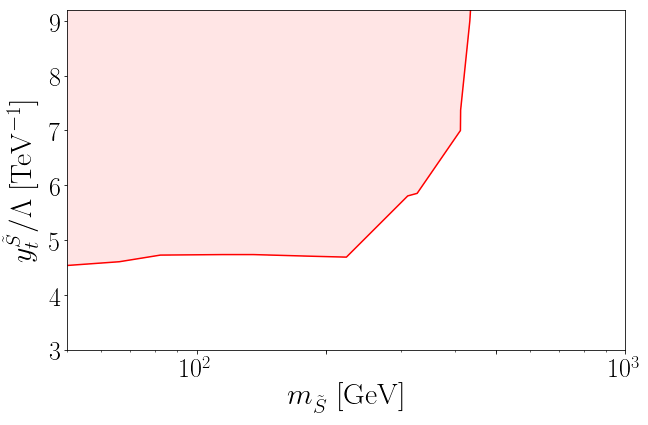}
 \end{center}
 \caption{\footnotesize Exclusion limits from the ATLAS mono-jet search \cite{Aaboud:2017phn} in the $m_{\tilde S} -c_{G}^{\tilde S}$ plane (left) and $m_{\tilde S} -y_t^{\tilde S}$ plane (right) for the pseudoscalar mediator. In both cases $y_{\tilde S}=1$, $m_\chi=10$ GeV and all other couplings are equal to zero.}
 \label{fig:monojpseudo}
\end{figure}

These kinds of signatures can be complemented by mono-jet signals associated to the invisible decays of $\tilde{S}$ for $m_{\tilde{S}}\geq 2 m_\chi$. Again, similarly to the case of the scalar mediator, we have recast the results from LHC searches \cite{Aaboud:2017phn} employing the CheckMATE package. For illustration we show the excluded regions in the $(m_{\tilde{S}},c_G^{\tilde{S}})$ and $(m_{\tilde{S}},y_t^{\tilde{S}})$ planes, assuming $m_\chi=10\,\mbox{GeV}$, in fig.~\ref{fig:monojpseudo}. As expected, we find that the limits on the gauge-portal are indistinguishable from the scalar case, while the bounds on the Yukawa-portal improve by a factor $\approx 1.5$ at low $m_{\tilde{S}}$.

Mixing with the Higgs is absent; however, decays of $h$ to pseudoscalar pairs are important for $m_{\tilde S} \leq m_h/2$ in the presence of $\lambda_{H\tilde{S}}$.
We combine the corresponding limits on decays of the SM-like Higgs into two light pseudoscalars derived in \cite{Aad:2015bua,Bauer:2017ota} (see also e.g.~\cite{Tunney:2017yfp} for related studies) and find stringent constraints on $\lambda_{H\tilde{S}}$.

\subsection{Combined results}

To obtain a global picture, we will follow the same strategy as in the previous section and perform the analysis of the pseudoscalar mediator by considering increasingly refined scenarios.

\subsubsection{Basic portals in isolation}

In order to illustrate the individual effects of the different interactions we first discuss again basic portal scenarios that form a subset of the Lagrangian in \eqref{eq:pseudo_lag}. In the pseudoscalar model only two portals are relevant: the gauge and the Yukawa portal. The mixing portal is forbidden by the assumed CP symmetry. The effective Higgs portal is in principle present but its phenomenology is identical to the one discussed in the previous section, and we will not recapitulate it here.

\begin{figure}
    \centering
    \subfloat{\includegraphics[width=0.43\linewidth]{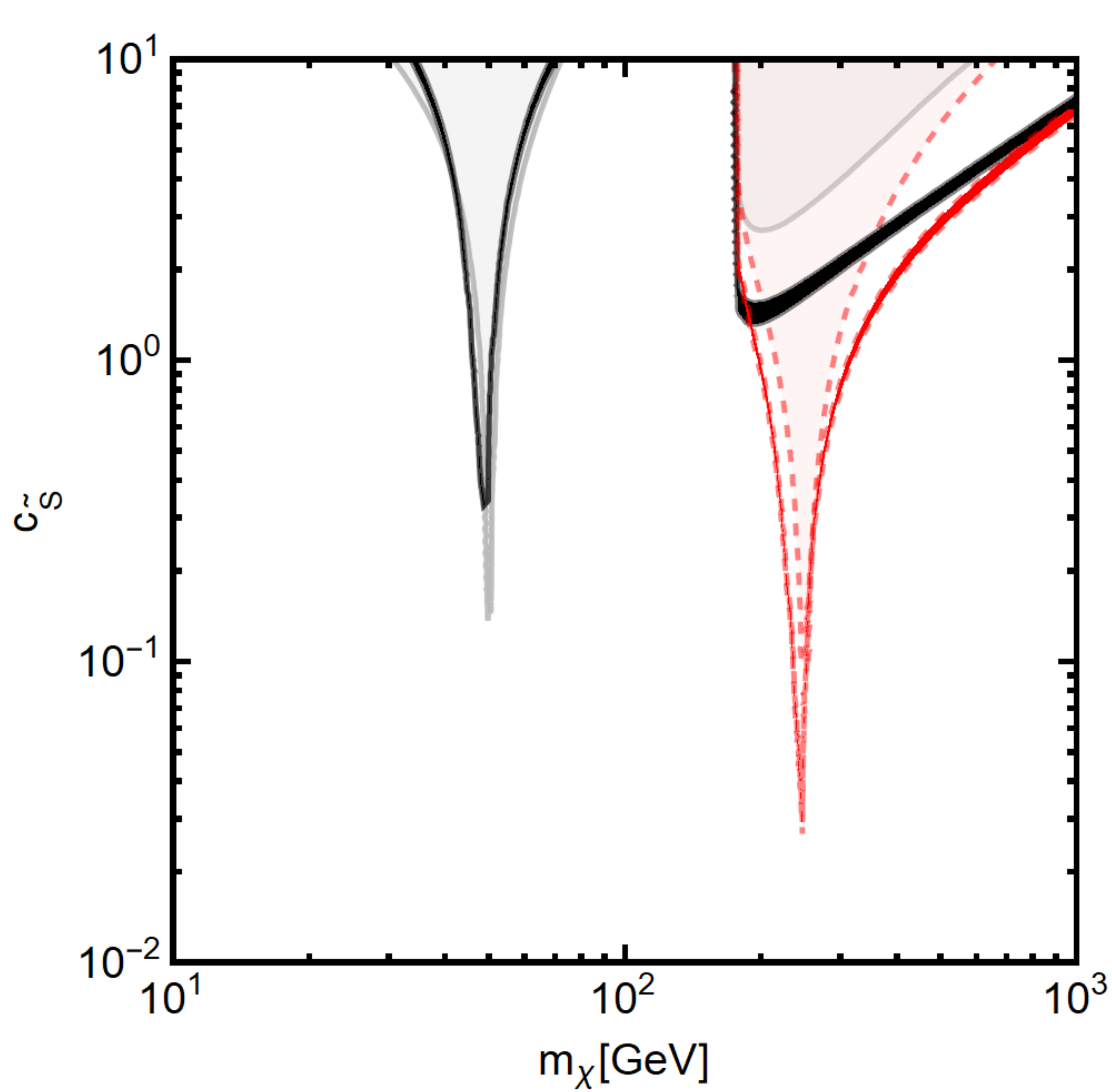}}
    \subfloat{\includegraphics[width=0.42\linewidth]{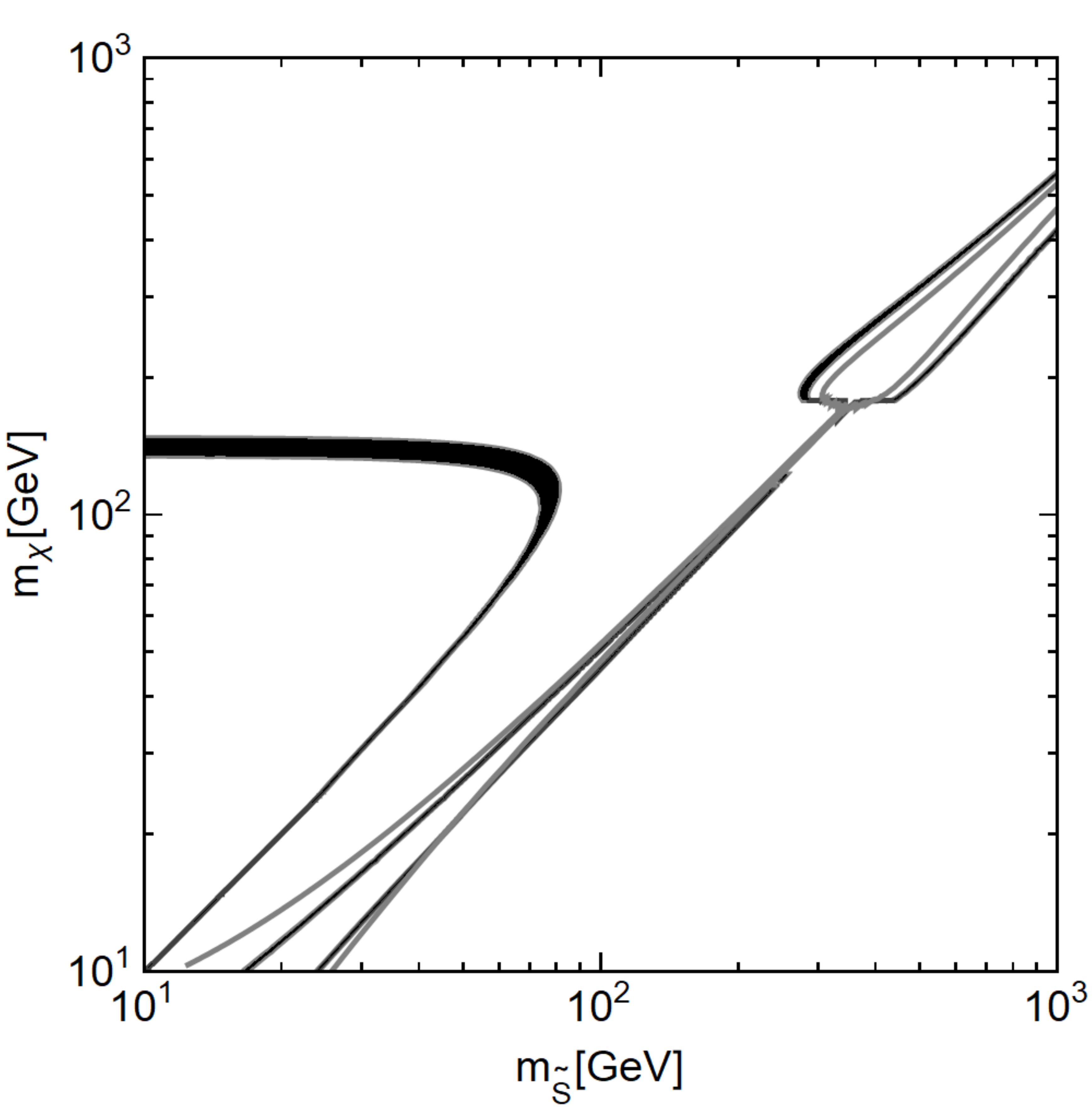}}\\
    \subfloat{\includegraphics[width=0.43\linewidth]{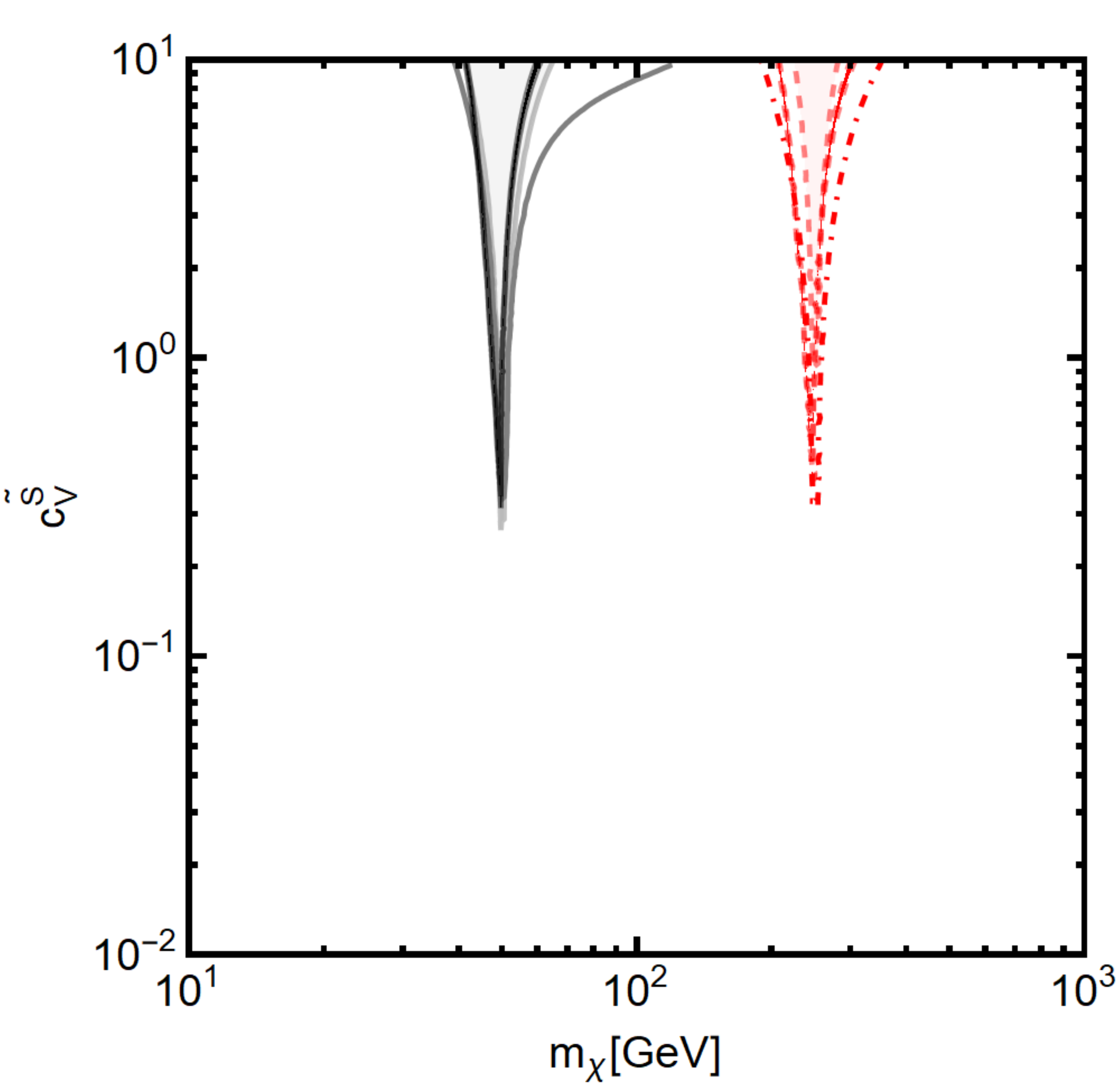}}
    \subfloat{\includegraphics[width=0.42\linewidth]{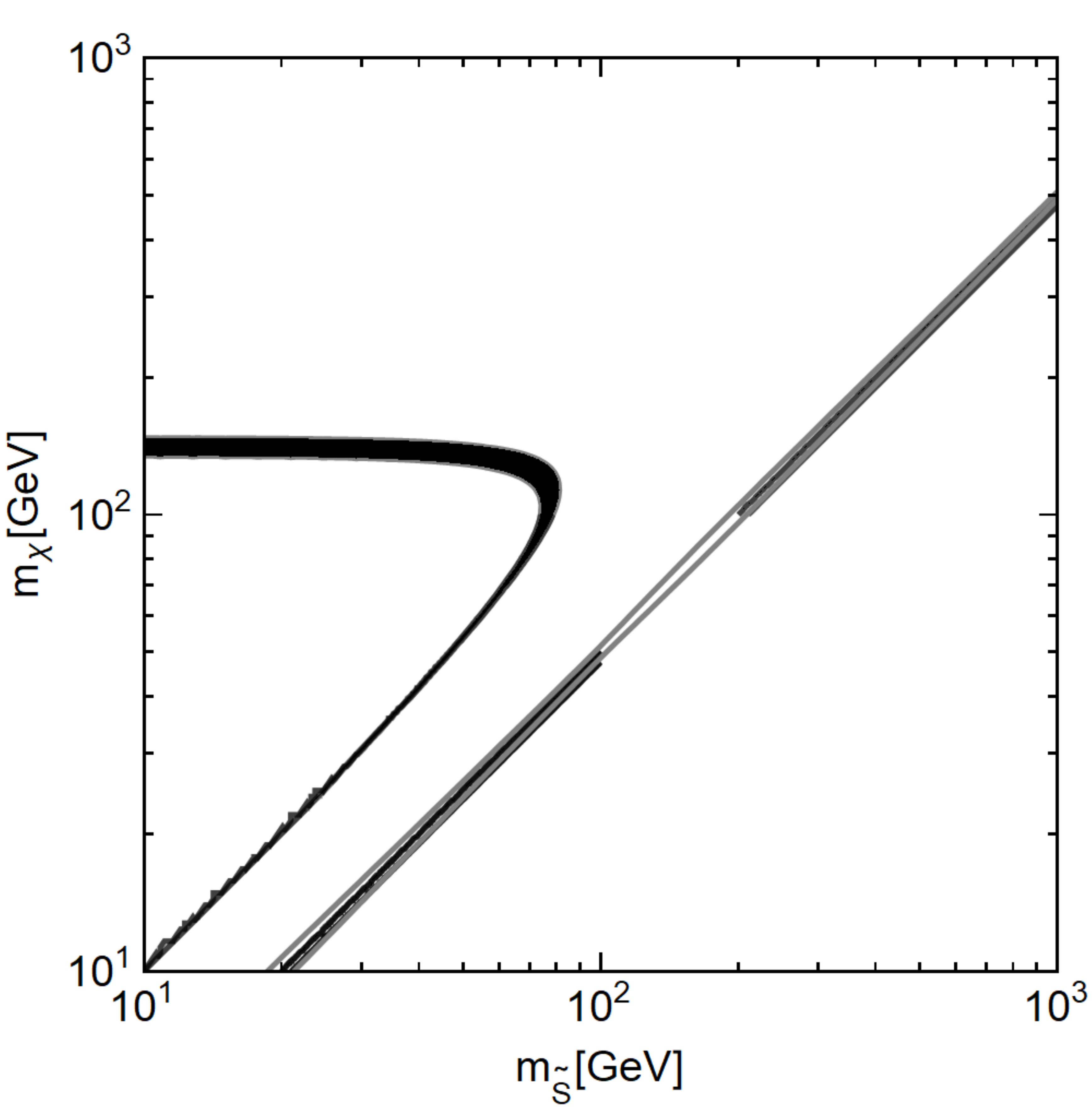}}\\
    \caption{\footnotesize Isocontours of correct relic abundance for the Yukawa (top), and gauge (bottom) portals with a pseudoscalar mediator. The left panels display the $m_\chi-\mbox{coupling}$ plane for $m_S=100$\,GeV (black, solid) and $m_S=500$\,GeV (red, dashed). The shaded regions correspond to the exclusion bounds from ID constraints. The plots in the right column show instead the $m_S-m_\chi$ plane for $y_{\tilde{S}}=1$ and $c_{\tilde{S}}=1$ (top) and $c_{V}^{\tilde{S}}=1$ (bottom). The regions inside the gray contours are excluded by ID. For all plots we have set $\Lambda=3\,\mbox{TeV}$.}
    \label{fig:pseudo1}
\end{figure}

A simple visualization of the cosmologically preferred parameter space in these two scenarios is presented in fig.~\ref{fig:pseudo1}, where we display isocontours corresponding to the observed relic density for fixed $m_{\tilde{S}} =100$\,GeV (black) and $m_{\tilde{S}}=500$\,GeV (red), as well as in the $m_{\tilde{S}}-m_\chi$ plane for fixed couplings. In all cases the $\lambda_{H\tilde{S}}$ coupling has been set to zero. The plots show that the preferred regions for the relic density are again the resonance, i.e. $m_\chi \simeq m_{{\tilde{S}}}/2$, and the secluded one. By comparing fig.~\ref{fig:pseudo1} with the analogous plots for the scalar mediator, we notice that the isocontours corresponding to the correct relic density cover a wider region of parameter space. This is a consequence of the $s$-wave cross section into fermion and gauge boson pairs. Having set $\lambda_{H\tilde{S}}=0$, the impact from DD is negligible, so no related contours appear in the figure. In contrast, ID bounds, both from $\gamma$-ray continuum and lines, should be taken into account. These limits are indicated by the shaded regions in fig.~\ref{fig:pseudo1}.

\begin{figure}
    \begin{flushleft}
    \hspace*{-1.2cm}\subfloat{\includegraphics[width=0.38\linewidth]{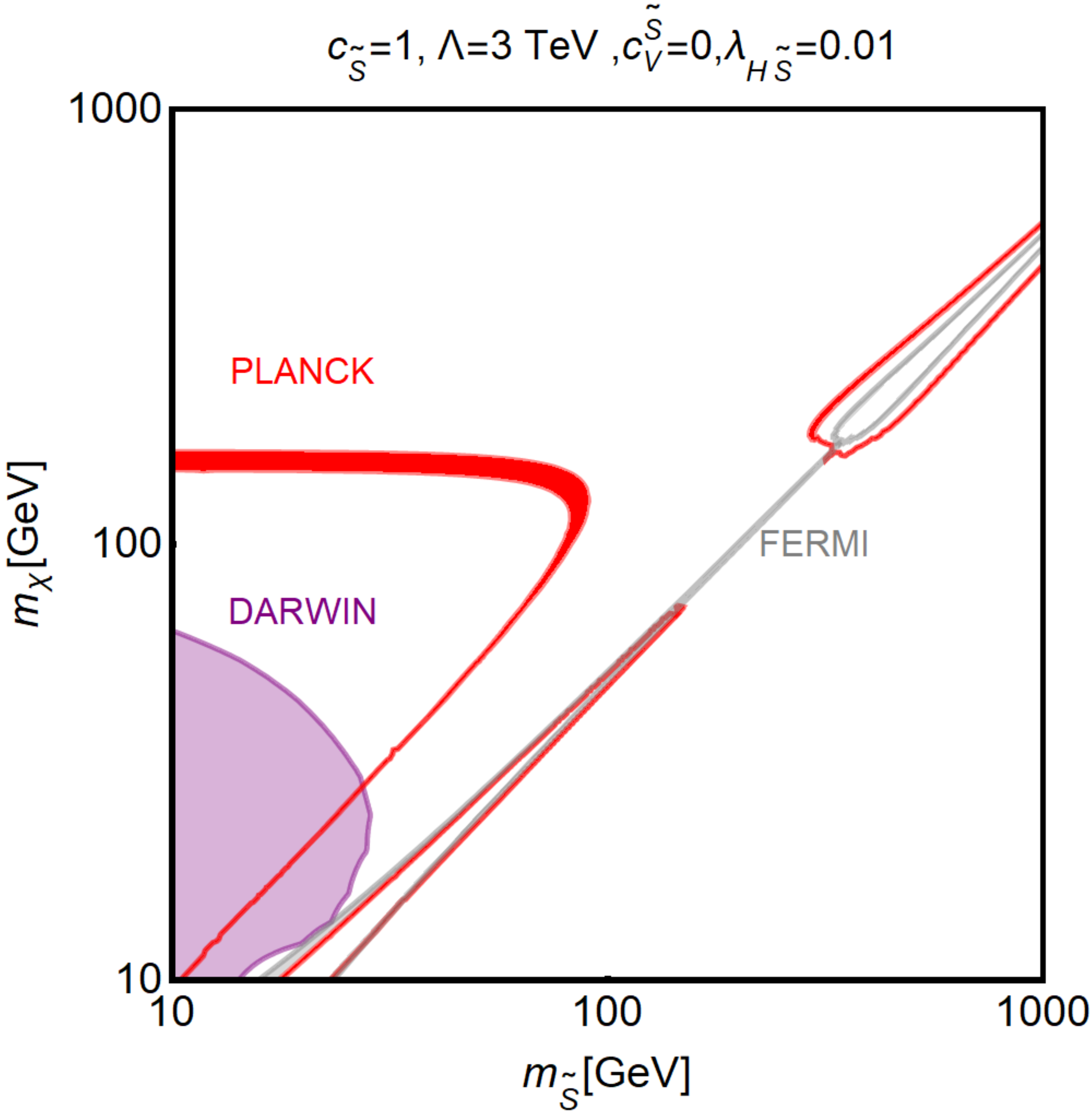}}
    \subfloat{\includegraphics[width=0.38\linewidth]{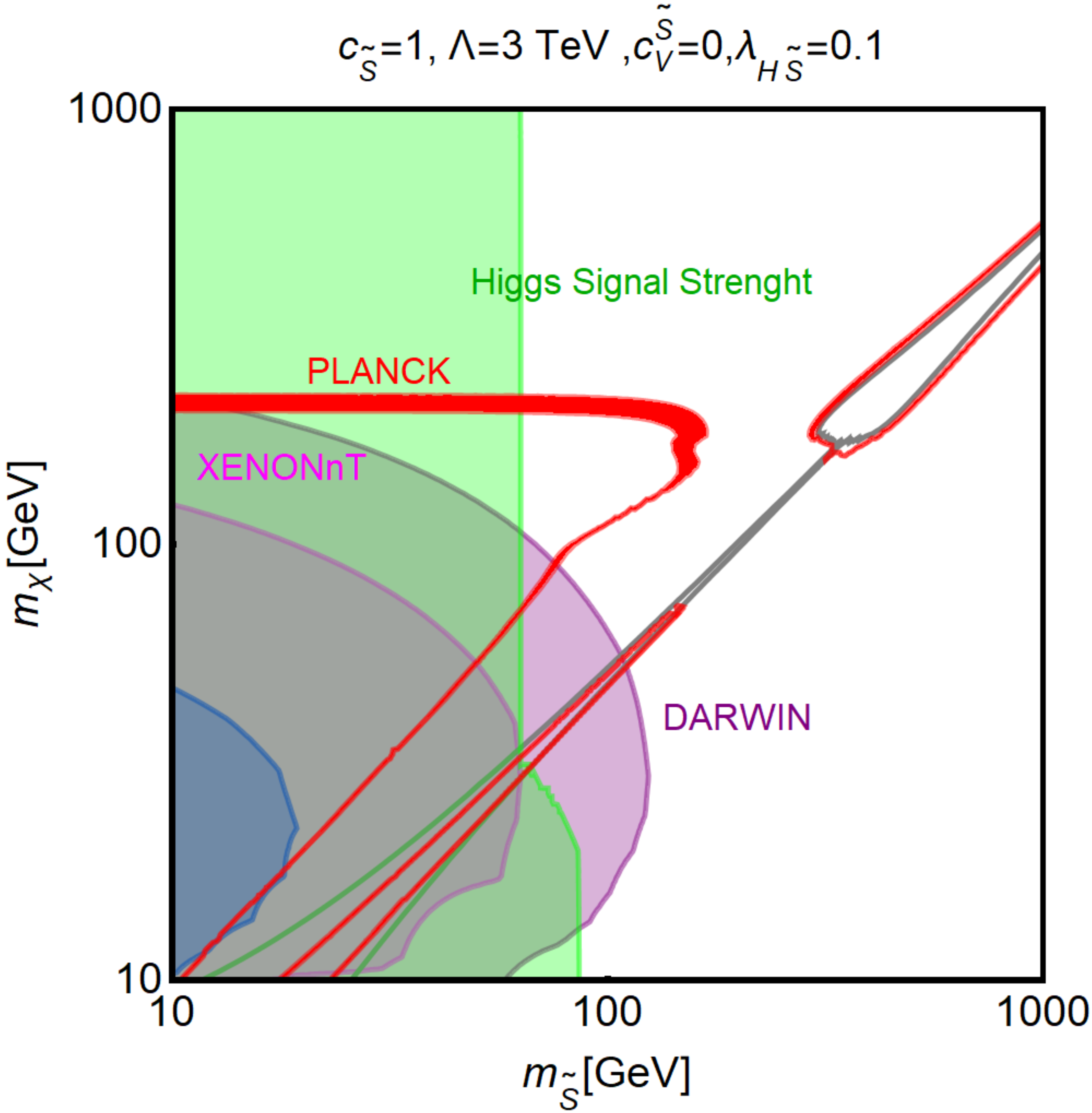}}
    \subfloat{\includegraphics[width=0.38\linewidth]{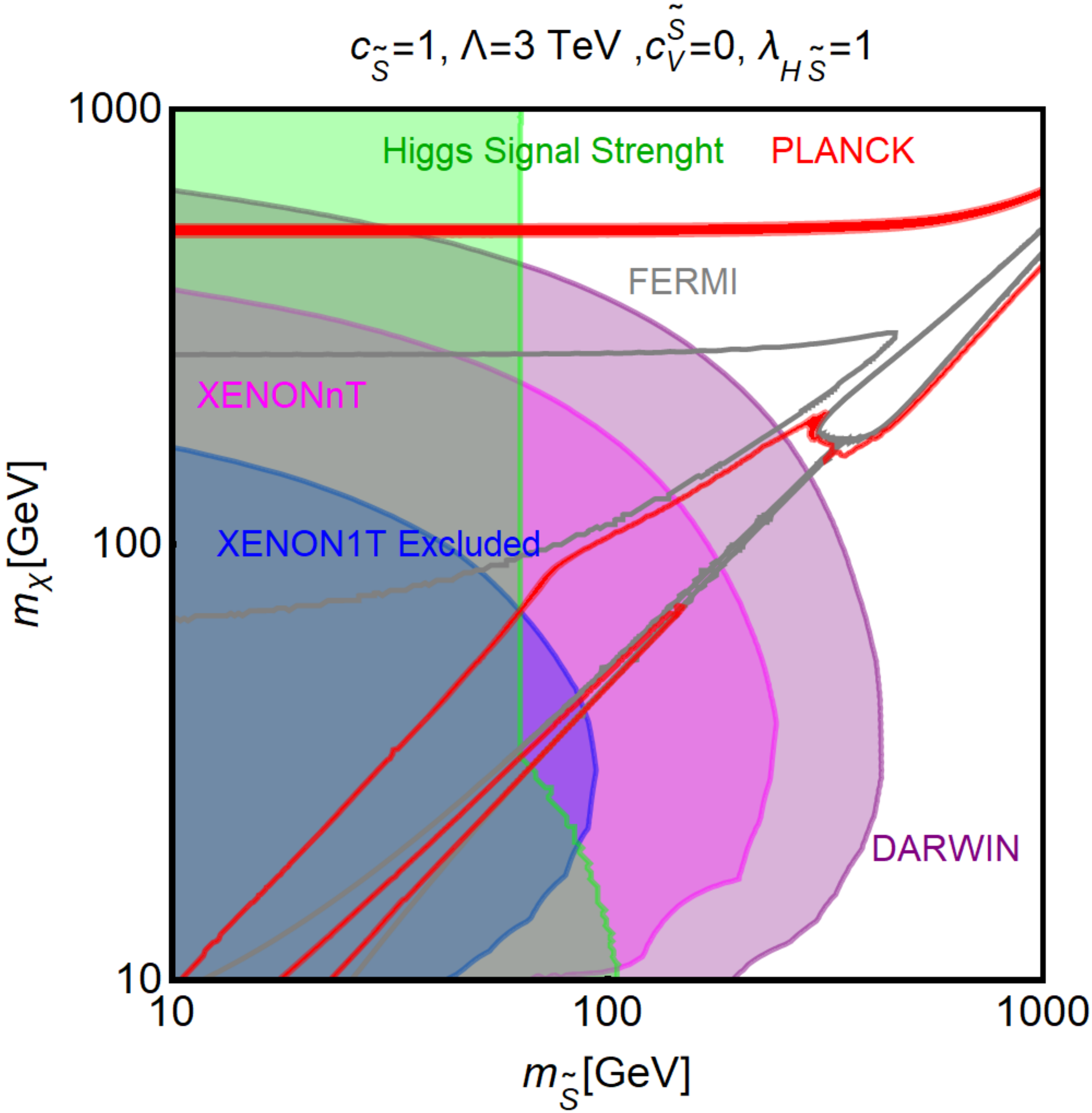}}
    \end{flushleft}
    \caption{\footnotesize{Constraints in the $m_{\tilde{S}}-m_\chi$ plane for the Yukawa portal with $c_{\tilde{S}}=1,y_{\tilde{S}}=1$ and three different assignations of $\lambda_{\tilde{H}S}=0.01,\, 0.1,\,1$. The isocontours of correct relic density are shown in red while the current (projected) limits from XENON1T (XENONnT/DARWIN) are given in blue (magneta/purple). The gray isocontours indicate limits from DM ID (viable regions of parameter space are outside these contours), while bounds from the $h\rightarrow \tilde{S}\tilde{S}$ decay are shown in green.}}
    \label{fig:pseudoDD}
\end{figure}

In order to illustrate the impact of the loop-induced DM-nucleon interactions discussed in the previous subsections we have reconsidered in fig.~\ref{fig:pseudoDD} the Yukawa portal fixing both $c_{\tilde{S}}$ and $y_{\tilde{S}}$ to 1 and varying instead the coupling $\lambda_{H\tilde{S}}$ from 0.01 to 1. 
As can be seen, the value of $\lambda_{H\tilde{S}}$, which controls the strength of the triangle contribution to DD, has a strong impact. Indeed, for small $\lambda_{H\tilde{S}}=0.01$ even a highly sensitive future experiment like DARWIN can only probe a quite limited region of the parameter space. However, the testable region becomes significant for higher values of $\lambda_{H\tilde{S}}$. Current constraints from XENON1T are only sensitive to masses of the mediator below $m_{\tilde S}\approx100$\,GeV even for $\lambda_{H\tilde{S}}=1$ and most of the DD region is also excluded by collider searches for $h\rightarrow \tilde{S}\tilde{S}$, marked in green in fig.~\ref{fig:pseudoDD}. Concerning the latter region we see that it extends, at light DM mass, at values of $m_S > m_h/2$, up to around 100 GeV. This because, as pointed in~\cite{Abe:2018bpo}, the decay process $h \rightarrow \tilde{S}\tilde{S}^{\star}\rightarrow \tilde{S}\chi \chi$, with $\tilde{S}^{\star}$ being an off-shell mediator, is important as well. However, with future generations of experiments this situation is expected to change, and scalar masses of up to several hundreds of GeV will be in reach of DD for sizable $\lambda_{H\tilde{S}}$.

\begin{figure}[h]
    \begin{flushleft}
    \hspace*{-1.8cm}\subfloat{\includegraphics[width=0.4\linewidth]{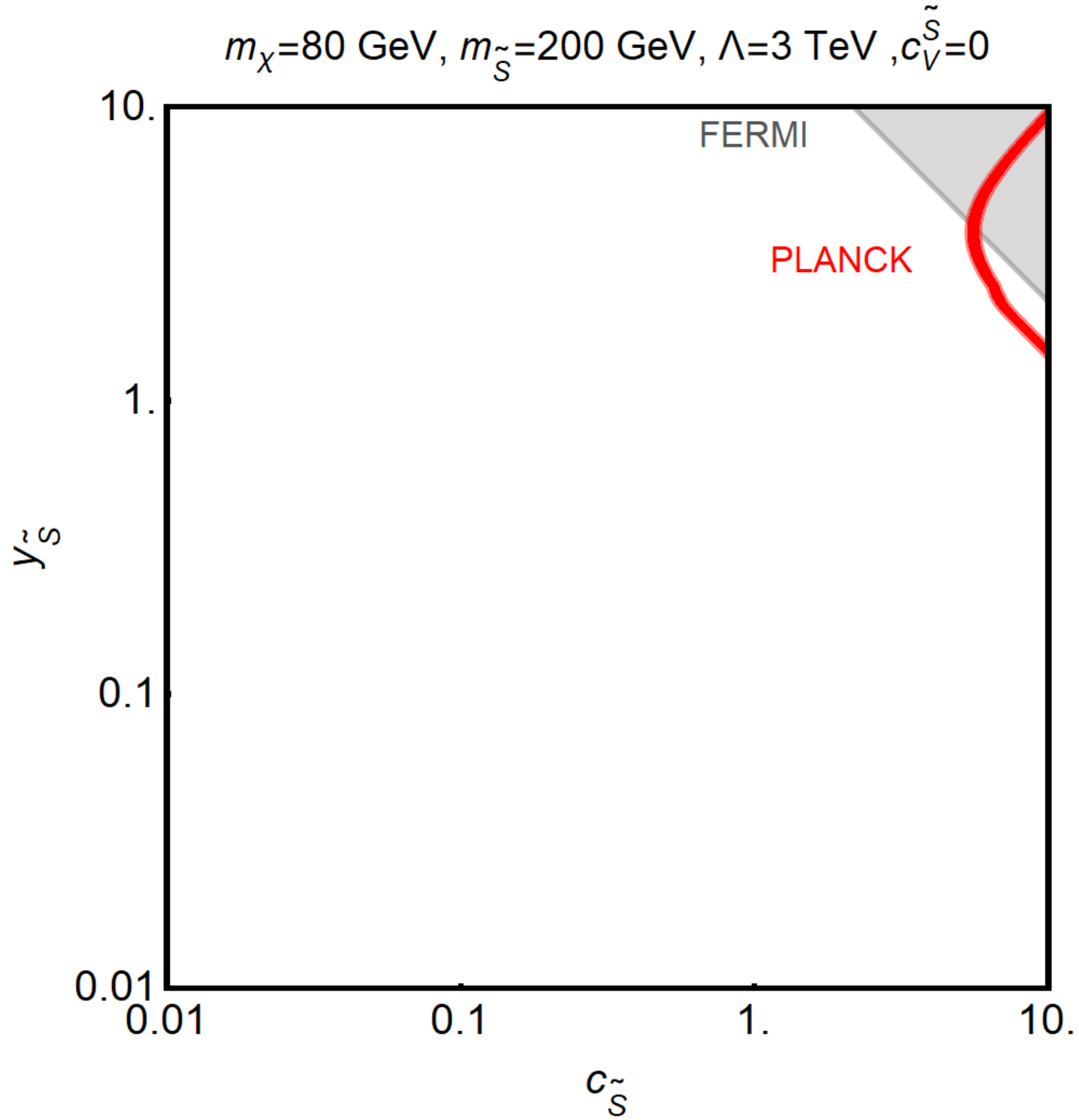}}
    \subfloat{\includegraphics[width=0.4\linewidth]{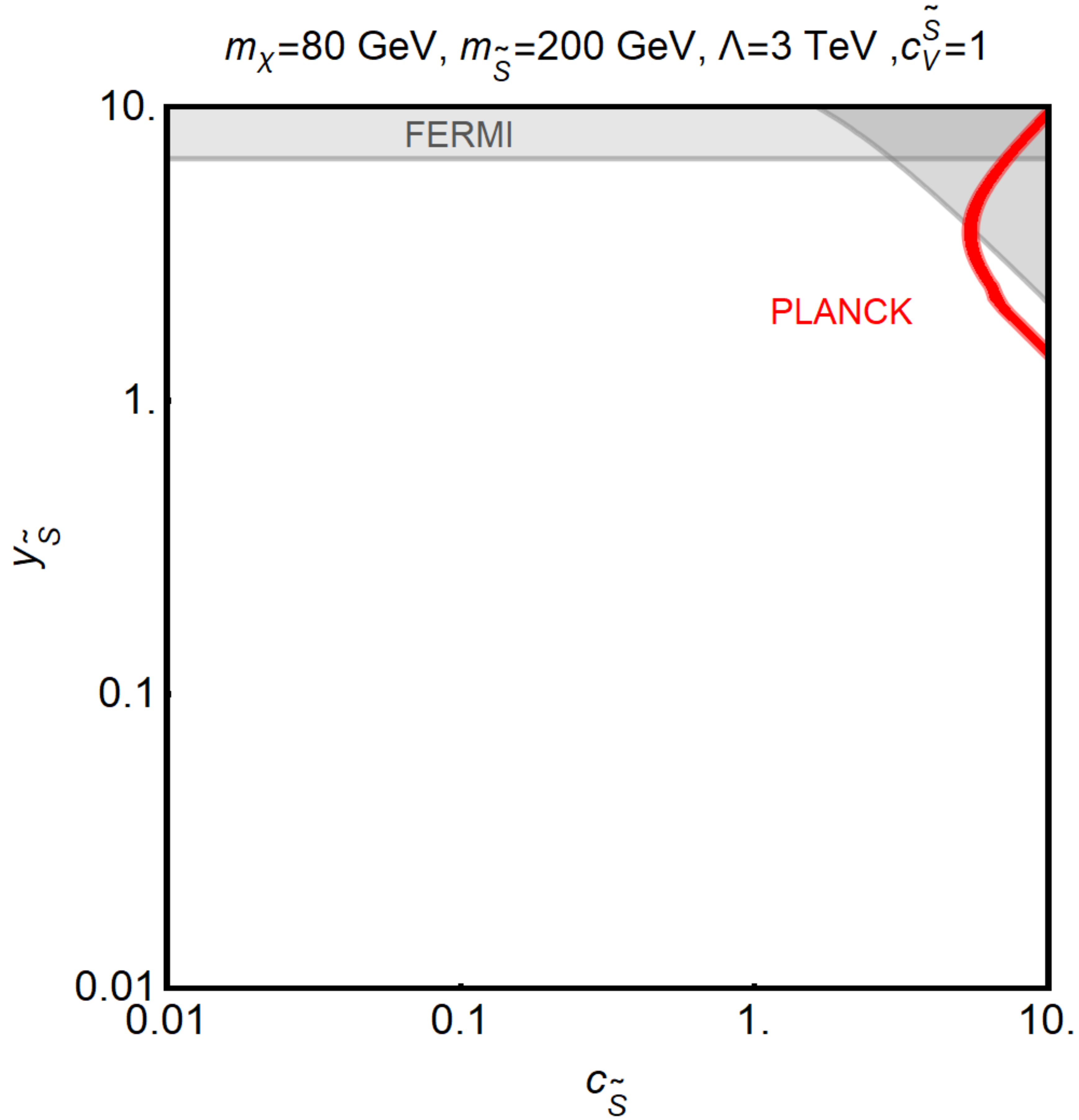}}
    \subfloat{\includegraphics[width=0.4\linewidth]{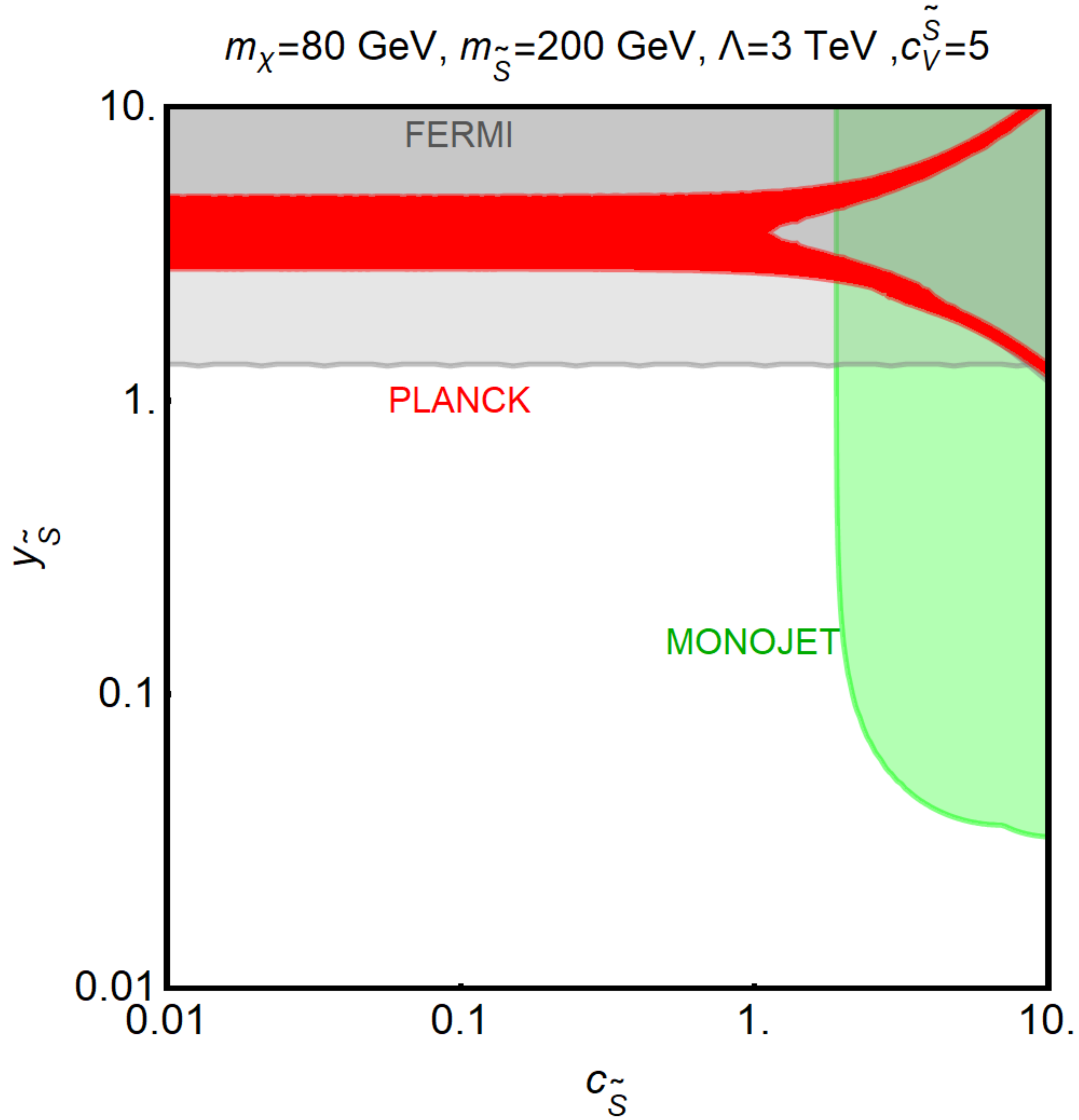}}\\
    \hspace*{-1.8cm}\subfloat{\includegraphics[width=0.4\linewidth]{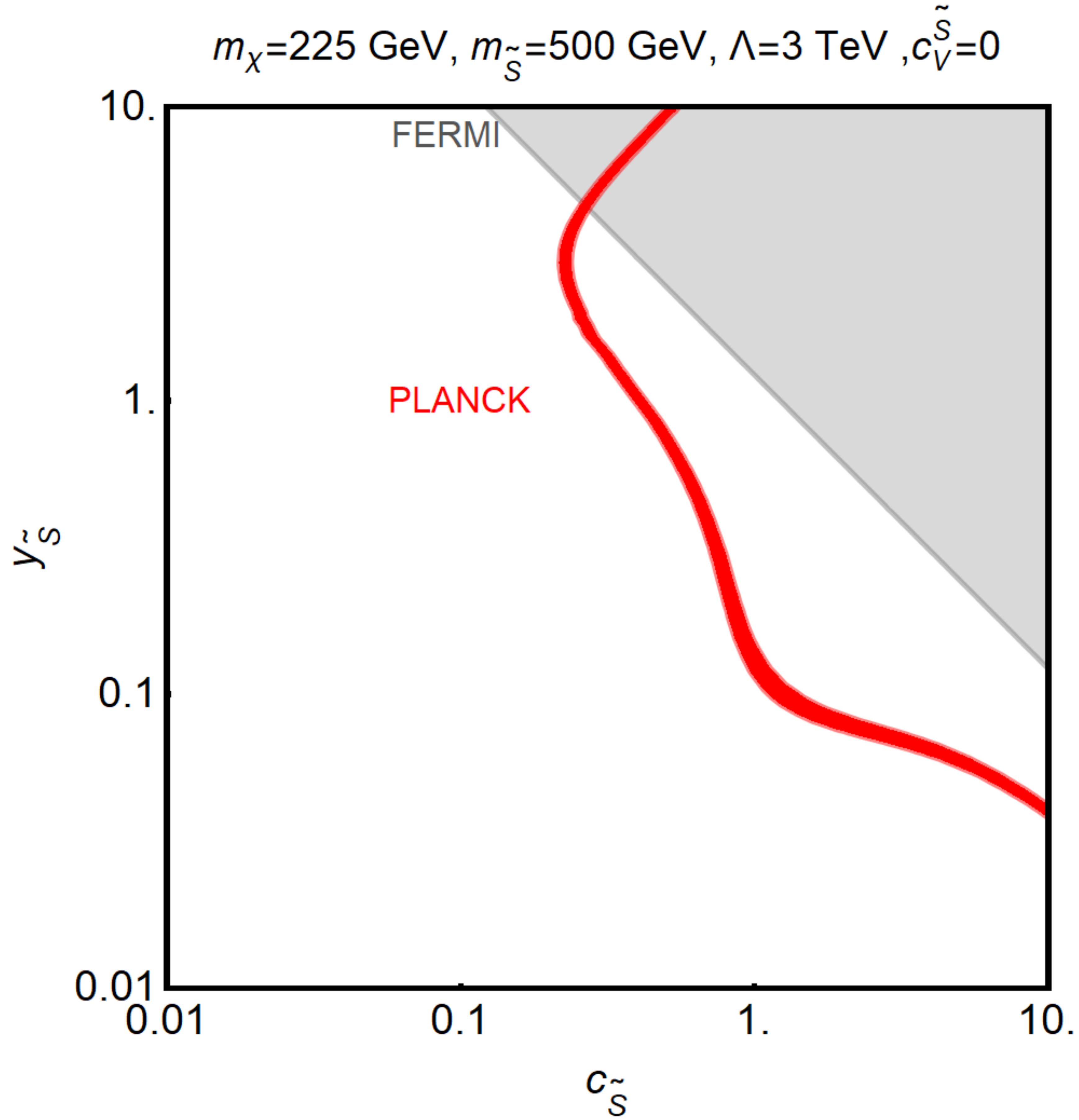}}
    \subfloat{\includegraphics[width=0.4\linewidth]{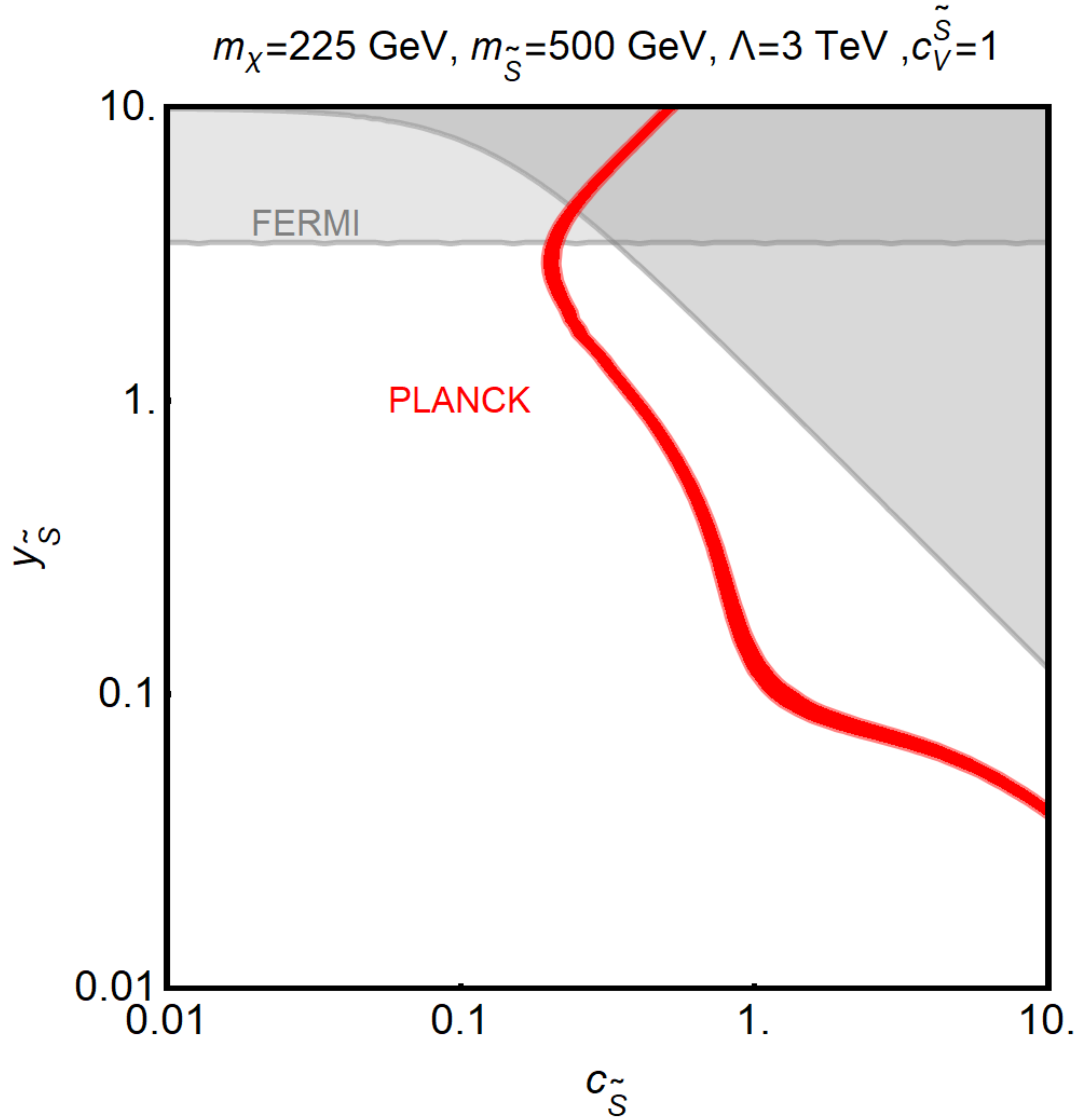}}
    \subfloat{\includegraphics[width=0.4\linewidth]{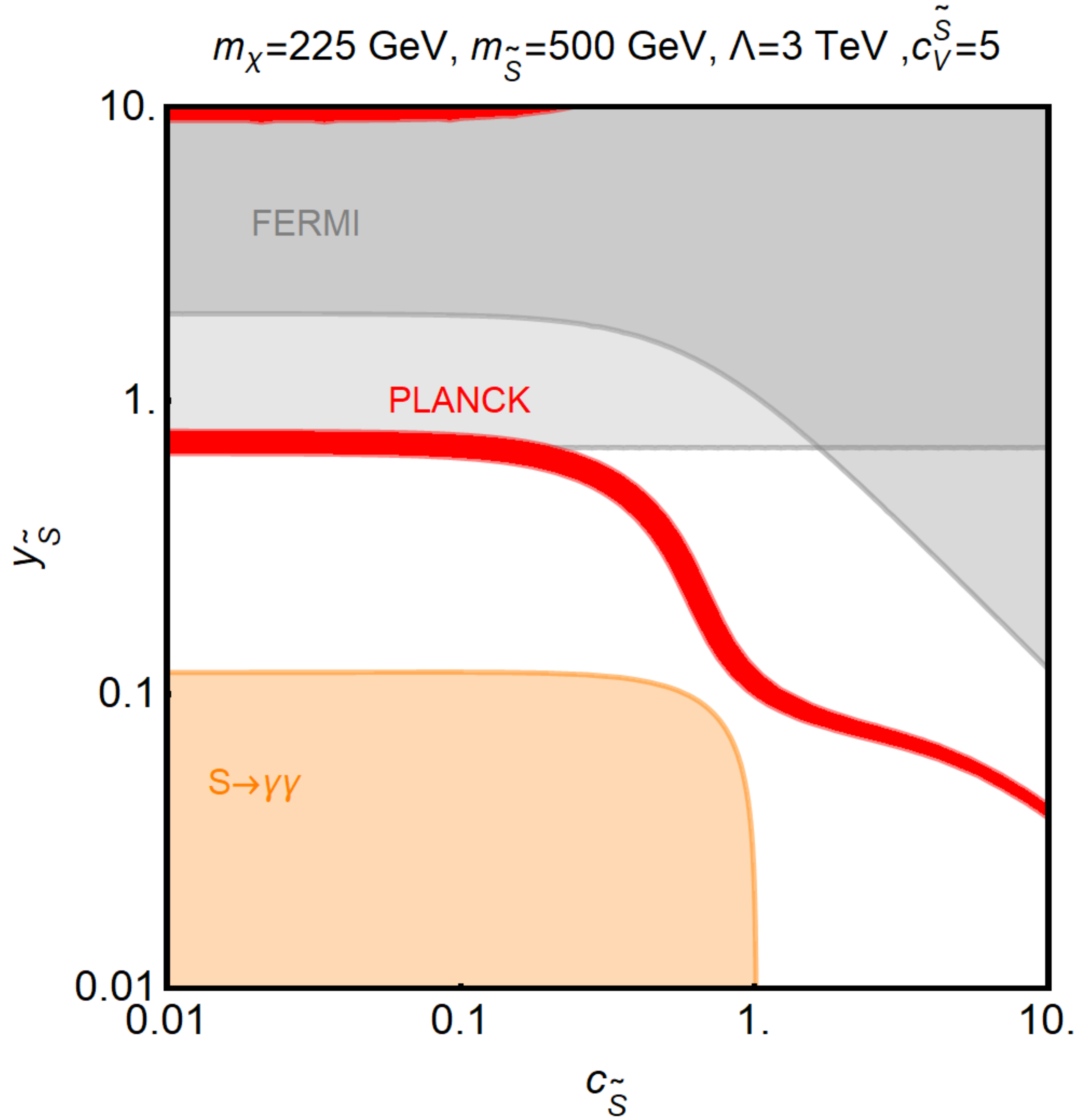}}\\
    \hspace*{-1.8cm}\subfloat{\includegraphics[width=0.4\linewidth]{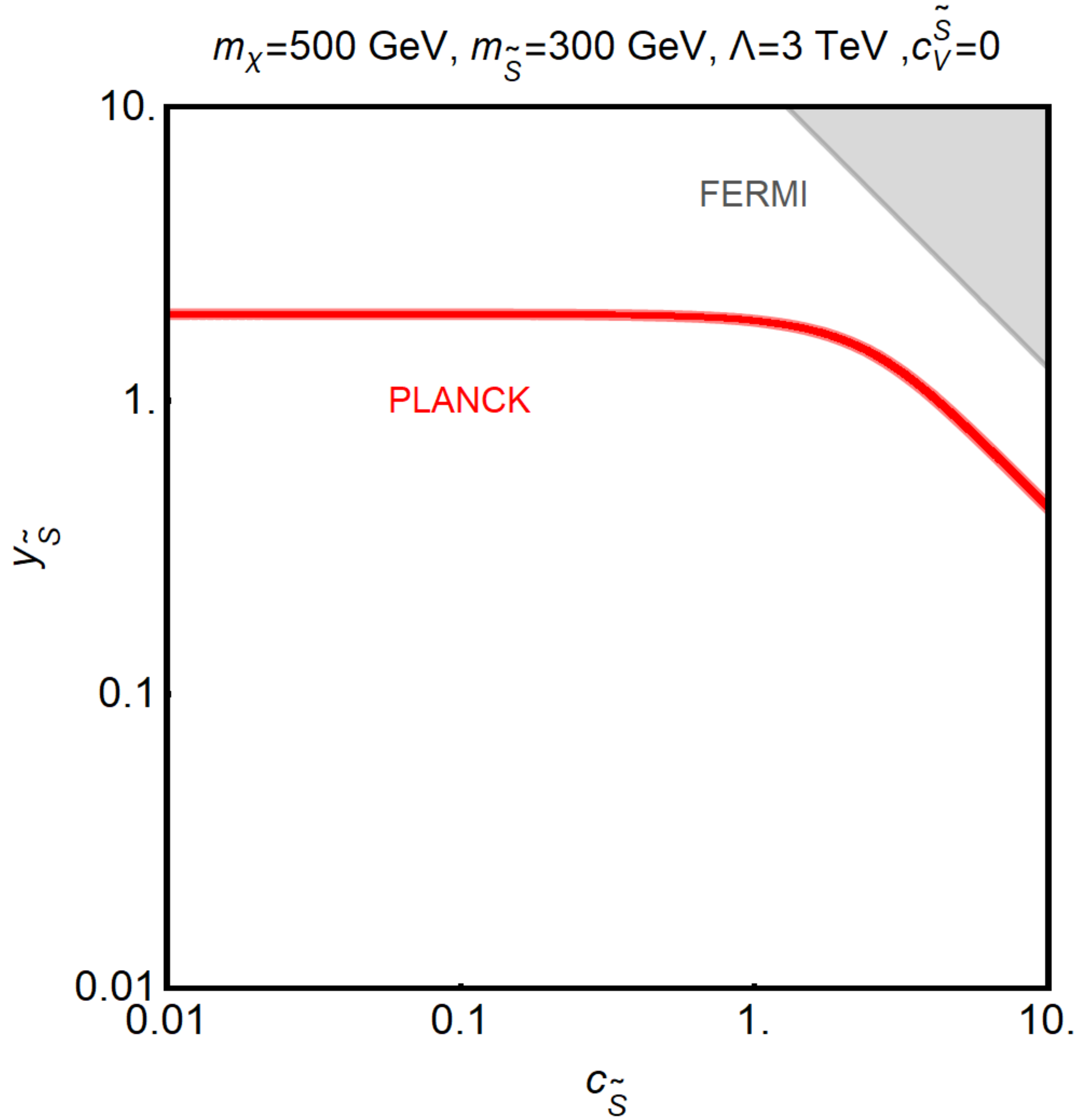}}
    \subfloat{\includegraphics[width=0.4\linewidth]{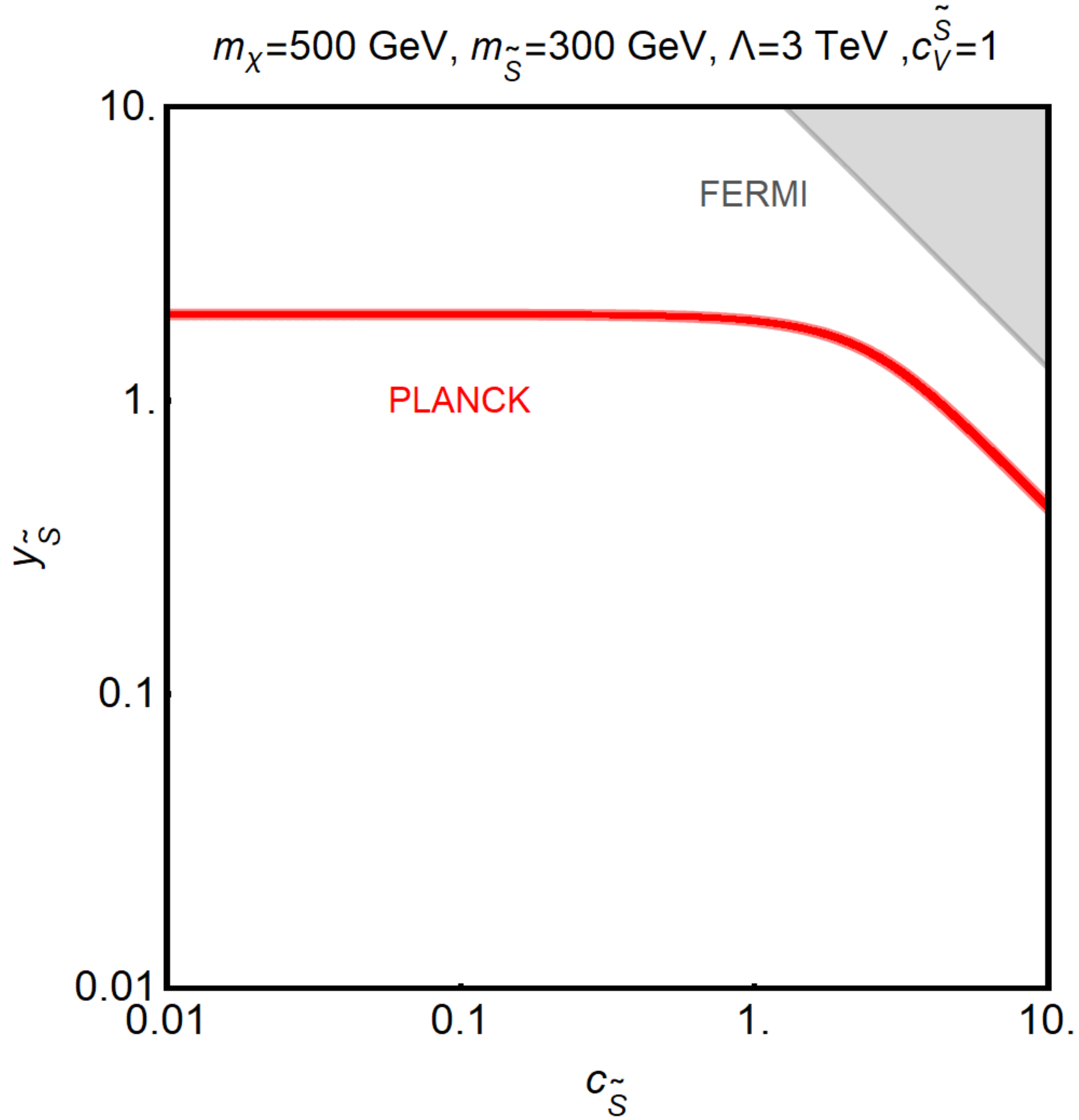}}
    \subfloat{\includegraphics[width=0.4\linewidth]{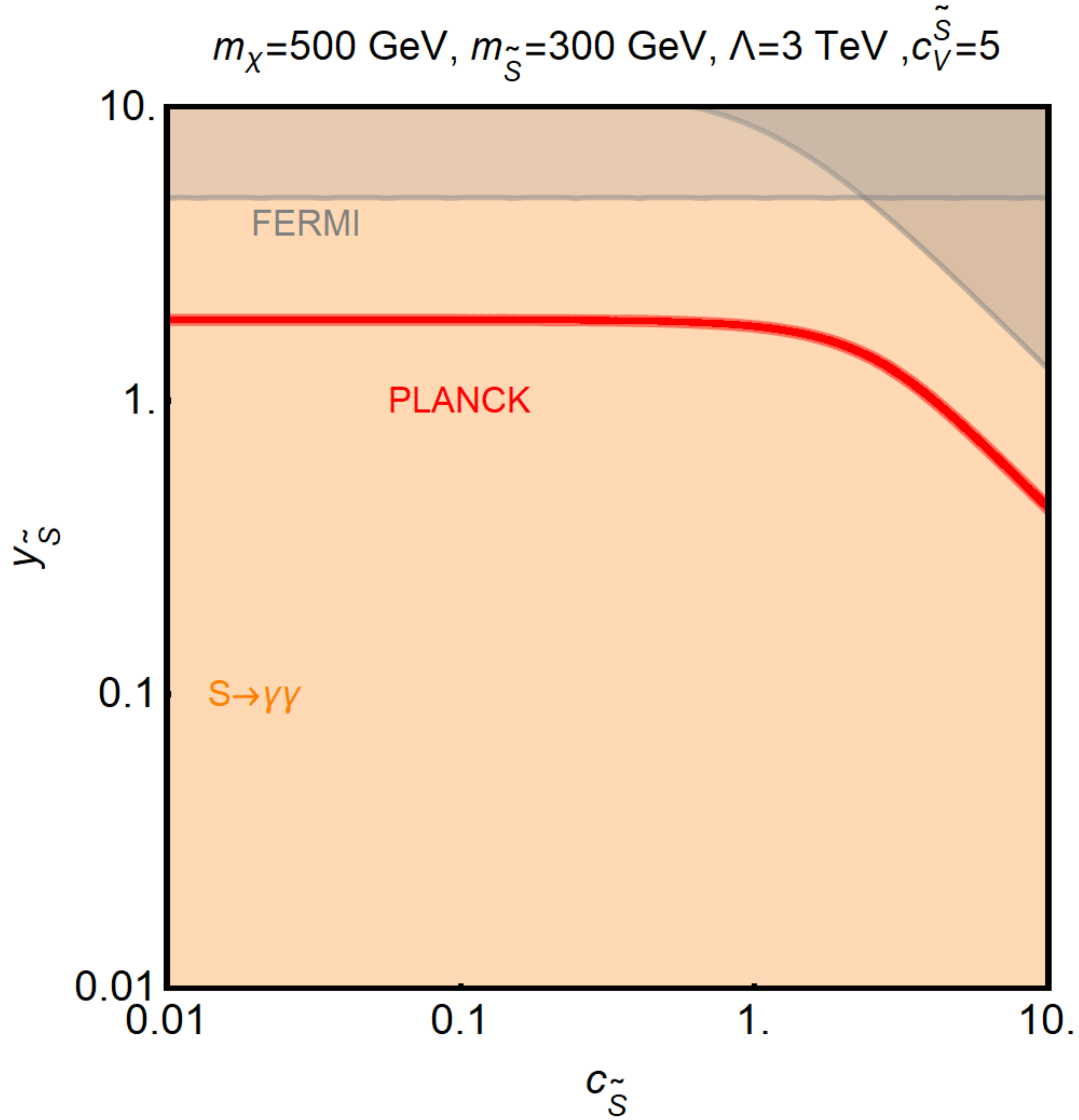}}
    \end{flushleft}
    \caption{\footnotesize{Summary of constraints in the $c_{\tilde{S}}-y_{\tilde{S}}$ plane for $c_V^{\tilde{S}}=0$  (left column), $c_V^{\tilde{S}}=1$ (center column), and $c_V^{\tilde{S}}=5$ (right column) and three different mass assignations as indicated in the individual panels. The red contours correspond to the correct relic density while the gray regions are excluded by limits from ID. Orange regions are excluded by collider searches for resonances decaying into photons and green region by mono-jet searches.}}
    \label{fig:pseudo_2D}
\end{figure}

\subsubsection{Combining portals}

As the next step we consider the simultaneous presence of the two aforementioned portals. More specifically we assume that both couplings $c_V^{\tilde{S}}$ and $c_{\tilde{S}}$ are different from zero while keeping $\lambda_{H\tilde{S}}=0$.
We will consider the same benchmarks for the DM and mediator masses as the for the scalar mediator, i.e. $(m_\chi,m_{\tilde{S}})=(80,200)\ {\rm GeV},\ (225,500)\ {\rm GeV},$ $(500,300)\ {\rm GeV}$, and show the combined constraints from DM and collider phenomenology in fig.~\ref{fig:pseudo_2D} considering the $(c_{\tilde{S}},y_{\tilde{S}})$ plane for three assignations of $c_V^{\tilde{S}}=0,\,1,\,5$.

\FloatBarrier
In each panel, the red isocontours represent the correct relic density,  while the orange regions are excluded by searches for diphoton resonances. The absence of mixing between the Higgs and the mediator reduces the decay branching fraction of the latter into massive gauge bosons so that we found no appreciable constraints from searches of diboson resonances for the considered benchmarks, while for the 200\,GeV mediator mono-jet searches (green) can exclude large values of $c_{\tilde{S}}$ and $c_V^{\tilde{S}}$.
The exclusions from ID are shown as gray regions. The panels display distinct regions corresponding to bounds from $\gamma$-ray continuum and $\gamma$-ray lines, respectively. Line signals depend only on  $c_V^{\tilde{S}}$ (and $y_{\tilde{S}}$), hence they appear as horizontal bands in the plots. As expected from  the discussion of the previous section, DD has no impact on benchmarks with $\lambda_{H\tilde{S}}=0$. The interplay of the different operators is most evident in the relic density contours.
The two benchmarks with $m_\chi \leq m_{\tilde{S}}$ show a substantial change of the relic density isocontours when $c_V^{\tilde{S}}\neq 0$. For the third benchmark, i.e. the one in the secluded regime, the relic density lines are mostly determined by the value of the $y_{\tilde{S}}$ coupling, indicating that they are primarily fixed by the annihilation into $\tilde{S}\tilde{S}$ pairs. Contrary to the case of the scalar mediator, we see however a change for $c_{\tilde{S}}>1$. In this regime, annihilations into fermion pairs contribute significantly to the relic density since this channel is not velocity suppressed for a pseudoscalar mediator.

\subsubsection{Scanning the parameter space}

\begin{figure}[t]
    \begin{flushleft}
    \hspace*{-1.8cm} \subfloat{\includegraphics[width=0.4\linewidth]{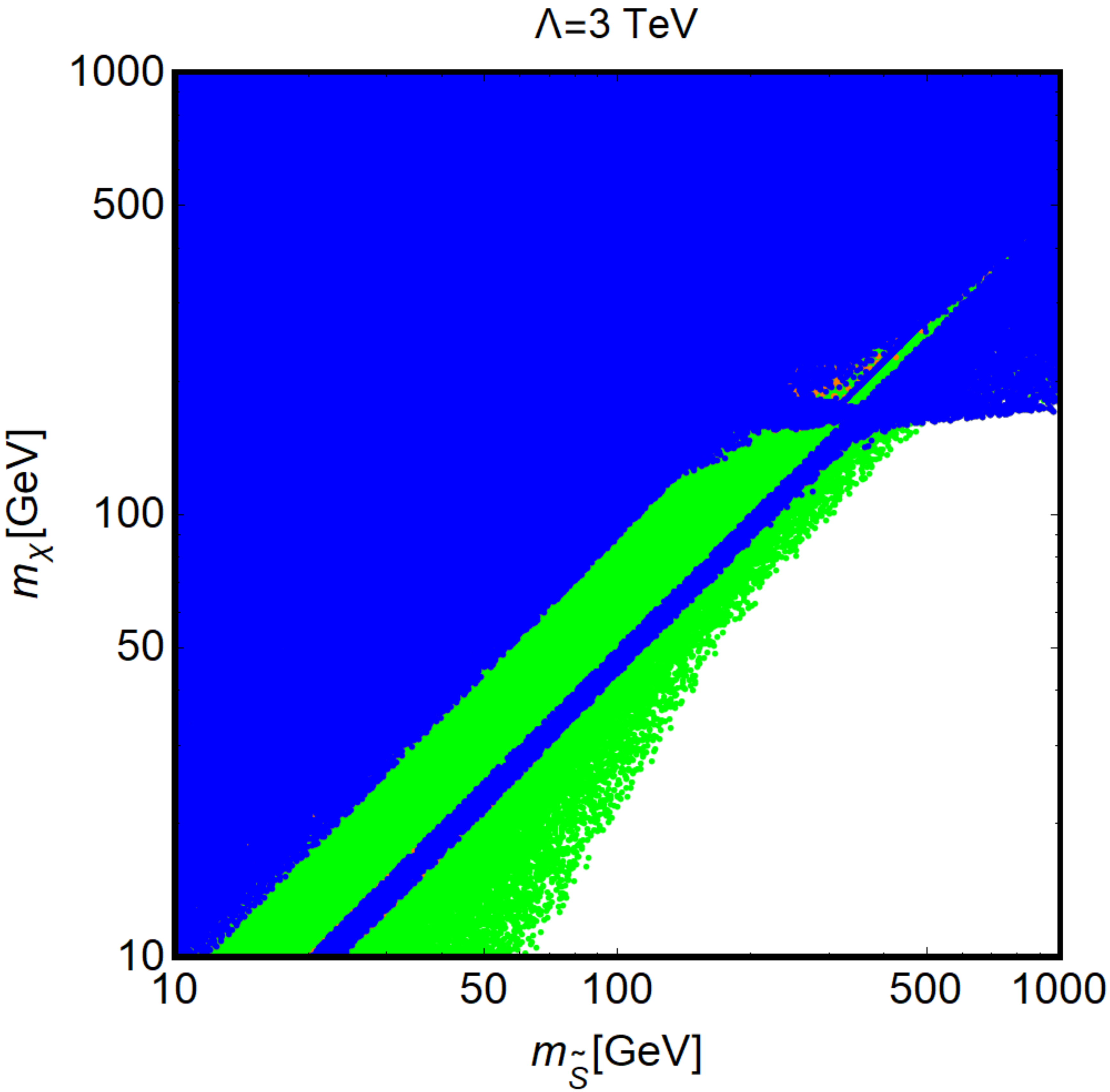}}
    \subfloat{\includegraphics[width=0.4\linewidth]{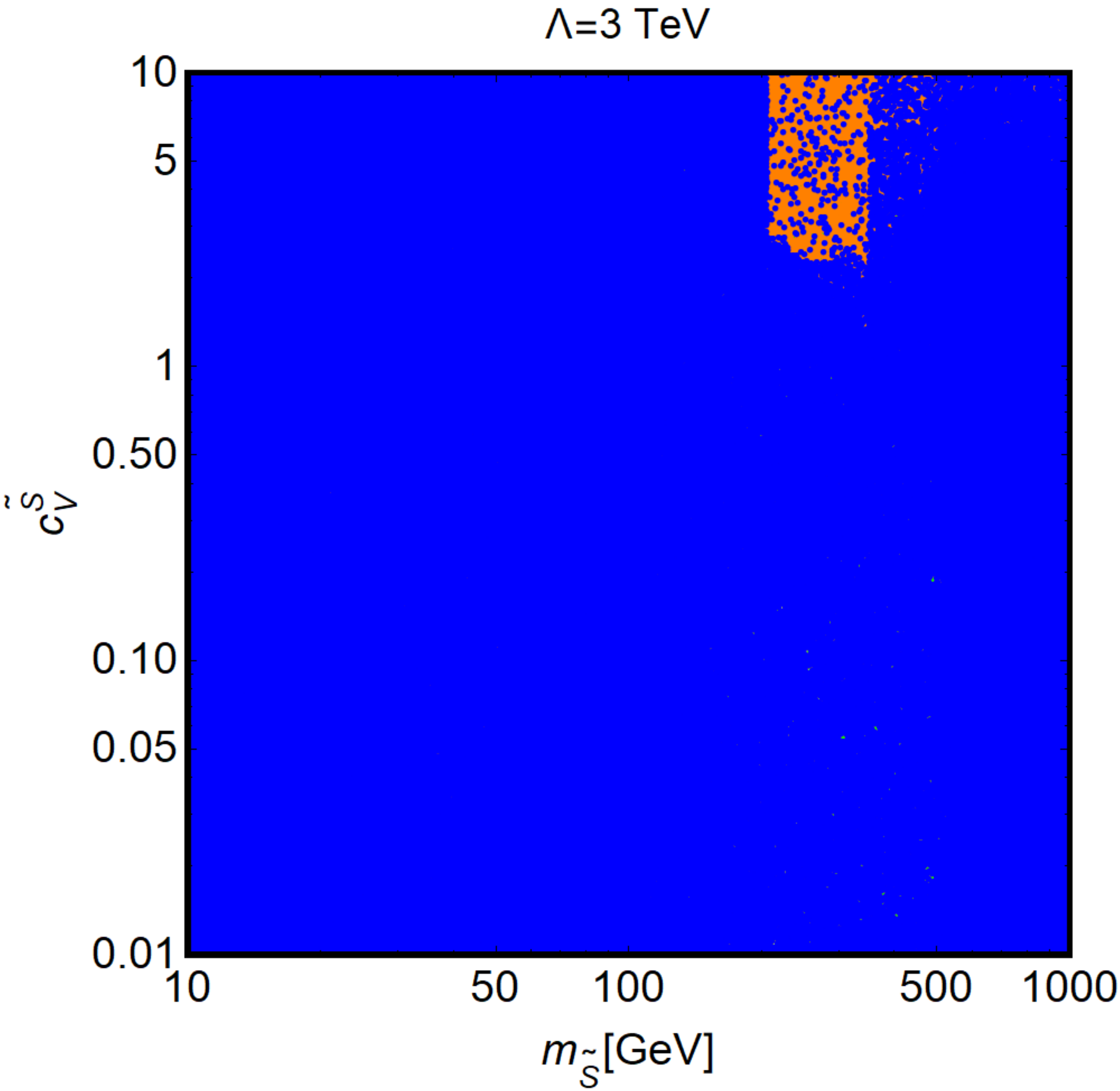}}
    \subfloat{\includegraphics[width=0.4\linewidth]{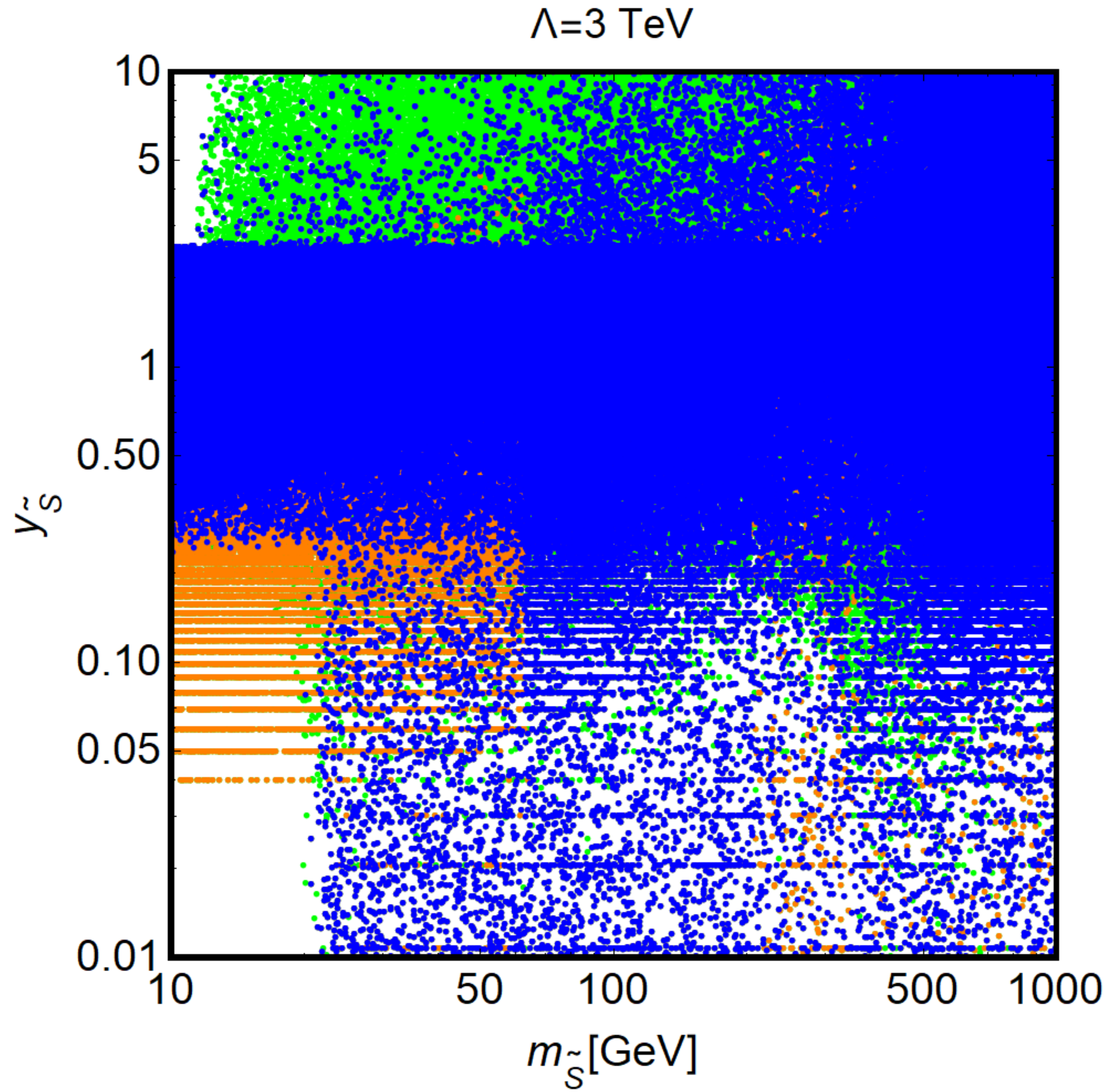}}\\
    \hspace*{-1.8cm} 
    \subfloat{\includegraphics[width=0.4\linewidth]{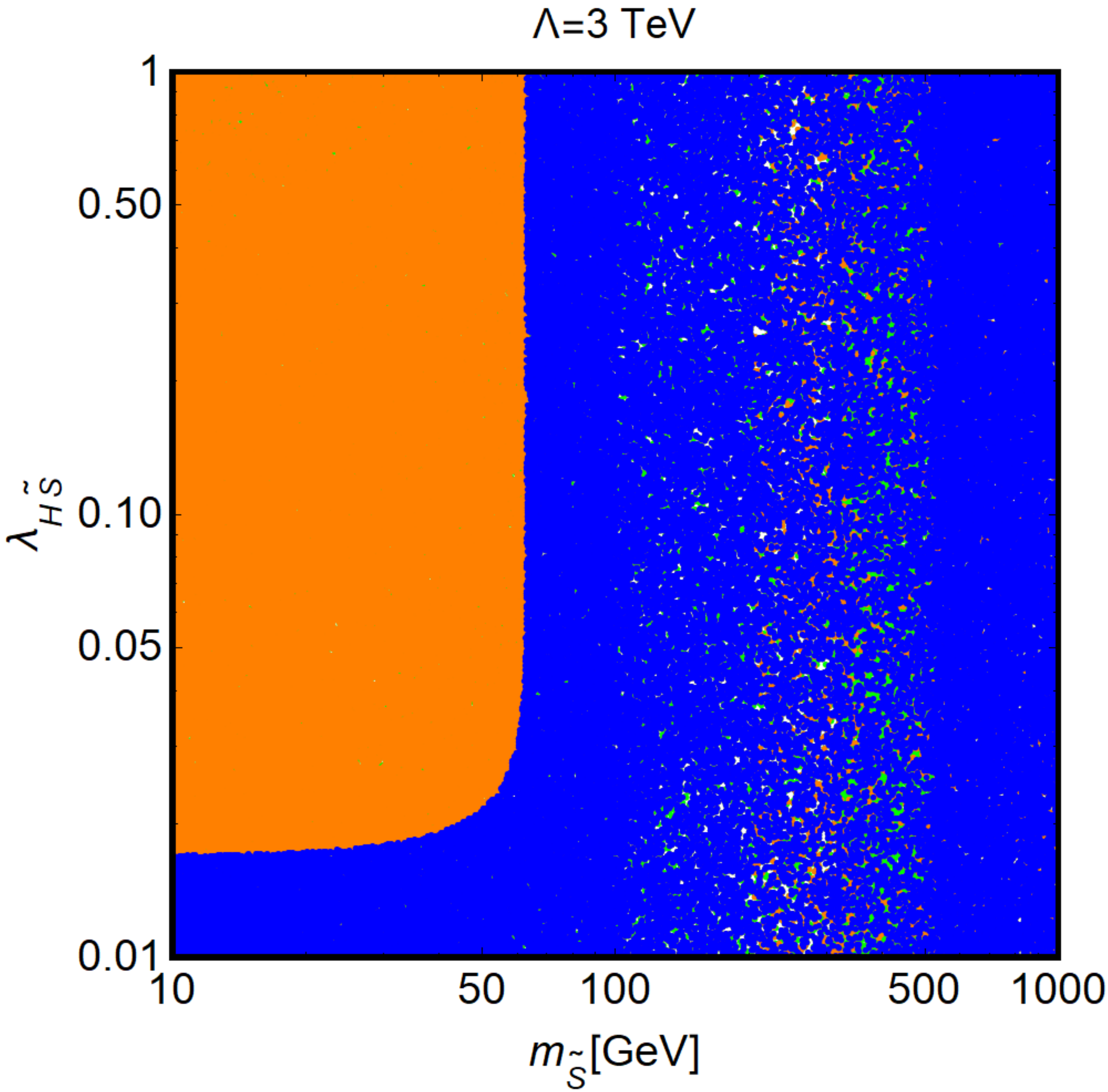}}
    \subfloat{\includegraphics[width=0.4\linewidth]{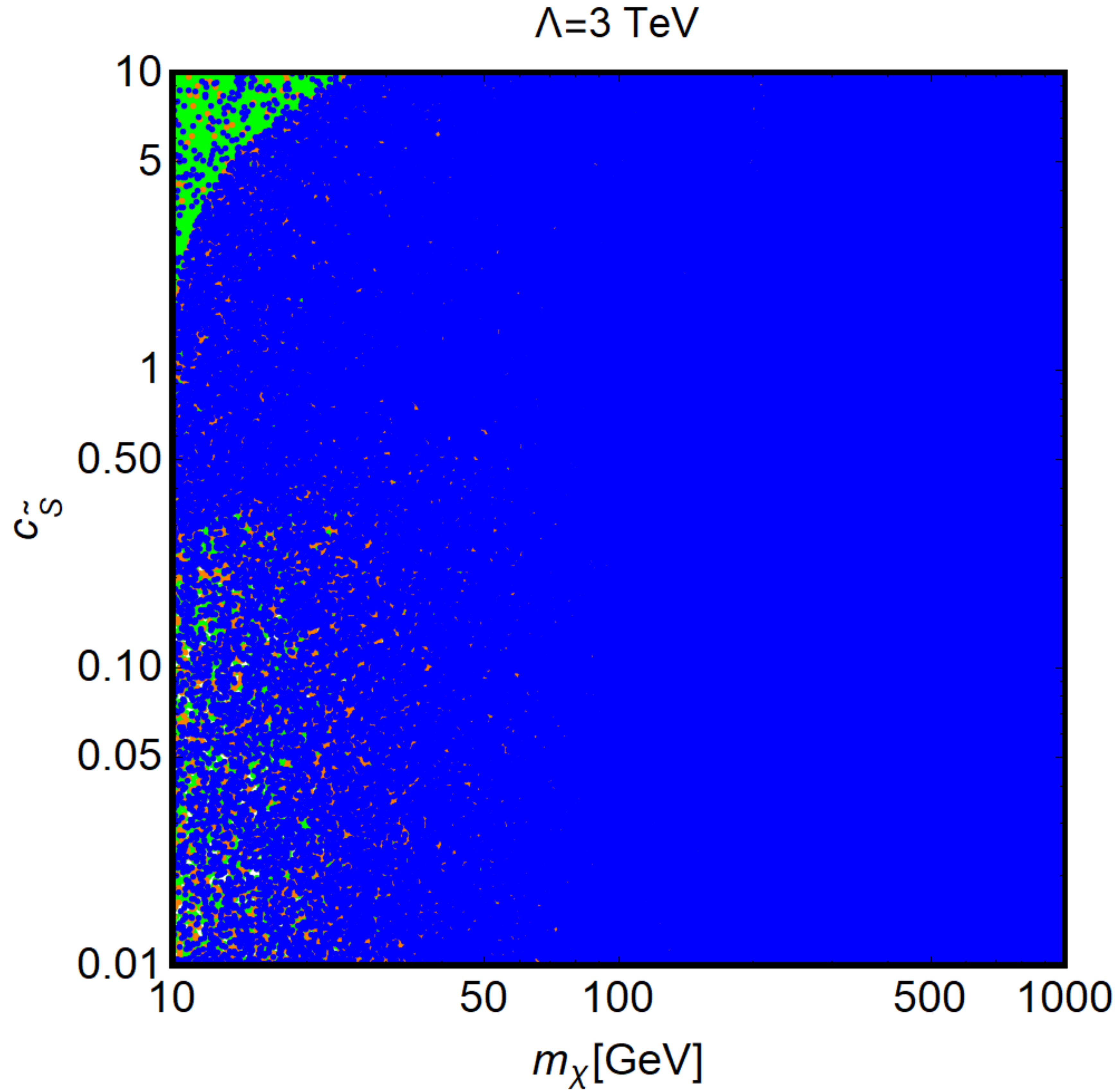}}
    \subfloat{\includegraphics[width=0.4\linewidth]{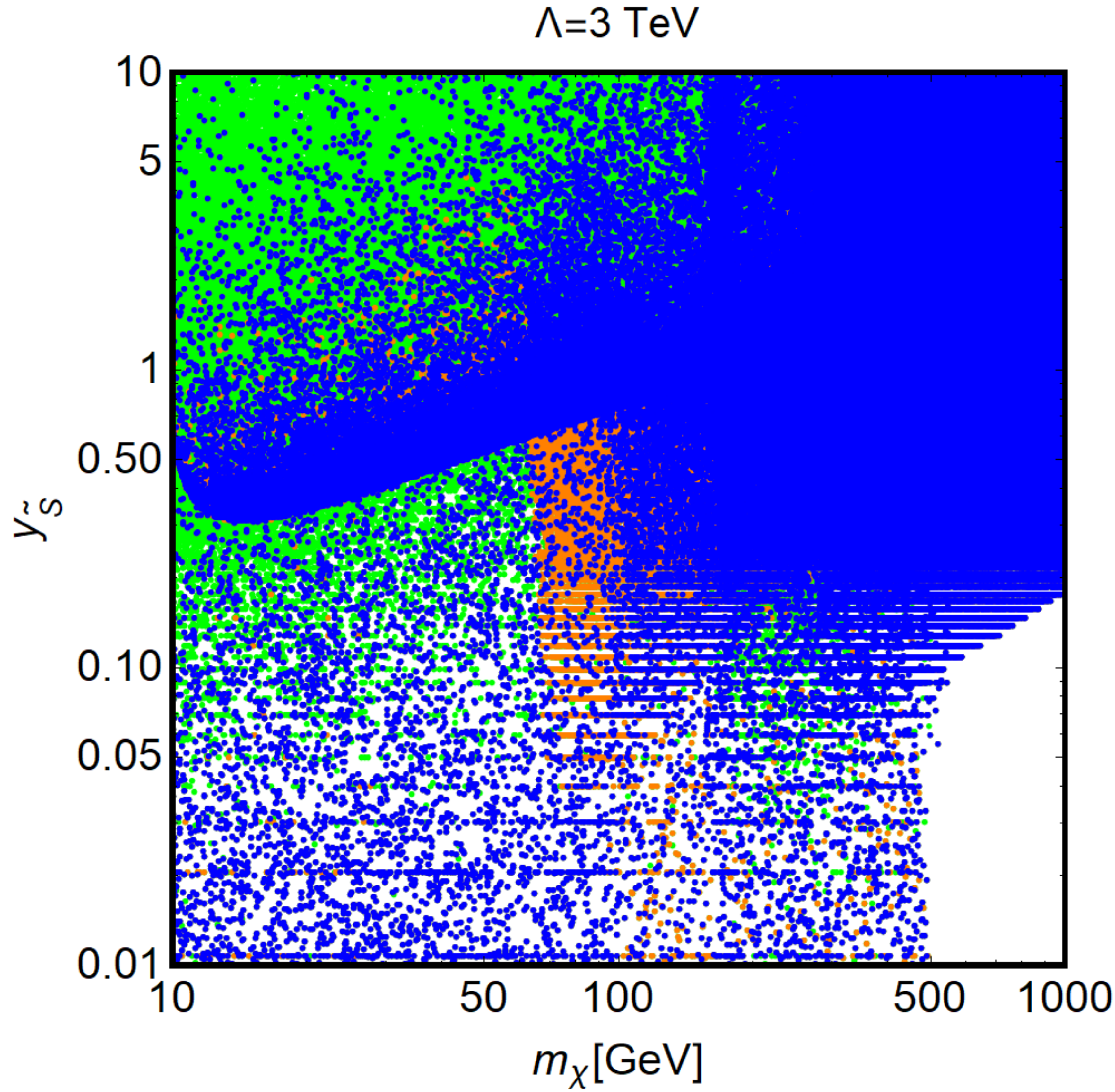}}
    \end{flushleft}
    \caption{\footnotesize{
    Results of the parameter scan for the pseudoscalar \eDMEFT\ setup in various planes, analogous to the scalar case presented in figs.~\ref{fig:scan1} and \ref{fig:scan1_bis}, see text for details. Contrary to the case of the scalar mediator we have shown the $(m_\chi,c_{\tilde{S}})$ plane, rather than the $(m_{\tilde{S}},c_{\tilde{S}})$ plane since the latter is not impacted by the constraints considered in our study.}}
    \label{fig:scan_pseudo}
\end{figure}

We can now conclude our survey with a general parameter-space scan. Following the analysis of the scalar case, we consider the six free parameters $(m_\chi,\, m_{\tilde{S}},\, y_{\tilde{S}},\, \lambda_{H\tilde S},\, c_{\tilde{S}},\, c_{V}^{\tilde{S}})$, and vary them within the ranges
    \vspace{-0.4cm}
\begin{align}
     m_\chi &\in \left[10,1000\right]\mbox{GeV} \nonumber\\
    m_{\tilde{S}} &\in \left[10,1000\right]\mbox{GeV} \nonumber\\
    \lambda_{H\tilde{S}} &\in \left[10^{-2},1\right] \\
    y_{\tilde{S}} &\in \left[10^{-2},10\right] \nonumber \\
     c_{\tilde{S}} &\in \left[10^{-2},10\right] \nonumber\\
    c_V^{\tilde{S}} &\in \left[10^{-2},10\right]   \,,\nonumber
\end{align}
with again $\Lambda=3$\,TeV. The corresponding model points are shown in fig.~\ref{fig:scan_pseudo} and distinguished through a similar color code as in the scalar case, namely:
\begin{itemize}
    \item {\it green points}: account only for the observed relic density;
    \item {\it orange points}: comply with DD and ID but are excluded by colliders;
    \item {\it blue points}: pass all the applied constraints.
\end{itemize}

Various slices in the higher-dimensional parameter space considered in our scan are shown in fig.~\ref{fig:scan_pseudo}. Focusing on the first plot, i.e. the $m_{\tilde{S}}-m_\chi$ plane, we notice a broader viable parameter space with respect to the case of a scalar mediator. 
While for $m_\chi \lesssim 150\,\mbox{GeV}$, the $m_\chi \sim m_{\tilde S}/2$ and the secluded regime are the only viable regions, points compatible with all the considered constraints are present also for ${m_\chi < m_{\tilde S}/2}$ for higher $m_\chi$. This feature, which is absent in the case of scalar mediator, is due to the much weaker DD limits. ID is putting relevant constraints on the parameter space but its impact is localized at small DM masses since current experimental sensitivity reaches at most $m_\chi \approx 150\,\mbox{GeV}$. As can be seen in the second and fourth plot of fig.~\ref{fig:scan_pseudo}, collider searches impact a small region of parameter space, in particular for $c_V^{\tilde{S}} > 1$ and $m_{\tilde{S}} \gtrsim 100\,\mbox{GeV}$ and for $m_{\tilde{S}} \leq m_h/2$ and $\lambda_{H\tilde{S}} \gtrsim 0.02$. In this last region the exclusion bound stems from the searches for the $h \rightarrow \tilde{S}\tilde{S}$ decay. Moving to the $y_{\tilde{S}}$ and $c_{\tilde{S}}$ parameters, we see from fig.~\ref{fig:scan_pseudo} that these are weakly constrained, with current exclusions of $y_{\tilde{S}},c_{\tilde{S}} \gtrsim 1$ for low values of $m_\chi,m_{\tilde{S}}$. From the lower panel of fig.~\ref{fig:scan_pseudo} we finally notice the absence of model points for $y_{\tilde{S}}\lesssim 0.1$ and $m_\chi \geq 500\,\mbox{GeV}$. With the $y_{\tilde{S}}^{(2)}$ and $y_H^{(2)}$ couplings  set to zero, the only way to achieve a viable DM phenomenology at small $y_{\tilde{S}}$ is via the $m_\chi \sim m_{\tilde{S}}/2$ pole. As the considered parameter space is limited to $m_{\tilde{S}}\leq 1$ TeV, the resonant regime is not included in our analysis for $m_\chi \geq 500\,\mbox{GeV}$.

\FloatBarrier

\section{Conclusions}
\label{sec:Conc}

The search for dark matter is one of the most important tasks in high energy physics today. This is reflected by the large number of experiments that probe various aspects of DM physics. Combining the data accumulated by the ongoing experimental efforts is challenging and a versatile framework that allows for a consistent theoretical interpretation is called for. The \eDMEFT, which combines effective field theory with the simplified model approach provides such an analysis tool.
In this article, we have performed a comprehensive survey of the phenomenology of the \eDMEFT\ with scalar and pseudoscalar mediator taking into account constraints
from the measured DM abundance,  
direct- and indirect-detection bounds, as well as the most relevant limits from collider searches. After presenting analytical and numerical results for the various annihilation cross sections and the DM-nucleon scattering in the presence of $D=5$ operators, we turned to an analysis of missing energy signatures at the LHC and discussed searches for scalar resonances decaying to vector-boson, di-jet, and Higgs-pair final states. Afterwards we approached a survey of the full EFT parameter space, first exploring a set of minimal portal scenarios that are realized within the \eDMEFT, before turning to interference effects between various operators. Interestingly, these allow for 
cancellations in the DD cross section that lead to allowed `blind-spots' for scalar mediators that could be missed in a naive simplified-model approach. Finally, we delivered comprehensive scans including all the essential operators, that show which parameter-space regions survive the constraints from dark matter phenomenology and from collider searches, pointing out new viable regions emerging in the \eDMEFT. In particular, we demonstrated how future XENONnT limits could corner conventional scalar portals to the dark sector, while in the \eDMEFT\ larger parts in different mass regions in the $(m_S,m_\chi)$ plane remain open. We find that a significant part of this parameter space  could however be in reach of the high-luminosity LHC thus highlighting the complementarity between the different experimental search strategies.

We have repeated our analysis considering a CP-odd mediator. While the collider phenomenology is largely similar to the one of scalar mediator there are striking differences in DD and ID. The DD cross section of a DM candidate that interacts with the SM via a pseudoscalar particle is loop suppressed and, therefore, the experimental limits provide a much looser constraint in this case. In contrast, ID is much more sensitive to pseudoscalar mediators since the annihilation cross section into a number of relevant final states corresponds to $s$-wave. Consequently, this scenario is most constrained for $m_\chi \leq 100$ GeV while the parameter space for heavier DM is largely open. Future experimental data will be crucial to test this kind of scenario.

All in all, the \eDMEFT\  proved to be a very versatile tool for the analysis of a very broad range of experimental signals and shows promise as a flexible and easy to use interface between theory and experiment.

\section*{Acknowledgments}

We are grateful to Giorgio Busoni and Thomas Hugle for useful discussions. VT acknowledges support by the IMPRS-PTFS.

\FloatBarrier

\bibliographystyle{JHEP}
\bibliography{eDMEFTPheno}

\end{document}